\documentclass[11pt,twoside,openright]{book}
\pdfoutput=1
\usepackage{appendix}
\usepackage{widebar}
\usepackage[colorlinks=true,hyperfootnotes=false]{hyperref}
\usepackage{morewrites}
\usepackage[paperwidth=8.5in, paperheight=11in,bindingoffset=.75in]{geometry}

\usepackage[english]{babel}
\usepackage[exerciseaslist=true,copyexercisesinsolutions=true,exercisesfontsize=small]{exsol}
\usepackage{amsmath}
\usepackage{amssymb}
\usepackage{slashed}
\usepackage{array}

\usepackage{tikz}
\usetikzlibrary{arrows, intersections,calc,shapes}
\usetikzlibrary{positioning}
\usetikzlibrary{decorations.pathmorphing}
\usetikzlibrary{decorations.markings}

\tikzset{
particle/.style={thick,draw=black, postaction={decorate},
    decoration={markings,mark=at position .7 with {\arrow[black]{triangle 45}}}},
photon/.style={decorate, draw=black,
    decoration={coil,aspect=0}},
gluon/.style={decorate, draw=black,
    decoration={coil,aspect=0.3,segment length=5pt,amplitude=3pt}}
}

\usepackage[shortlabels]{enumitem} 
\usepackage{mathtools} 
\usepackage{float} 
\usepackage[stable]{footmisc} 

\makeatletter
\newsavebox\myboxA
\newsavebox\myboxB
\newlength\mylenA
\newcommand*\xoverline[2][0.75]{%
    \sbox{\myboxA}{$\m@th#2$}%
    \setbox\myboxB\null
    \ht\myboxB=\ht\myboxA%
    \dp\myboxB=\dp\myboxA%
    \wd\myboxB=#1\wd\myboxA
    \sbox\myboxB{$\m@th\overline{\copy\myboxB}$}
    \setlength\mylenA{\the\wd\myboxA}
    \addtolength\mylenA{-\the\wd\myboxB}%
    \ifdim\wd\myboxB<\wd\myboxA%
       \rlap{\hskip 0.5\mylenA\usebox\myboxB}{\usebox\myboxA}%
    \else
        \hskip -0.5\mylenA\rlap{\usebox\myboxA}{\hskip 0.5\mylenA\usebox\myboxB}%
    \fi}
\makeatother

\newcommand\Htilde{\widetilde{H}}
\newcommand{\fracup}[2]{\frac{\raisebox{0.25ex}{$\displaystyle #1$}}{#2}}
\newcommand{\pvec}[1]{\vec{#1}\mkern2mu\vphantom{#1}}
\newcommand{\vev}[1]{\langle{#1}\rangle}

\title{Lectures on Flavor Physics and CP Violation}
\author{Benjam\'\i{}n Grinstein}
\date{April 2015} 

\begin{document}
\frontmatter 

\let\cleardoublepage\clearpage

\maketitle
\tableofcontents

\chapter{Preface}
I created this document in preparation for lectures I am to present at the 8th CERN Latin American School of High Energy Physics (CLASHEP) during the Winter (Summer?) of 2015. These lectures are intended for graduate students of experimental particle physics.  I aim at pedagogy, so don't look here for a complete list of topics, nor a complete set of references. Plainly, this document is not intended as a reference work.  It is not complete, but rather introductory.  My hope is that a physics student who has been exposed to the Standard Model of electroweak interactions will come out with an idea of why flavor physics remains one of the most vibrant areas in particle physics, both in theory and particularly in experiment. She or he will hopefully have an appreciation of the main aspects of the field and the crucial interconnections between theory and experiment that characterize it.

I started preparing this course as an adaptation of lectures I presented at TASI in 2013 and at Schladming in 2014. But because of the difference in scope and in audience I had to make major adjustments, definite choices on what to retain and what to omit. While some old hats may disagree with my choices, I am satisfied with the outcome and reasonably confident that the product will satisfy my customers.  Of course, the jury is out.  If you, the reader, happens to be one of those customers, I would really appreciate some feedback: email me, text me, call me, whatever (but beware, I don't Tweet). I hope to get invited to lecture somewhere again in the future, and your valuable opinion can help me improve as a lecturer.

Particle Physics has just entered an era of great excitement. You may not appreciate this if you live and work in the US,  as government funding of the discipline erodes there, but its palpable in  Physics departments of universities and laboratories  around the world. This bodes well for the future of the field. I need not explain why it is that much of the excitement is coming from CERN. But CERN has not only become the leading laboratory of high energy physics in the world, it has also taken a leadership role in education, at least in areas that pertain the lab's disciplines.   This makes sense. It is the youngsters of today that will be the researchers of that tomorrow. And these youngsters need training. The CLASHEP is but one of CERN's contribution to this effort. It gives students in Latin America a rare opportunity to study topics that are unlikely found in the curriculum at their institutions and to meet with other students from Latin America and researcher-instructors from around the world. 
 I feel privileged and honored  that I have been given the opportunity to present these lectures on Flavor Physics and CP Violation and hope that the writeup of these lectures can be of use to many current and future students that may not have the good fortune of attending a CLASHEP.

Being lectures, there are lots of exercises that go with these. The exercises are interspersed in
the material rather than collected at the end of chapters. The problems tend to expand or check on
one point and I think it's best for a student to solve the exercises in context.  I have many ideas
for additional exercises, but only limited time. I hope to add some more in time and keep an update
accesible on the web. Some day I will publish the solutions.  Some are already typed into the TeX
source and I hope to keep adding to it.\footnote{For the arXiv version I have
    \hyperlink{chap:sols}{included those solutions} bellow the Bibliography.}

No one is perfect and I am certainly far from it. I would appreciate
alert readers to send me any typos, errors or any other needed
corrections they may find. Suggestions for any kind of improvement are
welcome. I will be indebted if you'd  send them to me at bgrinstein@ucsd.edu

\bigskip

{\raggedleft Benjam\'\i{}n Grinstein\\
 San Diego, February 2015\\
}
\mainmatter
\chapter{Flavor Theory}

\section{Introduction: What/Why/How?}

\paragraph{WHAT:} There are six different types of quarks: $u$ (``up''), $d$ (``down''),\break
$s$ (``strange''), $c$ (``charm''), $b$ (``bottom'') and $t$
(``top''). Flavor physics is the study of different types of quarks,
or ``flavors,'' their spectrum and the transmutations among
them. More generally different types of leptons, ``lepton flavors,''
can also be included in this topic, but in this lectures we
concentrate on quarks and the  hadrons that contain them. 

\paragraph{WHY:} Flavor physics is very rich. You should have a copy of the \href{http://pdg.lbl.gov}{PDG},
or at least a bookmark to \href{http://pdg.lbl.gov}{pdg.lbl.gov} on your computer. A quick inspection
of the PDG reveals that a great majority of content gives transition
rates among hadrons with different quark content, mostly decay
rates. This is tre realm of flavor physics. We aim at understanding this wealth
of information in terms of some simple basic principles. That we may
be able to do this is striking endorsement of the validity of our theoretical model of
nature, and gives stringent constraints on any new model of nature you
may invent. Indeed, many models you may have heard about, in fact many
of the most popular models, like gauge mediated SUSY breaking and
walking technicolor, were invented to address the strong constraints
imposed by flavor physics. Moreover, all observed CP violation (CPV)
in nature is tied to flavor changing interactions, so understanding of
this fundamental phenomenon is the domain of flavor physics. 

\paragraph{HOW:} The richness of  flavor physics comes at a price: while flavor
transitions occur intrinsically at the quark level, we only observe
transitions among hadrons. Since quarks are bound in hadrons by the
strong interactions we face the problem of confronting theory with
experiment in the context of mathematical models that are not
immediately amenable to simple analysis, like perturbation
theory. Moreover, the physics of flavor more often than not involves
several disparate time (or energy) scales, making even dimensional analysis
somewhere between difficult and worthless. Many tools have been
developed to address these issues, and these lectures will touch on several
of them. Among these:
\begin{itemize}
\item Symmetries allow us to relate different processes and sometimes
  even to predict the absolute rate of a transition.
\item Effective Field Theory (EFT) allows to systematically
disentangle the effects of disparate scales. Fermi theory is an EFT
for electroweak interactions at low energies. Chiral Lagrangians
encapsulate the information of symmetry relations  of  transitions
among pseudo-Goldstone bosons. Heavy Quark Effective Theory (HQET)
disentangles the scales associated with the masses of heavy quarks
from the scale associated with hadron dynamics and makes explicit spin
and heavy-flavor symmetries. And so on.
\item Monte-Carlo simulations of strongly interacting quantum field theories on the lattice
can be used to compute some quantities of basic interest that cannot be computed using perturbation theory.  
\end{itemize}

\section{Flavor in the Standard Model}
Since the Standard Model of Strong and Electroweak interactions (SM) works so well, we will adopt it as our standard (no pun intended) paradigm. All alternative theories that are presently studied build on the SM; we refer to them collectively as Beyond the SM (BSM). Basing our discussion on the SM is very useful:
\begin{itemize}
\item It will allow us to introduce concretely the methods used to think about and quantitatively analyze Flavor physics. It should be straightforward to extend the  techniques introduced in the context of the SM to specific BSM models. 
\item Only to the extent that we can make precise calculations in the SM and confront them with comparably precise experimental results can we meaningfully study effects of other (BSM) models. 
\end{itemize}
So let's review the SM. At the very least, this allows us to agree on
notation. The SM is a gauge theory, with gauge group $SU(3)\times
SU(2) \times U(1)$. The $SU(3)$ factor models the strong interactions of ``colored'' quarks and gluons,
$SU(2)\times U(1)$ is the famous Glashow-Weinberg-Salam model of the
electroweak interactions. Sometimes we will refer to these as
$SU(3)_\text{c}$ and $SU(2)_\text{W}\times U(1)_\text{Y}$ to
distinguish them from other physical transformations characterized by the same mathematical groups. The matter content of the model consists of
color triplet quarks: left handed spinor doublets $q^i_L$ with $U(1)$
``hypercharge'' $Y=1/6$ and  right handed spinor singlets $u^i_R$ and
$d^i_R$ with $Y=2/3$ and $Y=-1/3$. The color ($SU(3)$),
weak ($SU(2)$), and Lorentz-transformation indices are implicit. The
``$i$'' index runs over $i=1,2,3$ accounting for three copies, or ``generations.'' A more concise description is $q^i_L=(3,2)_{1/6}$, meaning that $q^i_L$ transforms as a $\mathbf{3}$ under $SU(3)$, a $\mathbf{2}$ under $SU(2)$ and has $Y=1/6$ (the $U(1)$ charge). Similarly, $u^i_R=(3,1)_{2/3}$ and $d^i_R=(3,1)_{-1/3}$. The leptons are color singlets: $\ell^i_L=(1,2)_{-1/2}$ and $e^i_R=(1,1)_{-1}$. 

We give names to the quarks in different generations:
\begin{equation}
q^i_L=\left(\begin{pmatrix}u_L\\d_L\end{pmatrix}, \begin{pmatrix}c_L\\s_L\end{pmatrix}, \begin{pmatrix}t_L\\b_L\end{pmatrix}\right), \qquad u^i_R=(u_R, c_R, t_R), \qquad d^i_R=(d_R,s_R,b_R).
\end{equation}
Note that we have used the same symbols, ``$u$'' and ``$d$,'' to denote the collection of quarks in a generation and the individual elements in the first generation. When the superscript $i$ is explicit this should give rise to no confusion. But soon we will want to drop the superscript to denote collectively the generations as vectors $q_L$, $u_R$ and $d_R$, and then we will have to rely on the context to figure out whether it is the collection or the individual first element that we are referring to. For this reason some authors use the capital letters $U_R$ and $D_R$ to denote the vectors in generation space. But I want to reserve $U$ for unitary transformations, and I think you should have no problem figuring out what we are talking about from context.

Similarly, for leptons we have
\begin{equation}
\ell^i_L=\left(\begin{pmatrix}\nu_{eL}\\e_L\end{pmatrix}, \begin{pmatrix}\nu_{\mu L}\\\mu_L\end{pmatrix}, \begin{pmatrix}\nu_{\tau L}\\\tau_L\end{pmatrix}\right), \qquad e^i_R=(e_R, \mu_R,\tau_R).
\end{equation}

The last ingredient of the SM is the Brout-Englert-Higgs (BEH) field, $H$, a collection of complex scalars transforming as $(1,2)_{1/2}$. The BEH field has an expectation value, which we take to be
\begin{equation}
\label{eq:higgsvev}
\langle H\rangle =\frac1{\sqrt2}\begin{pmatrix}0\\v\end{pmatrix}.
\end{equation}
The hermitian conjugate field $\Htilde=i\sigma^2 H^*$ transforms as
$(1,2)_{-1/2}$ and is useful in constructing Yukawa interactions
invariant under the electroweak group.
The covariant derivative is
\begin{equation}
D_\mu =\partial_\mu+ig_s T^a A^a_\mu+i g_2\frac{\sigma^j}2 W^j_\mu +i g_1YB_\mu.
\end{equation}
Here we have used already the Pauli $\sigma^i$ matrices as generators
of $SU(2)$, since the only fields which are non-singlets under this
group are all doublets (and, of course, one should replace zero for
$\sigma^j$ above in the case of singlets). It should also be clear
that we are using the generalized Einstein convention: the repeated
index $a$ is summed over $a=1,\ldots, N_c^2-1$, where $N_c=3$ is the
number of colors, and $j$ is summed over $j=1,2,3$. The generators
$T^a$ of
$SU(3)$ are normalized so that in the fundamental representation
$\text{Tr}(T^aT^b)=\tfrac12 \delta^{ab}$. With this we see that
$\langle H\rangle $ is invariant under $Q=\tfrac12\sigma^3+Y$, which
we identify as the generator of an unbroken $U(1)$ gauge group,
the electromagnetic charge. The field strength tensors for $A^a_\mu$,
$W^j_\mu$ and $B_\mu$ are denoted as $G^a_{\mu\nu}$, $W^j_{\mu\nu}$,  and
$B_{\mu\nu}$, respectively, and that of electromagnetism by
$F_{\mu\nu}$. 

The Lagrangian of the SM is the most general combination of monomials (terms) constructed out of these fields constrained by (i)~giving a hermitian Hamiltonian, (ii)~Lorentz invariance, (iii)~Gauge invariance, and (iv)~renormalizability. This last one implies that these monomials, or ``operators,'' are of dimension no larger than four.\footnote{The action integral $S=\int d^4x\,\mathcal{L}$ has units of $\hbar$, and since we take $\hbar=1$, the engineering dimensions of the  Lagrangian density $\mathcal{L}$ must be $-4$. } Field redefinitions by linear transformations that preserve Lorentz and gauge invariance bring the kinetic terms to canonical form. The remaining terms are potential energy terms, either Yukawa interactions or BEH-field self-couplings. The former are central to our story: 
\begin{equation} 
  -\mathcal{L}_{\text{Yuk}}=\sum_{i,j}\left[\lambda_{U}{}^i_j\Htilde  \widebar q_{Li} u_R^j+\lambda_D{}^i_{j} H \widebar q_{Li} d_R^j +\lambda_E{}^i_{j} H \widebar \ell_{Li} e_R^j+\text{h.c.}  \right]
\end{equation}
We will mostly avoid explicit index notation from here on. The reason for upper and lower indices will become clear below. The above equation can be written more compactly  as  
\begin{equation} 
  \label{eq:yuk} -\mathcal{L}_{\text{Yuk}}=\Htilde \widebar q_{L}\lambda_{U} u_R+ H \widebar q_{L}\lambda_D d_R +H \widebar \ell_{L}\lambda_E e_R+\text{h.c.} 
\end{equation}

\paragraph{Flavor ``symmetry.''} In the absence of Yukawa interactions
({\it i.e.}, setting $\lambda_{U}=\lambda_{D}=\lambda_{E}=0$ above)
the SM Lagrangian has a large global symmetry. This is because the
Lagrangian is just the sum of covariantized kinetic energy therms,
$\sum_n\widebar\psi_n i\slashed{D}\psi_n$, with the sum running over all
the fields in irreducible representations of the the SM gauge group, and
one can make linear unitary transformations among the fields in a
given SM-representation  without altering the
Lagrangian:
\[
q_L\to U_q\; q_L\;, \quad u_R\to U_u\; u_R \;,\quad\ldots \quad e_R\to U_e\; e_R~,
\]
where $U^\dagger_qU_q^{\phantom{\dagger}}=\cdots=U^\dagger_eU_e^{\phantom{\dagger}}=1$. 
 Since there are
$N_f=3$ copies of each SM-representation this means these are $N_f\times
N_f $ matrices, so that for each SM-representation the redefinition freedom is by elements of the group  $U(N_f)$.  Since there are five distinct SM-representations (3
for quarks and 2 for leptons), the full symmetry group is
$U(N_f)^5=U(3)^5$.\footnote{Had we kept indices explicitly we would have written \(
q_L^i\to U_q{}^i{}_j\: q_L^j\;,  u_R^i\to U_u{}^i{}_j\: u_R^j \;,\ldots ,  e_R^i\to U_e{}^i{}_j\: e_R^j
\). The fields transform in the fundamental representation of $SU(N_f)$. We use upper indices for this. Objects, like the hermitian conjugate of the fields, that transform in the anti-fundamental representation, carry lower indices. The transformation matrices have one upper and one  lower indices, of course. } In the quantum theory each of the $U(1)$ factors
(corresponding to a redefinition of the $N_f$ fields in a given
SM-representation by multiplication by a common phase) is anomalous, so
the full symmetry group is smaller. One can make non-anomalous 
combinations of these $U(1)$'s, most famously $B-L$, a symmetry that
rotates quarks and leptons simultaneously, quarks by $-1/3$ the
phase of leptons. For our purposes it is the non-abelian factors that
are most relevant, so we will be happy to restrict our attention to
the symmetry group $SU(N_f)^5$.

The flavor symmetry is broken explicitly by the Yukawa
interactions. We can keep track of the pattern of symmetry breaking by
treating the Yukawa couplings as ``spurions,'' that is, as constant
fields. For example, under $SU(N_f)_q\times SU(N_f)_u$ the first term in 
\eqref{eq:yuk} is invariant if we declare that $\lambda_U$ transforms as  a bi-fundamental, 
$\lambda_U\to U_q\lambda_U U_u^\dagger$; check:   
\[
\widebar q_{L}\lambda_{U} u_R \to  \widebar q_{L}U_q^\dagger (U_q^{\phantom{\dagger}}\lambda_{U}U_u^\dagger)U_u u_R= \widebar q_{L}\lambda_{U} u_R.
\]
So this, together with  $\lambda_D\to U_q^{\phantom{\dagger}}\lambda_D U_d^\dagger$  and $\lambda_E\to U_\ell^{\phantom{\dagger}} \lambda_E U_e^\dagger$  renders the whole Lagrangian invariant. 

Why do we care? As we will see, absent tuning or large parametric suppression, {\it new interactions that break this  ``symmetry'' tend to produce rates of flavor transformations that are inconsistent with observation.} This is not an absolute truth, rather a statement about the generic case. 

In these lectures we will be mostly concerned with hadronic flavor, so  from here on
we focus on the $G_F\equiv SU(3)^3$ that acts on quarks. 

\section{The KM matrix and the KM model of CP-violation}
Replacing  the BEH field by its VEV, Eq.~\eqref{eq:higgsvev},  in the Yukawa terms  in \eqref{eq:yuk} 
we obtain  mass terms for  quarks and leptons:
\begin{equation}
 \label{eq:qmass} 
-\mathcal{L}_{\text{m}}=\frac{v}{\sqrt2} \widebar u_{L}\lambda_{U} u_R+\frac{v}{\sqrt2}    \widebar d_{L}\lambda_D d_R +\frac{v}{\sqrt2}  \widebar e_{L}\lambda_E e_R+\text{h.c.} 
\end{equation}
For simpler computation and interpretation of the model it is best to make further field redefinitions that render the  mass terms diagonal while maintaining the canonical form of the kinetic terms (diagonal, with unit normalization). The field redefinition must be linear (to maintain explicit renormalizability of the model) and commute with the Lorentz group and the part of the gauge group that is unbroken by the electroweak VEV (that is, the $U(1)\times SU(3)$ of electromagnetism and color). This means the linear transformation can act to mix only quarks with the same handedness and electric charge (and the same goes  for leptons):
\begin{equation}
\label{eq:quarkV}
u_R \to V_{u_R} u_R,\quad u_L \to V_{u_L} u_L,\quad  d_R \to V_{d_R} d_R,\quad  d_L \to V_{d_L} d_L.
\end{equation}
Finally, the linear transformation will preserve the form of the kinetic terms, say, $\widebar u_L i \slashed{\partial} u_L
\to 
(\widebar u_L V_{u_L}^\dagger)  i \slashed{\partial} (V_{u_L} u_L)=\widebar u_L (V_{u_L}^\dagger V_{u_L}) i \slashed{\partial} u_L$, if $V_{u_L}^\dagger  V_{u_L} =1$, that is, if they are unitary. 

Now, choose to make these field redefinitions by matrices that diagonalize the mass terms, 
\begin{equation}
\label{eq:massDtransf}
V_{u_L}^\dagger \lambda_{U} V_{u_R}^{\phantom{\dagger}}= \lambda_{U}^\prime, \quad V_{d_L}^\dagger \lambda_{D} V_{d_R}^{\phantom{\dagger}}= \lambda_{D}^\prime\;.
\end{equation}
Here the matrices with a prime, $\lambda_{U}^\prime$ and $\lambda_{D}^\prime$, are diagonal, real and positive. 

\begin{exercises}
\begin{exercise}
Show that this can always be done. That is, that an arbitrary matrix $M$ can be transformed  into a real, positive diagonal matrix $M'=P^\dagger MQ$ by a pair of unitary matrices, $P$ and $Q$. 
\end{exercise}
\begin{solution}
I'll give you a physicist's proof. If you want to be a mathematician, and use Jordan Normal forms, be my guest. Consider the matrices $M^\dagger M $ and $ M M^\dagger$. They are both hermitian so they can each be diagonalized by a unitary transformation. Moreover, they both obviously have real non-negative  eigenvalues. And they  have the same eigenvalues: using the properties of the determinant you can see that the characteristic polynomial is the same, $\det(M^\dagger M -x)=\det(MM^\dagger-x)$. So we have matrices $P$ and $Q$ such that $P^\dagger ( MM^\dagger) P=Q^\dagger ( M^\dagger M) Q=D=$ real, non-negative, diagonal. We can rewrite $D= P^\dagger ( MM^\dagger) P= ( P^\dagger  M Q)( P^\dagger  M Q)^\dagger= XX^\dagger$, where $X=  P^\dagger  M Q$. Similarly, we also have $D=X^\dagger X$, and multiplying this by X on the left we combine the two results into $XD=DX$. Let's assume all the entries in $D$ are all different and non-vanishing  (I will leave out the special cases, you can feel in the details). Then $DX-XD=0$ means, in components $(D_{ii}-D_{jj})X_{ij}=0$ which means that $X_{ij}=0$ for $j\ne i$. So $X$ is diagonal, with $|X_{ii}| = \sqrt{D_{ii}}$. We can always take $ P^\dagger  M Q=\sqrt{D}=M'$, by further transformation by a diagonal unitary matrix on the  left or right.
\end{solution}
\end{exercises}

Then from 
\begin{equation}
-\mathcal{L}_{\text{m}}=\frac{v}{\sqrt2} \Big( \widebar u_{L}\lambda'_{U} u_R+ \widebar d_{L}\lambda'_D d_R +  \widebar e_{L}\lambda_E e_R+\text{h.c.}\Big)
=\frac{v}{\sqrt2}\Big(\widebar u\lambda'_{U} u+   \widebar d\lambda'_D d +  \widebar e\lambda_E e\Big)
\end{equation}
we read off the diagonal mass matrices, $m_U= v \lambda'_{U}/\sqrt2$, $m_D=v \lambda'_D /\sqrt2$ and $m_E=v \lambda_E/\sqrt2 $. 

Since the field redefinitions in \eqref{eq:quarkV} are not symmetries of the Lagrangian (they fail to commute with the electroweak  group), it is not guaranteed that the Lagrangian is independent of the  matrices $V_{u_L}, \ldots, V_{d_R}$. We did choose the transformations to leave the kinetic terms in  canonical form. We now check the effect of~\eqref{eq:quarkV}  on the gauge interactions. Consider first the singlet fields $u_R$. Under the field redefinition we have
\[
\widebar u_R \,( g_s\slashed{A}^a T^a+ \tfrac23 g_1 \slashed{B}) u_R \to 
\widebar u_R V^\dagger_{u_R} \, ( g_s\slashed{A}^a T^a+\tfrac23 g_1 \slashed{B}) V_{u_R} u_R = \widebar u_R \, ( g_s\slashed{A}^a T^a+\tfrac23 g_1 \slashed{B}) u_R~.
\]
It remains unchanged (you can see this by making explicit the so-far-implicit indices for color and for spinor components). Clearly the same happens with the $d_R$ fields. The story gets more interesting with the left handed fields, since they form doublets. First let's look at the terms that are diagonal in the doublet space:
\begin{multline*}
\widebar q_L ( g_s\slashed{A}^a
T^a+\tfrac12g_2\slashed{W}^3\sigma^3+\tfrac16 g_1\slashed{B})q_L \\
= \widebar u_L ( g_s\slashed{A}^a T^a+\tfrac12g_2\slashed{W}^3+\tfrac16 g_1\slashed{B}) u_L+
\widebar d_L ( g_s\slashed{A}^a T^a-\tfrac12g_2\slashed{W}^3+\tfrac16 g_1\slashed{B})d_L
\end{multline*}
where in going to the second line we have expanded out the doublets in their components. The result is invariant under \eqref{eq:quarkV} very much the same way that the $u_R$ and $d_R$ terms are.  Finally we have the off-diagonal terms. For these let us introduce
\[\sigma^\pm = \frac{\sigma^1\pm i\sigma^2}{\sqrt2}, \quad
\text{and}\quad W^\pm= \frac{W^1\mp i W^2}{\sqrt2}
\]
so that $\sigma^1W^1+\sigma^2 W^2 = \sigma^+W^++\sigma^- W^-$ and $(\sigma^+)_{12}=\sqrt2$,   $(\sigma^-)_{21}=\sqrt2$, and all other elements vanish. It is now easy to expand:
\begin{multline}
\label{eq:chargedCurr}
\widebar q_L\tfrac12 g_2(\sigma^1W^1+\sigma^2 W^2 ) q_L = \tfrac{1}{\sqrt2}g_2\widebar u_L\slashed{W}^+ d_L +\tfrac{1}{\sqrt2}g_2
 \widebar d_L\slashed{W}^- u_L\\ \to \tfrac{1}{\sqrt2}g_2\widebar
 u_L(V_{u_L}^\dagger V_{d_L}^{\phantom{\dagger}})\slashed{W}^+ d_L + \tfrac{1}{\sqrt2}g_2 \widebar d_L(V_{d_L}^\dagger V_{u_L\vphantom{d_L}}^{\vphantom{\dagger}})\slashed{W}^- u_L
\end{multline}
A relic of our field redefinitions has remained in the form of the unitary matrix $V=V_{u_L}^\dagger V_{d_L}$. We call this the Kobayashi-Maskawa (KM) matrix. You will also find this as the Cabibbo-Kobayashi-Maskawa, or CKM, matrix in the literature. Cabibbo figured out the $2\times 2$ case, in which the matrix is orthogonal and given  in terms of a single angle, the {\it Cabibbo angle}. Because Kobayashi and Maskawa were first to introduce the $3\times3$ version with an eye to incorporate CP violation in the model (as we will study in detail below), in these notes we refer to it as as the KM matrix. 

A general unitary $3\times 3$ matrix has $3^2$ complex entries, constrained by $3$ complex plus $3$ real conditions. So the KM matrix is in general parametrized by 9 real entries.  But not all are of physical consequence. 
We can perform further transformations of the form of
\eqref{eq:quarkV} that leave the mass matrices in
\eqref{eq:massDtransf}  diagonal and
non-negative if the unitary matrices are diagonal with $V_{u_L}=V_{u_R}=\text{diag}(e^{i\alpha_1},e^{i\alpha_2},e^{i\alpha_3}) $ and  $V_{d_L}=V_{d_R}=\text{diag}(e^{i\beta_1},e^{i\beta_2},e^{i\beta_3})$. Then  $V$ is redefined by $V_{ij}\to e^{i(\beta_j-\alpha_i)}V_{ij}$. These five independent phase differences reduce the number of independent parameters in $V$ to $9-5=4$. It can be shown that this can in general be taken to be 3 rotation angles and one complex phase. It will be useful to label the matrix elements by the quarks they connect:
\[
V= \begin{pmatrix}
V_{ud} & V_{us} & V_{ub}\\
V_{cd} & V_{cs} & V_{cb}\\
V_{td} & V_{ts} & V_{tb}
\end{pmatrix}~.
\]

\smallskip

Observations:
\begin{enumerate}
\item That there is one irremovable phase in $V$ impies that CP is not a symmetry of the SM Lagrangian. It is broken by the terms $\widebar u_L V \slashed {W}^+ d_L + \widebar d_L V^\dagger \slashed{W}^- u_L$. To see this, recall that under CP 
$\widebar u_L \gamma^\mu  d_L \rightarrow -\widebar d_L \gamma_\mu  u_L$ and $W^{+\mu}\to -W^-_\mu$. Hence CP invariance requires $V^\dagger =V^T$. 
\begin{exercises}
\begin{exercise}
In QED, charge conjugation is $\widebar e\gamma^\mu e\to -\widebar e \gamma^\mu e$ and $A^\mu \to - A^\mu$.  So $\widebar e \slashed {A} e$ is invariant under $C$.\\
So what about QCD? Under charge conjugation  $\widebar q T^a \gamma^\mu q \to \widebar q (-T^a)^T \gamma^\mu q$, but $ (-T^a)^T=(-T^a)^*$ does not equal $-T^a$ (nor $T^a$). So what does charge conjugation mean in QCD? How does the gluon field, $A^a_\mu$, transform?
\end{exercise}
\begin{solution}
I have never seen this discussed in a textbook, or elsewhere. Maybe one of the readers will write a nice article for AJP (don't forget to include me!). If you think of the ``transformation arrow'' more properly as the action by a unitary operator on the Hilbert space, $C$, 
so that $\widebar e\gamma^\mu e\to -\widebar e \gamma^\mu e$ really means $C(\widebar e\gamma^\mu e)C^{-1}= -\widebar e \gamma^\mu e$, then it is clear that $T^a$ is not changed since it is a $c$-number that commutes with $C$. What we need is
$A_\mu^a T^a\to A_\mu^a (-T^a)^T$. This is accomplished by a transformation $ A_\mu^a\to R^{ab}A_\mu^b$ with a real matrix $R$ that must take $T^a$ into minus its transpose: $R^{ba} T^b = -T^{aT}$. Since the matrices $T^a$ are in the fundamental representation of $SU(3)$ we have $R^{ca}=2\text{Tr}[T^c(R^{ba} T^b)] =-2\text{Tr}( T^c T^{aT})$. $R$ is indeed real:
$(R^{ca})^* =-2\text{Tr}( T^{c*} T^{a\dagger})$, then using  $T^{a\dagger}=T^a$, $\text{Tr}(A^T)=\text{Tr}(A)$ and cyclicity of trace, it follows that $R$ is a real symmetric matrix. Notice that $R^2=1$ for consistency (the negative transpose of the negative transpose is the identity). You can check this using the identity $2T^a_{ij}T^a_{mn}=\delta_{in}\delta_{mj}-\frac13\delta_{ij}\delta_{mn}$.

In physical terms this means that under charge conjugation the, say, blue-antigreen gluon is transformed into minus the green-antiblue gluon, and so on. 

I have seen in places  an explanation for charge conjugation in QCD along these lines: first take the quark field $q$ and rewrite in terms of a left- and a right-handed fields, $q_L$ and $q_R$. Then replace $q_R$ by its charge-conjugate, which is also a left-handed field, $q^c_L$. Now $q_L$ is a triplet under color while $q_L^c$ is an antitriplet under color. So charge conjugation is simply $q_L\leftrightarrow q^c_L$. This is incomplete (and therefore wrong). If you were to ignore the transformation of the gluon field the resulting Lagrangian would not be gauge invariant since now the covariant  derivative acting on $q_L$ has a generator for an anti-triplet, $-T^{aT}$, while the covariant derivative acting on $q_L^c$ has generator $T^a$ appropriate for a triplet. It is only after you transform the gluon field that everything works as it should!
\end{solution}
\begin{exercise}
If two entries  in $m_U$ (or in $m_D$) are equal show that $V$ can be brought into a real matrix and hence is an  orthogonal transformation (an element of $O(3)$). 
\end{exercise}
\begin{solution}
Without loss of generality we may assume the first two entries in
$m_U$ are equal. This means that the remnant freedom to redefine
quark fields without changing neither the kinetic nor the mass terms
is not just by individual phases on all flavors but also by a
$2\times2$ unitary matrix acting on the degenerate quarks. Let
$u_{L,R}\to Uu_{L,R} $, then $U$ is of the form
\[\left(\begin{array}{c|c}A & 0\\ \hline \\[-2.5ex] 0 & e^{i\alpha_3}\end{array}
\right)\]
where $A$ is a $2\times2$ unitary matrix and ``0'' stands for a
2-component zero vector. 
Let also $W$ be the diagonal matrix with entries $e^{i\beta_i}$,
$i=1,2,3$, and redefine $d_{L,R}\to W d_{L,R}$. This has the effect of
redefining $V\to U^\dagger VW$. To see what is going on let's write
$V$ in terms of a $2\times2$ submatrix, $X$, two 2-component column
vectors, $\psi$ and $\eta$, and a complex number, $z$:
\[\left(\begin{array}{c|c}X & \psi\\ \hline \\[-2.5ex]\eta^T & z\end{array}.
\right)\]
Then $V$ is transformed into
\begin{equation}
\label{eqsol:vtrans}
V=\left(\begin{array}{c|c}A^\dagger X\begin{pmatrix}e^{i\beta_1}&0\\0& e^{i\beta_2}\end{pmatrix}  & e^{i\beta_3}A^\dagger\psi\\[1.8ex]
    \hline \\[-1.9ex] e^{-i\alpha_3}\eta^T\begin{pmatrix}e^{i\beta_1}&0\\0& e^{i\beta_2}\end{pmatrix} & e^{i(\beta_3-\alpha_3)}z\end{array}
\right).\end{equation}
Now we can choose $A^\dagger$ so that $e^{i\beta_3}A^\dagger\psi$ has
vanishing lower component and real upper component. This still leaves
freedom in $A$ to make a rotation by a phase of the (vanishing) lower
component. So we may take
\[
\psi= \begin{pmatrix}|\psi|\\0\end{pmatrix}\qquad\text{and}\qquad
e^{i\beta_3}A^\dagger=
\begin{pmatrix}1 &0\\0&e^{i(\gamma+\beta_3)}\end{pmatrix}.
\]
At this point it is worth making the trivial observation that for
fixed $\beta_3$ one can make the third row of the new $V$ matrix in
\eqref{eqsol:vtrans} real by choosing $\beta_1,\beta_2$
and~$\alpha_3$. We are left with the $2\times2$ block,
\[
A^\dagger X\begin{pmatrix}e^{i\beta_1}&0\\0& e^{i\beta_2}\end{pmatrix} =\begin{pmatrix}e^{-i\beta_3} &0\\0&e^{i\gamma}\end{pmatrix}X\begin{pmatrix}e^{i\beta_1}&0\\0& e^{i\beta_2}\end{pmatrix} 
\]
Now choose $\beta_3$ and $\gamma$ to make real the first column. This
means that the only entries of $V$ with a phase are the top two
entries of the second column, $V_{21}$ and $V_{22}$. But unitarity of
$V$ requires $V_{2i}V_{3i}^*=0$. Since $V_{32}=0$ this can only be
satisfied if $V_{21}$ is real. Then $V_{2i}V_{1i}^*=0$ can only be
satisfied if $V_{22}$ is real. Hence all elements in $V$ are real.
\end{solution}
\end{exercises}
\item Precise knowledge of the elements of $V$ is necessary to  constrain new physics (or to test the validity of the SM/CKM theory). We will describe below how well we know them and how. But for now it is useful to have a sketch that gives a rough order of magnitude of the magnitude of the elements in $V$:
\begin{equation}
\label{eq:Veps}
V\sim \begin{pmatrix}\epsilon^0 & \epsilon^1& \epsilon^3\\\epsilon^1 & \epsilon^0& \epsilon^2\\\epsilon^3 & \epsilon^2& \epsilon^0\end{pmatrix}, \qquad\text{with $\epsilon\sim 10^{-1}$.}
\end{equation}
\item Since $VV^\dagger=V^\dagger V=1$ the rows as well as the columns  of $V$ are orthonormal vectors. In particular,  $\sum_k V^{\phantom{*}}_{ik}V^*_{jk}=0$ for $j\ne i$. Three complex numbers that  sum to zero are represented on the complex plane as a triangle. As the following table shows, the resulting triangles are very different in shape. Two of them are very squashed, with one side much smaller than the other two, while the third one has all sides of comparable size. As we shall see, this will play a role in understanding when CP asymmetries in decay rates can be sizable. 

\medskip

\begin{tabular}[c]{l|l|c|m{4cm}}
$ij$ & $\sum V_{ik}^{\phantom{*}}V_{jk}^*=0$ & $\sim \epsilon^n$& \parbox[t]{5cm}{shape\\ (normalized to unit base)}\\ \hline\hline
12 & $V_{ud}^{\phantom{*}}V_{cd}^*+V_{us}^{\phantom{*}}V_{cs}^*+V_{ub}^{\phantom{*}}V_{cb}^*=0$ & $\epsilon + \epsilon +\epsilon^5=0$ & \parbox[c][1cm]{5cm}{\begin{tikzpicture}
\draw (0,0) -- (4,0) -- node[right]{$\epsilon^4$} (4,0.2) -- (0,0);
\end{tikzpicture}} \\ \hline
23 & $V_{cd}^{\phantom{*}}V_{td}^*+V_{cs}^{\phantom{*}}V_{ts}^*+V_{cb}^{\phantom{*}}V_{tb}^*=0$ & $\epsilon^4 + \epsilon^2 +\epsilon^2=0$ & \parbox[c][1.2cm]{5cm}{\begin{tikzpicture}
\draw (0,0) -- (4,0) -- node[right]{$\epsilon^2$} (4,0.5) -- (0,0);
\end{tikzpicture}} \\ \hline
13 & $V_{ud}^{\phantom{*}}V_{td}^*+V_{us}^{\phantom{*}}V_{ts}^*+V_{ub}^{\phantom{*}}V_{tb}^*=0$ & $\epsilon^3 + \epsilon^3 +\epsilon^3=0$ & \parbox[c][2.5cm]{5cm}{\begin{tikzpicture}
\draw (0,0) -- (4,0) -- node[right]{$1$} (3,2) -- (0,0);
\end{tikzpicture}} \\\hline
\end{tabular}

\bigskip

These are called ``unitarity triangles.'' The most commonly discussed is in the 1-3 {\it columns}, 
\[
V_{ud}^{\phantom{*}}V_{ub}^*+V_{cd}^{\phantom{*}}V_{cb}^*+V_{td}^{\phantom{*}}V_{tb}^*=0\quad\Rightarrow\quad \text{ \parbox[c][2.5cm]{5cm}{\begin{tikzpicture}[scale=0.5]
\draw (0,0) --node[below]{1}  (4,0) -- node[right]{$\sim\!1$} (1.6,2) -- node[left]{$\sim\!1$}(0,0);
\end{tikzpicture}} }
\]
Dividing by the middle term we can be more explicit as to what we mean by the unit base unitarity triangle:
\[
\frac{V_{ud}^{\phantom{*}}V_{ub}^*}{V_{cd}^{\phantom{*}}V_{cb}^*}+1+\frac{V_{td}^{\phantom{*}}V_{tb}^*}{V_{cd}^{\phantom{*}}V_{cb}^*}=0
\]
We draw this on the complex plane and introduced some additional notation:  the complex plane is $z=\widebar\rho + i \widebar \eta$
and the internal angles of the triangle are\footnote{This convention is popular in the US, while in Japan  a different convention is more common:   $\phi_1=\beta$, $\phi_2=\alpha$ and $\phi_3=\gamma$.}  $\alpha$, $\beta$ and $\gamma$; see Fig.~\ref{fig:UTsketch}.

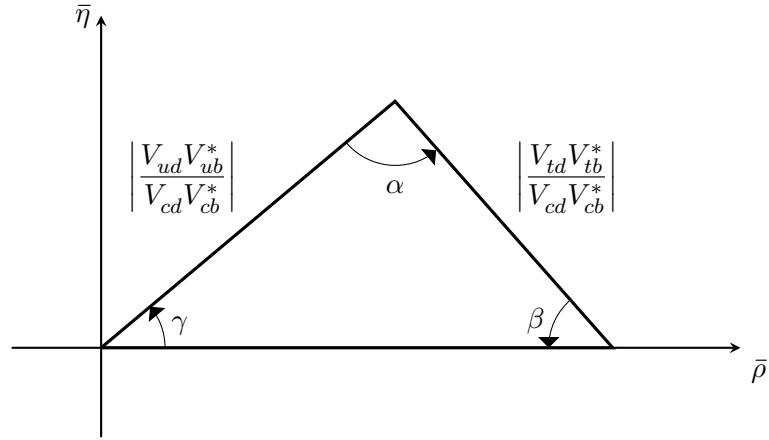
\begin{figure}
\begin{center}
\begin{tikzpicture}[scale=1.7,>=triangle 90]
\coordinate (A) at (0,0);
\coordinate (B) at  (4,0); 
\coordinate (C) at  (40:3cm) ;

\draw[->,>=stealth,thick] ($(A)+(-0.7,0)$) -- ($(B)+(1,0)$) node[below right]{$\widebar\rho$};
\draw[->,>=stealth,thick] ($(A)+(0,-0.7)$) -- +(0,3.3) node[left]{$\widebar\eta$};

\draw[very thick]   (A) -- (B) -- node[above right]{$\displaystyle\left|\frac{V_{td}^{\phantom{*}}V_{tb}^*}{V_{cd}^{\phantom{*}}V_{cb}^*}\right|$} (C)  node[below=0.9cm]{ $\alpha$} -- node[above left]{$\displaystyle\left|\frac{V_{ud}^{\phantom{*}}V_{ub}^*}{V_{cd}^{\phantom{*}}V_{cb}^*}\right|$}(0,0);

\draw[black]  ($(B)+(155:0.5cm)$)  node[left]{$\beta$};

\draw[->] ($(A)+(0.5,0)$) arc (0:40:0.5cm)  ;
\draw  ($(A)+(20:0.5cm)$)  node[right]{$\gamma$};

\draw[->] ($(C)+(220:0.5)$) arc (220:310:0.5);
\draw[->] ($(B)+(130:0.5)$) arc (130:180:0.5);

\end{tikzpicture}
\end{center}
\caption{\label{fig:UTsketch} Unitarity triangle in the $\xoverline\rho$-$\xoverline\eta$ plane. The base is of unit length. The sense of the angles is indicated by arrows.}
\end{figure}

The angles of the unitarity triangle, of course, are completely determined by the KM matrix, as you will now explicitly show:

\begin{figure}
\begin{center}
\includegraphics[width=4in]{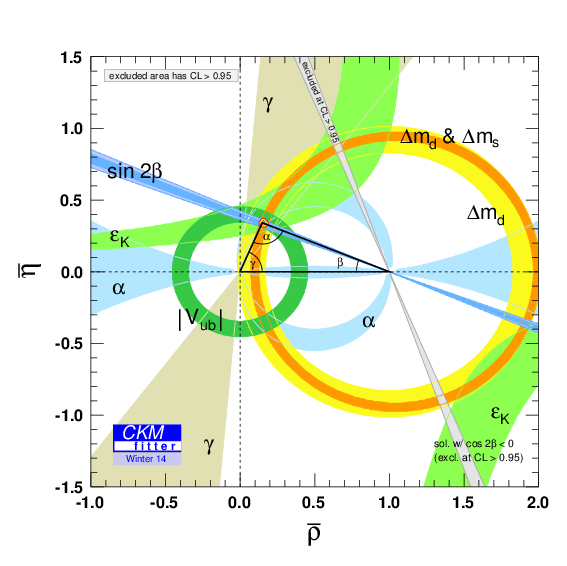}\hfill
\includegraphics[width=4in]{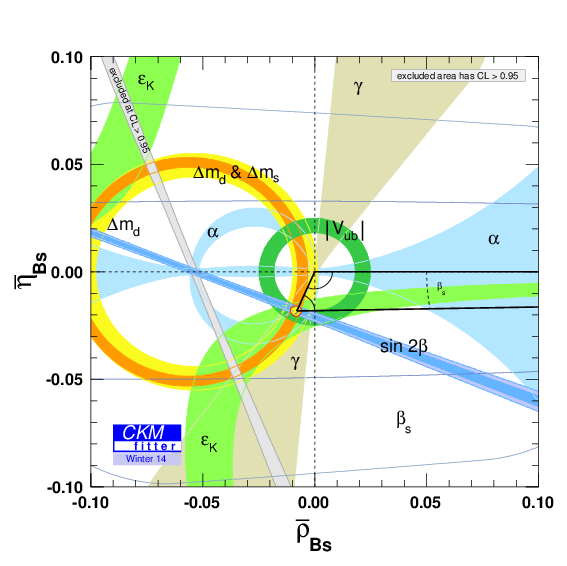}
\end{center}
\caption{\label{fig:CKMfitterTRIANGLES}  Experimentally determined unitarity triangles~\cite{ckmfitter}. Upper pane:  ``fat''  1-3  columns triangle.  Lower pane: ``skinny''  2-3 columns triangle.}
\end{figure}

\begin{exercises}
\begin{exercise}
\label{eqsol:alphabetagamma}
Show that 
\begin{enumerate}[(i)]
\item $\displaystyle \beta= \text{arg}\left(- \fracup{V_{cd}^{\phantom{*}}V_{cb}^*}{V_{td}^{\phantom{*}}V_{tb}^*}\right)$, 
 $\displaystyle \alpha= \text{arg}\left(- \fracup{V_{td}^{\phantom{*}}V_{tb}^*}{V_{ud}^{\phantom{*}}V_{ub}^*}\right)$ and 
 $\displaystyle \gamma= \text{arg}\left(- \fracup{V_{ud}^{\phantom{*}}V_{ub}^*}{V_{cd}^{\phantom{*}}V_{cb}^*}\right)$.
\item These are invariant under phase redefinitions of quark fields (that is, under the remaining arbitrariness). Hence these are candidates for observable quantities.  
\item The area of the triangle is $-\frac12\,\text{Im}\frac{\raisebox{0.45ex}{$\scriptstyle V_{ud}^{\phantom{*}}V_{ub}^*$}}{V_{cd}^{\phantom{*}}V_{cb}^*}
=-\frac12\,\frac1{|V_{cd}^{\phantom{*}}V_{cb}^*|^2}\text{Im}\left(V_{ud}^{\phantom{*}}V_{cd}^*V_{cb}^{\phantom{*}}V_{ub}^*\right)$. 
\item The product $J=
\text{Im}\left(V_{ud}^{\phantom{*}}V_{cd}^*V_{cb}^{\phantom{*}}V_{ub}^*\right)$ (a ``Jarlskog invariant'') is also invariant under  phase redefinitions of quark fields. 
\end{enumerate}
Note that
$\text{Im}\left(V_{ij}^{\phantom{*}}V_{kl}^{\phantom{*}}V_{il}^*V_{kj}^*\right)=J(\delta_{ij}\delta_{kl}-\delta_{il}\delta_{kj})$
is the common area of all the un-normalized triangles. The area of a
normalized triangle is $J$ divided by the square of the magnitude of the side that is normalized to unity. 
\end{exercise}
\begin{solution}
\begin{enumerate}[(i)]
\item Take the equation that defines the triangle
\[
\frac{V_{ud}^{\phantom{*}}V_{ub}^*}{V_{cd}^{\phantom{*}}V_{cb}^*}+1+\frac{V_{td}^{\phantom{*}}V_{tb}^*}{V_{cd}^{\phantom{*}}V_{cb}^*}=0
\]
and depict it as a triangle in the complex plane:
\begin{center}
\begin{tikzpicture}[scale=1.7,>=triangle 90]
\coordinate (A) at (0,0);
\coordinate (B) at  (4,0); 
\coordinate (C) at  (40:3cm) ;

\draw[->,>=stealth,thick] ($(A)+(-0.7,0)$) -- ($(B)+(1,0)$) node[below right]{$\widebar\rho$};
\draw[->,>=stealth,thick] ($(A)+(0,-0.7)$) -- +(0,3.3) node[left]{$\widebar\eta$};

\draw[->,very thick]   (A) -- (B);
\draw[->,very thick] (B) -- node[above right]{$\displaystyle\frac{V_{td}^{\phantom{*}}V_{tb}^*}{V_{cd}^{\phantom{*}}V_{cb}^*}$} (C);
\draw[->,very thick] (C)  node[below=0.9cm]{ $\alpha$} -- node[above left]{$\displaystyle\frac{V_{ud}^{\phantom{*}}V_{ub}^*}{V_{cd}^{\phantom{*}}V_{cb}^*}$}(0,0);

\draw[black]  ($(B)+(155:0.5cm)$)  node[left]{$\beta$};
\draw[->] ($(A)+(0.5,0)$) arc (0:40:0.5cm)  ;
\draw  ($(A)+(20:0.5cm)$)  node[right]{$\gamma$};
\draw[->] ($(C)+(220:0.5)$) arc (220:310:0.5);
\draw[->] ($(B)+(130:0.5)$) arc (130:180:0.5);
\end{tikzpicture}
\end{center}
Note that the vector from the origin to $(\bar\rho,\bar\eta)$ is the opposite of $\frac{V_{ud}^{\phantom{*}}V_{ub}^*}{V_{cd}^{\phantom{*}}V_{cb}^*}$, so the angle $\gamma$ is the argument of minus this, $ \gamma= \text{arg}\left(- \frac{V_{ud}^{\phantom{*}}V_{ub}^*}{V_{cd}^{\phantom{*}}V_{cb}^*}\right)$. Next, the angle that $\frac{V_{td}^{\phantom{*}}V_{tb}^*}{V_{cd}^{\phantom{*}}V_{cb}^*}$ makes with the $\bar\rho$ axis is $\pi-\beta= \text{arg}\left( \frac{V_{td}^{\phantom{*}}V_{tb}^*}{V_{cd}^{\phantom{*}}V_{cb}^*}\right)$, from which $\beta= \text{arg}\left(- \frac{V_{cd}^{\phantom{*}}V_{cb}^*}{V_{td}^{\phantom{*}}V_{tb}^*}\right)$ follows. $\alpha$ is most easily obtained from $\alpha+\beta+\gamma=\pi$ using the two previous results and the fact that $\text{arg}(z_1)+\text{arg}(z_2)=\text{arg}(z_1z_2)$.
\item In the numerator or denominator of these expressions, the re-phasing of the charge-$+\frac23$ quarks cancel; for example, $V_{td}^{\phantom{*}}V_{tb}^*\to
(e^{i\phi}V_{td}^{\phantom{*}})(e^{i\phi}V_{tb}^{\phantom{*}})^*=V_{td}^{\phantom{*}}V_{tb}^*$. The re-phasing of the charge-$-\frac13$ quarks cancel between numerator and denominator; for example, for the $d$ quark  $\frac{V_{ud}^{\phantom{*}}V_{ub}^*}{V_{cd}^{\phantom{*}}V_{cb}^*}\to \frac{e^{i\phi}V_{ud}^{\phantom{*}}V_{ub}^*}{e^{i\phi}V_{cd}^{\phantom{*}}V_{cb}^*}=\frac{V_{ud}^{\phantom{*}}V_{ub}^*}{V_{cd}^{\phantom{*}}V_{cb}^*}$.
\item From question (i) we see that $\bar\eta=-\text{Im}\frac{V_{ud}^{\phantom{*}}V_{ub}^*}{V_{cd}^{\phantom{*}}V_{cb}^*}$, and this is the height of the triangle of unit base. The area is $1/2$ base time height from which the first result follows. The second expression is obtained from the first by multiplying  by
$1=\frac{V_{cd}^*V_{cb}^{\phantom{*}}}{V_{cd}^*V_{cb}^{\phantom{*}}}$.
\item In $J=
\text{Im}\left(V_{ud}^{\phantom{*}}V_{cd}^*V_{cb}^{\phantom{*}}V_{ub}^*\right)$ each $V_{ix}$ appears with one, and only one, other factor of $V_{iy}^*$, and one, and only one, factor of $V_{jx}^*$.
\end{enumerate}
\end{solution}
\end{exercises}

\item {\it Parametrization of $V$:} Since there are only four independent parameters in the matrix that contains $3\times3$ complex entries, it is useful to have a completely general parametrization in terms of four parameters. The standard parametrization can be understood as a sequence of rotations about the three axes, with the middle rotation incorporating also a phase transformation:
\begin{gather*}
V=CBA, \\
\intertext{where}
A=\begin{pmatrix} c_{12}& s_{12}&0\\  -s_{12}& c_{12}&0\\ 0& 0&1\end{pmatrix},\quad
B=\begin{pmatrix} c_{13}& 0& s_{13}e^{-i\delta}\\  0& 1&0\\ -s_{13}e^{i\delta}&0& c_{13} \end{pmatrix},\quad
C=\begin{pmatrix} 1&0& 0\\0& c_{23}& s_{23}\\ 0& -s_{23}& c_{23}\end{pmatrix}.
\end{gather*}
Here we have used the shorthand, $c_{ij}=\cos\theta_{ij}$, $s_{ij}=\sin\theta_{ij}$, where the angles $\theta_{ij}$ all lie on the first quadrant. From the phenomenologically observed rough order of magnitude of elements in $V$ in~\eqref{eq:Veps} we see that the angles $\theta_{ij}$ are all small. But the phase $\delta$ is large, else all triangles would be squashed.

An alternative and popular parametrization is due to Wolfenstein. It follows from the above by introducing parameters $A$, $\lambda$, $\rho$ and $\eta$ according to
\begin{equation}
\label{eq:Wpar}
s_{12}=\lambda, \quad s_{23}=A\lambda^2,\quad s_{13}e^{i\delta}=A\lambda^3(\rho+i\eta)
\end{equation}
The advantage of this parametrization is that if $\lambda$ is of the order of  $\epsilon$, while the other parameters are of order one, 
then the KM matrix elements have the rough order in~\eqref{eq:Veps}. It is easy to see
that $\rho$ and $\eta$ are very close to, but not quite, the coordinates of the apex of the unitarity triangle in Fig.~\ref{fig:UTsketch}. One can adopt the alternative, but tightly related parametrization in terms of $A$, $\lambda$, $\widebar\rho$ and~$\widebar\eta$:
\[
s_{12}=\lambda, \quad s_{23}=A\lambda^2,\quad s_{13}e^{i\delta}=A\lambda^3(\widebar\rho+i\widebar \eta)\frac{\sqrt{1-A^2\lambda^4}}{\sqrt{1-\lambda^2}[1-A^2\lambda^4(\widebar\rho+i\widebar \eta)]} .
\]
\begin{exercises}
\begin{exercise}
\begin{enumerate}[(i)]
\item Show that  \[
\widebar\rho+i\widebar \eta = - \fracup{V_{ud}^{\phantom{*}}V_{ub}^*}{V_{cd}^{\phantom{*}}V_{cb}^*},
\]
hence $\widebar\rho$ and $\widebar \eta$ are indeed the coordinates of the apex of the unitarity triangle and are invariant under quark phase redefinitions. 
\item Expand in $\lambda\ll1$ to show
\[
V=\begin{pmatrix} 1-\tfrac12\lambda^2 &\lambda& A\lambda^3(\rho-i\eta)\\ -\lambda&  1-\tfrac12\lambda^2 & A\lambda^2\\
A\lambda^3(1-\rho-i\eta)& - A\lambda^2 &1\end{pmatrix} +\mathcal{O}(\lambda^4)
\]

\end{enumerate}
\end{exercise}
\begin{solution}
\begin{enumerate}[(i)]
\item  We are not looking for a graphical representation solution, as was done in  Exercise \ref{eqsol:alphabetagamma}. Instead, we want to show this form the definitions of the parameters $\lambda, A, \bar\rho$ and~$\eta$.  This is just plug in and go. First, 
\[V=
\begin{pmatrix}
 c_{12} c_{13} & c_{13} s_{12} & s_{13}e^{-i\delta} \\
 -c_{12} s_{23} s_{13}e^{i\delta}-c_{23}
   s_{12} & c_{12} c_{23}-s_{12} s_{23}
   s_{13}e^{i\delta} & c_{13} s_{23} \\
 s_{12} s_{23}-c_{12} c_{23} s_{13}e^{i\delta}
   & -c_{12} s_{23}-c_{23} s_{12}
   s_{13}e^{i\delta} & c_{13} c_{23}
\end{pmatrix}.
\]
You can break the computation into
  smaller steps. For example,
  $V_{ud}=c_{12}c_{13}=\sqrt{1-\lambda^2}c_{13}$ and
  $V_{c,b}=c_{13}s_{23}=A\lambda^2 c_ {13}$ so that 
\[ \frac{V_{ud}}{V^*_{cb}}=\frac{\sqrt{1-\lambda^2}}{A\lambda^2}.\]
Similarly,
\[ \frac{V^*_{ub}}{V_{cd}}=-\frac{A\lambda^2\sqrt{1-A^2\lambda^4}z}{\sqrt{(1-\lambda^2)(1-A^2\lambda^4)}},\]
where $z=\bar\rho+i\bar\eta$. The result follows.
\item Again plug in and go. But you can be clever about it. For
  example, since $s_{13}\sim\lambda^3$, we have
  $c_{13}=\sqrt{1-s_{13}^2}=1+\mathcal{O}(\lambda^6)$. Similarly
  $c_{23}=1+\mathcal{O}(\lambda^4)$ and
  $c_{12}=1-\frac12\lambda^2+\mathcal{O}(\lambda^4)$.  Plugging these,
  and \eqref{eq:Wpar} into the explicit form of $V$ above the result follows.
  and the 
\end{enumerate}
\end{solution}
\end{exercises}
\end{enumerate}

\section{Determination of KM Elements}
Fig.~\ref{fig:CKMfitterTRIANGLES}  shows the state of the art in our knowledge  of the angles of the unitarity triangles for the 1-3 and 2-3 columns of the KM matrix. How are these determined? More generally, how are KM elements measured? Here we give a tremendously compressed description. 

The relative phase between elements of the KM matrix is associated with possible CP violation. So measurement of rates for processes that are dominated by one entry in the KM are insensitive to the relative phases. Conversely, CP asymmetries directly probe relative phases.

\subsection{Magnitudes}
The magnitudes of elements of the KM matrix are measured as follows:
\begin{enumerate}[(i)]
\item $|V_{ud}|$ is measured through allowed nuclear transitions. The theory is fairly well understood (even if it is nuclear physics) because the transition matrix elements are constrained by symmetry considerations. 
\item $|V_{us}|$, $|V_{cd}|$, $|V_{cs}|$, $|V_{ub}|$, $|V_{cb}|$, are
  primarily probed through semi-leptonic decays of mesons, $M\to M'
  \ell\nu$ ({\it e.g.}, $K^+\to \pi^0 e^+\nu$).  
\item $|V_{tq}|, (q=d,s,b)$ are inferred from  processes that proceed at 1-loop through a virtual top-quark. It is also  possible to measure some of these directly from single top production (or decay). 
\end{enumerate}

The  theoretical
  difficulty is to produce a reliable estimate of the rate, in terms
  of the KM matrix elements, in light of the quarks being strongly
  bound in hadrons. Moreover, theorists have to produce a good estimate for a quantity that experimentalists can measure. There is some tension between these. We will comment on this again below, but let me give one example. The inclusive rate for semileptonic decay of $B$ mesons can be reliably calculated. By inclusive we mean $B$ decays to a charged lepton, say $\mu$, plus a neutrino, plus other stuff, and the rate is measured regardless of what the other stuff is. The decay rate is then the sum over the rates of decays into any particular type of whatever makes up the ``stuff.'' Sometimes the decay product is a $D$ meson, sometimes a $D^*$ meson and other times seven pions or whatever,  always plus $\mu\nu$. Now these decays sometimes involve $b\to c\mu\nu$ which comes in the rate with a factor of $|V_{cb}|^2$ that we would like to determine, and sometimes involves $b\to u\mu\nu$ with a factor of $|V_{ub}|^2$  that we also want to determine. But the total semileptonic rate does not allow us to infer separately $|V_{cb}|^2$ and $|V_{ub}|^2$. Knowing that $|V_{cb}|^2\gg |V_{ub}|^2$ means we can measure well $|V_{cb}|$ from the inclusive semileptonic rate. But then how do we get at $|V_{ub}|$? One possibility, and that was the first approach at this measurement, is to measure the rate of inclusive semileptonic $B$ decays only  for large $\mu$ energy. Since hadrons containing charm are far heavier than those containing up-quarks, there is a range of energies for the $\mu$ resulting from the decay that is not possible if $B$  decayed into charm. These must go through $b\to u\mu\nu$ and therefore their rate  is proportional to $|V_{ub}|^2$. But this is not an inclusive rate, because it does not sum over all possible decay products. It is difficult to get an accurate theoretical prediction for this. 

The determination of magnitudes is usually done from semi-leptonic decays because the theory is more robust than for hadronic decays. Purely leptonic decays, as in $B^-\to\mu^-\bar\nu$ are also under good theoretical control, but their rates are very small because they are helicity suppressed in the SM (meaning that the ``$V-A$'' nature of the weak interactions, $V=\text{vector}$, $A=\text{axial}$,  gives a factor of $m_\mu/m_b$ in the decay amplitude). We lump them into the category of  ``rare'' decays and use them, with an independent determination of the KM elements, to test the accuracy of the SM and put bounds on new physics. We distinguish exclusive from inclusive semileptonic decay measurements:

\subsubsection{Exclusive semileptonic decays}
By an ``exclusive'' decay we mean that the final state is fixed as in, for example, $B\to D\pi e\nu$. To appreciate the theoretical challenge  consider
  the decay of a pseudoscalar meson to another pseudoscalar meson. The
  weak interaction couples to a $V-A$  hadronic current, $\widebar \psi' (\gamma^\mu-\gamma^\mu\gamma_5) \psi$, and a corresponding leptonic current; see Eq.~\eqref{eq:chargedCurr}. The probability amplitude for  the transition is given by 
\[
\mathcal{A}=\langle M' \ell\nu| \frac{g_2^2{V_{ij}}}{M_W^2} { \bar u^i_L\gamma^\mu d^j_L}\bar e_L\gamma_\mu\nu_L|M\rangle.
\]
The leptonic current, being excluded from the strong interactions, offers no difficulty and we can immediately  compute its contribution to the amplitude. The contribution to the amplitude from the hadronic side then involves
\begin{equation}
\label{eq:ffsdefd}
\langle \pvec{p}'|V^\mu | \pvec p\rangle = f_+(q^2)(p+p')^\mu + f_-(q^2)q^\mu,
\end{equation}
where $V^\mu=\bar u^i\gamma^\mu d^j$ and $q=p-p'$. The bra and ket stand for the meson final and initial states,
characterized only by their momentum and internal quantum numbers,
which are implicit in the formula.  The    matrix element is to be
computed non-perturbatively with regard to the strong
interactions. Only the vector current (not the axial) contributes, by
parity symmetry of the strong interactions.  The expression on the right-hand-side of \eqref{eq:ffsdefd} is the most general function of $p$ and $p'$  that is co-variant under Lorentz transformations ({\it i.e.}, transforms as a four vector). It involves the coefficients $f_\pm$, or ``form factors,'' that are a function of $q^2$ only, since the other invariants are fixed ($p^2=m_M^2$  and $p^{\prime2}=m_{M'}^2$).   In the 3-body decay, $p=p'+q$ so $q$ is the sum of the momenta of the leptons.  It is conventional to write the form factors as functions of $q^2$. When the term $f_-(q^2) q^\mu$  is contracted with the leptonic current one gets a negligible contribution, $q\cdot (V-A)\sim m_\ell$, when $\ell=e$ or $\mu$. So the central problem is to determine $f_+$. Symmetry considerations can produce good estimates of $f_+$ at specific kinematic points, which is sufficient for the determination of the magnitude of the KM matrix elements. Alternatively one may determine the form factor using Monte Carlo simulations of QCD on the lattice. 

\begin{exercises}
\begin{exercise}
Show  that $q\cdot (V-A)\sim m_\ell$ for the leptonic charged
current. Be more
precise than ``$\sim$.''
\end{exercise}
\end{exercises}

To see how this works, consider a simpler example first. We will show that the electromagnetic form factor for the pion is determined by the charge of the pion at $q^2=0$. Take $J^\mu$ to be the electromagnetic current of light quarks, $J^\mu(x) =\frac23\bar u(x)\gamma^\mu u(x) -\frac13\bar d(x) \gamma^\mu d(x)$. Charge conservation means $\partial_\mu J^\mu=0$. Now, the matrix element of this between pion states is
\begin{equation}
\label{eq:piff}
\vev{\pi(\pvec{p}')|J^\mu(0)|\pi(\pvec p)}=f_+(q^2)(p+p')^\mu+f_-(q^2)q^\mu
\end{equation}
Restoring the $x$ dependence in $J^\mu$ is easy, $J^\mu(x)=e^{i\hat P\cdot x} J^\mu(0) e^{-i\hat P\cdot x} $ where $\hat P^\mu$ is the 4-momentum operator. This just gives the above times $\exp(-i q\cdot x)$. Hence  the matrix element of the divergence of $J^\mu$ is just the above contracted with $q^\mu$. But $\partial_\mu J^\mu=0$ so we have
\[
f_+(q^2)(p+p')\cdot q+f_-(q^2)q^2=0
\]
The first term has $(p+p')\cdot q=(p+p')\cdot (p-p')=p^2-p^{\prime2}=m_\pi^2-m_\pi^2=0$ so we have $f_-(q^2)=0$. Moreover, the electric charge operator is
\[
\hat Q=\int d^3x\,J^0(x)
\]
and we should have 
\begin{equation}
\label{eq:relNorm}
\vev{\pi(\pvec{p}')|\hat Q|\pi(\pvec p)}=Q_\pi\vev{\pi(\pvec{p}')|\pi(\pvec p)}
=Q_\pi(2\pi)^32E\delta^{(3)}(\pvec p -\vec{p}')
\end{equation}
where $Q_\pi$ is the charge of the $\pi$ state ($\pm1$ for a $\pi^\pm$ and $0$ for a $\pi^0$) and we have used the relativistic normalization of states. Integrating the time component of ~\eqref{eq:piff} to compute the matrix element of $\hat Q$ is the same as inserting a factor of 
\[
\int d^3x\, e^{-iq\cdot x}=(2\pi)^3\delta^{(3)}(\pvec p -\vec{p}')
 \]
into the left hand side of \eqref{eq:piff} and comparing both sides we have
\[
2EQ_\pi = f_+(q^2)(E+E')
\]
or $f_+(0)=Q_\pi$ since the condition $\pvec{p}'=\pvec p$ for equal mass particles gives $E'=E$ and therefore $q^\mu=0$. To recap, conservation of $J^\mu$ implies $f_-(q^2)=0$ and $f_+(0)=\pm1$ for charged pions,  $f_+(0)=0$ for neutral pions.

\underline{$K\to\pi \ell\nu$:} One can repeat this for kaons and pions, where the symmetry now is Gell-Mann's flavor-$SU(3)$. Let me remind you of this, so you do not confuse this ``flavor'' symmetry with the ``flavor'' symmetry we introduced earlier. If we want to understand the behavior of matter at  energies sufficiently high that kaons are produced but still too low to produce  charmed states,   we can use for the Lagrangian
\[
\mathcal{L}=\bar u i\slashed{D} u +\bar d i\slashed{D} d +\bar s i\slashed{D} s 
\]
where the covariant derivative only contains the gluon field. Electromagnetic and weak interactions have to be added as perturbations. The Lagrangian is invariant under the $SU(3)$ group of transformations in which the $u$, $d$ and $s$ quarks form a triplet: if $q=(u,d,s)^T$, the symmetry is $q\to Uq$ with $U$ a unitary $3\times 3$ matrix. The pions and kaons, together with the $\eta$ particle form an octet of $SU(3)$: the $3\times 3$ traceless matrix
\[
M=\begin{pmatrix}
\frac{\pi^0}{\sqrt2}-\frac{\eta}{\sqrt6}&\pi^+ & K^+\\
\pi^-&-\frac{\pi^0}{\sqrt2}-\frac{\eta}{\sqrt6}&K^0\\
K^-&\widebar K^0&\frac{\eta}{\sqrt3}
\end{pmatrix}.
\]
The flavor quantum numbers of these are in 1-to-1 correspondance with the matrix $q\times \widebar{q}^T$. In particular note that the 2-3 element, the $K^0$, has content $q_2\bar q_3=d\bar s$: kaons have strangeness $-1$, while anti-kaons have strangeness $+1$.  Symmetry means that the quantum mechanical probability amplitudes (a.k.a. matrix elements) have to be invariant  under $M\to UMU^\dagger$.  The symmetry implies $f_-(q^2)=0$ and $f_+(0)=1$ for the form factors of the conserved currents associated with the $SU(3)$ symmetry transformations.  In reality, however,  this symmetry does not hold as accurately as isospin. A better Lagrangian includes masses for the quarks, and  masses vary among the quarks,  breaking the symmetry:
\[
\mathcal{L}=\bar u (i\slashed{D}-m_u) u +\bar d (i\slashed{D}-m_d) d +\bar s (i\slashed{D}-m_s) s 
\]
Since the largest source of symmetry breaking is the mass of the strange quark ($m_s\gg m_d\gtrsim m_u$), one expects corrections to  $f_+(0)-1$ of order $m_s$. But since $f_+$ is  dimensionless the correction must be relative to some scale,  $f_+(0)-1\propto m_s/\Lambda$, with $\Lambda$ a hadronic scale, say,  $\Lambda\sim 1$~GeV. This seems like bad news, an uncontrolled 10\% correction. Fortunately, by a theorem of Ademolo and Gatto, the symmetry breaking parameter appears at second order, $f_+(0)-1\propto (m_s/\Lambda)^2\sim1$\%. Combining  data for neutral and charged semi-leptonic $K$ decays the PDG gives $|V_{us}|f_+(0)=0.2163\pm0.0005$~\cite{pdg} which to a few percent can be read off as the value of the magnitude of the KM matrix element. Monte-Carlo simulations of QCD on a lattice give a fairly accurate determination of the form factor; the same section of the PDG reports $f_+(0)=0.960\pm0.005$ which it uses to give $|V_{us}|=0.2253\pm0.0008$. Note that the theoretical calculation of $f_+$ is remarkably accurate, about at the half per-cent level. The reason this accuracy can be achieved is that one only needs to calculate the deviation of $f_+(0)$ from unity, an order $(m_s/\Lambda)^2$ effect, with moderate accuracy.

\underline{$B\to D\ell\nu$:} We cannot extend this to the heavier quarks because then $m_c/\Lambda>1$ is a bad expansion parameter. Remarkably, for transitions among heavy quarks there is another symmetry, dubbed ``Heavy Quark Symmetry'' (HQS),  that allows similarly successful predictions; for a basic introduction see~\cite{Grinstein:1992ss}. For  transitions from a heavy meson (containing a heavy quark, like the $B$ or $D$ mesons) to a light meson (made exclusively of light quarks, like the $\pi$ or $K$ mesons) one requires other methods, like lattice QCD, to determine the remaining KM matrix elements.

A word about naming of mesons. Since $K^0$ by convention has strangeness $-1$, we take by analogy $B^0$ to have bottomness (or beauty, in Europe) $-1$. So the flavor quantum numbers of heavy mesons are $\widebar{B}^0=b\bar d$, $B^-=b\bar u$, $\widebar{B}_s=b\bar s$, $D^0=c\bar u$, $D^+=c\bar d$, $D_s=c\bar s$. 

Here is an elementary, mostly conceptual, explanation of how HQS works. The heavy mesons are composed of a quark that is very heavy compared to the binding energy of mesons, plus a light anti-quark making the whole thing neutral under color, plus a whole bunch of glue and quark-antiquark pairs. This ``brown muck'' surrounding and color-neutralizing the heavy quark is complicated and we lack good, let alone precise, mathematical models for it.   The interactions of this brown muck  have low energy compared to the mass of the heavy quark,  so that they do not change the state of motion of the heavy quark: in the rest frame of the meson, the heavy quark is at rest. The central observation of HQS is that all the brown muck  sees is a static source of color, regardless of the heavy quark mass. Hence there is a symmetry between $B$ mesons and $D$ mesons: they have the same brown muck, only different static color sources. A useful  analogy to keep in mind is from atomic physics: the chemical properties of different isotopes of the same element are the same to high precision because the electronic cloud (the atomic brown muck) does not change even as the mass of the atomic nucleus (the atomic heavy quark) changes. 

To put this into equations, we start by characterizing the heavy meson state by its velocity rather than its momentum, $v^\mu = p^\mu/m$. That is because we are considering the limit of infinite mass of the heavy quark, $m\to \infty$. Notice that infinite mass does not mean the meson is at rest. You can boost to a frame where it moves. More interestingly, even if both $b$ and $c$ quarks are infinitely heavy, the process $b\to c \ell\nu$ can produce a moving $c$ quark in the rest-frame of the decaying $b$-quark. Another trivial complication is that the relativistic normalization of states, as in \eqref{eq:relNorm}, includes a factor of energy, $E\to\infty$. So we take $|\pvec v\rangle=(1/\sqrt{m})|\pvec p\rangle$. For the application of the HQS it is more convenient (and natural) to parametrize the matrix element of the vector current in terms of the 4-velocities. Doing so, and using an argument analogous to that introduced  previously  to show  $f_-(q^2)=0$, we have
\[
\vev{\pvec{v}'|V^\mu|\pvec v}=\xi(v\cdot v')(v+v')^\mu.
\]
Comments: (i) the infinitely heavy states could be two same flavored mesons with a flavor diagonal current, e.g.,  $B^-\to B^-$  with $V^\mu = \widebar b\gamma^\mu b$, or two different flavors with an of diagonal current, {\it e.g.}   $B^-\to D^0$  with $V^\mu = \widebar c\gamma^\mu b$; (ii) the form factor, now labeled $\xi$ and called an ``Isgur-Wise'' function, is in principle a function of the three Lorentz invariants we can make out of the 4-vectors $v^\mu$ and $v^{\prime\mu}$, but since $v^2=v^{\prime2}=1$ it only depends on $v\cdot v'$;  (iii) rewriting this in terms of 4-momenta gives a relation between $f_+$ and $f_-$ (but not $f_-=0$); and, most importantly, (iv) the analogue to $f_+(0)=1$ is
\[
\xi(1)=1.
\]
Note that $v\cdot v'=1$ corresponds to the resulting meson not moving relative to the decaying one (in other words, remaining at rest in the rest frame of the decaying meson), so that the invariant mass of the lepton pair, $q^2$,  is as large as it can be: $v\cdot v'=1$ is $q^2 = q^2_{\rm max}=(m_B-m_D)^2$. 

The analogue of the theorem of Ademolo and Gato for HQS is Luke's theorem~\cite{Luke:1990eg}. It states that the corrections to the infinite mass predictions for form factors at $v\cdot v'=1$ first appear  at order $1/m^2$ rather than the na\"ively expected~$1/m$.

The prediction of the $B\to D$ form factors  at one kinematic point ($q^2=q^2_{\rm max}$) can be used to experimentally determine $|V_{cb}|$. Again a tension arises between theory and experiment: at the best theory point ($q^2=q^2_{\rm max}$) the decay rate vanishes. In practice this problem is circumvented by extrapolating from $q^2<q^2_{\max}$ and by including $B\to D^* \ell\nu$ in the analysis. The $D^*$ is the spin-1 partner of the $D$ meson. We have not explained this here, but HQS relates the $D$ to the $D^*$ mesons: they share a common  brown muck. The reason is simple, the spin of the heavy quark interacts with the brown muck via a (chromo-)magnetic interaction, but magnetic moments are always of the form charge-over-mass, $g/m$, so they vanish at infinite mass. We can combine the spin-$\frac12$ heavy quark with the spin-$\frac12$ brown muck in a spin-0 or a spin-1 state, and since the spin does not couple,  they have the same mass and the same matrix elements (form factors). 

\begin{exercises}
\begin{exercise}
For $B\to D\ell\nu$ write the form factors $f_\pm(q^2)$ in terms of the Isgur-Wise function. What does $\xi(1)=1$ imply for $f_\pm$? Eliminate the Isgur-Wise function to obtain a relation between $f_+$ and $f_-$. 
\end{exercise}
\begin{solution}
In 
\[
\vev{\pvec{v}'|V^\mu|\pvec v}=\xi(v\cdot v')(v+v')^\mu
\]
we need to (i) write $v=p/m_b$ and $v'=p'/m_c$ and (ii) replace $|\pvec v\rangle\to (1/\sqrt{m_b})|\pvec p\rangle $ and $|\pvec{v}'\rangle\to (1/\sqrt{m_c})|\pvec{p}'\rangle $, thus:
\[ 
\frac1{\sqrt{m_bm_c}}\vev{\pvec{p}'|V^\mu|\pvec p}=\xi(v\cdot
v')\left(\frac{p^\mu}{m_b}+\frac{p^{\prime\mu}}{m_c}\right) 
\]
Comparing with \eqref{eq:ffsdefd}, we read off
\[
f_\pm(q^2)=\frac12\sqrt{m_bm_c}\left(\frac1{m_b}\pm\frac1{m_c}\right)\xi((q^2-m_b^2-m_c^2)/2m_bm_c).
\] 
The relation between form factors is $f_-/f_+=(m_c-m_b)/(m_c+m_b)$
and we note that this correctly gives $f_-=0$ when the two quarks are
identical. Finally, at $q^2=q^2_{\rm max}$ we have  
\[
f_\pm(q_{\rm max}^2)=\frac12\sqrt{m_bm_c}\left(\frac1{m_b}\pm\frac1{m_c}\right).
\] 
\end{solution} 
\end{exercises}

\subsubsection{Inclusive semileptonic decays}
As we have said, the inclusive semileptonic decay rate
$\Gamma(\widebar B\to X\ell\nu)$  means the rate of decay of a
$\widebar B$ to $\ell\nu$ plus anything. We further distinguish
$\Gamma(\widebar B\to X_c\ell\nu)$ when the anything contains a charm
quark and therefore the underlying process at the quark level is
$b\to c\ell\nu$ and similarly $\Gamma(\widebar B\to X_u\ell\nu)$ from
$b\to u\ell\nu$. 

\begin{figure}
\begin{center}
\includegraphics[width=0.6\textwidth]{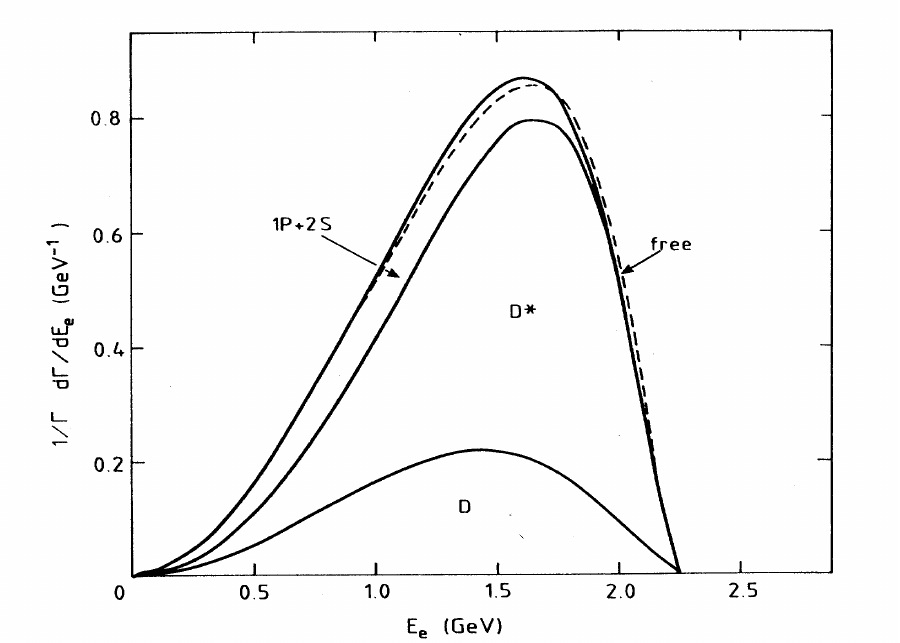}
\end{center}
\caption{\label{fig:isgw1}Quark-hadron duality in $B\to X_c e\nu$ in
  a non-relativistic model of mesons. The figure, taken from
  \cite{Isgur:1988gb},  shows how the
  spectrum with respect to the electron energy normalized to the
  total semileptonic width,  $\frac{1}{\Gamma} \frac{d\Gamma}{dE_e}$, is
  built up from exclusive decays. The lowest solid line is the
  contribution from $B\to De\nu$, the next higher one includes the
  $D^*$ final state and the highest one is the total contribution from
all 1S, 1P and 2S states. The dashed line corresponds to the free
quark $b\to c\ell\nu$ rate. }
\end{figure}

\begin{figure}
\begin{center}
\includegraphics[width=0.7\textwidth]{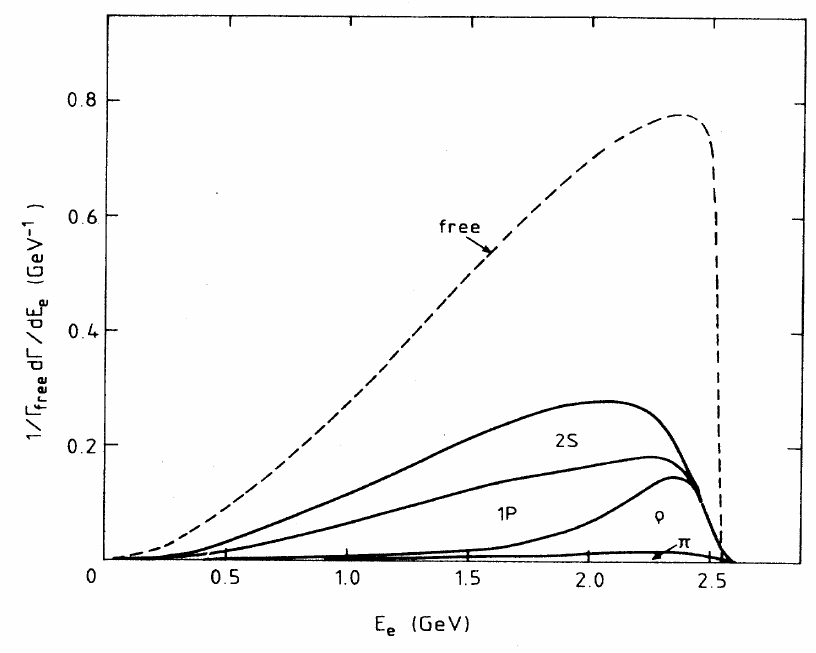}
\end{center}
\caption{\label{fig:isgw2} As in Fig.~\ref{fig:isgw1} but for $b\to u
  e \nu$, from~\cite{Isgur:1988gb}.  }
\end{figure}

There is good reason to believe that {\it quark-hadron duality} holds
for these quantities. Quark-hadron duality means that instead of
computing the rate for the transition between hadrons, in this case
mesons, we can compute the rate for the transition between quarks and
the answer is the same, $\Gamma(\widebar B\to X_c\ell\nu)=\Gamma(b\to
c\ell\nu)$. Fig.~\ref{fig:isgw1} shows in solid curves how the  spectrum with respect to
the electron energy,  $d\Gamma(B\to Xe\nu)/dE_e$, builds up from
exclusive modes, starting with $B\to D e\nu$ and adding to it $B\to
D^* e \nu$ and then the sum of all 1S, 1P and 2S states. By comparison
the $b\to c e\nu$ spectrum is shown as a dashed line. The agreement
between the sum over exclusives and the free quark decay is apparent.
By comparison Fig.~\ref{fig:isgw2} shows the $b\to u e \nu$ case. To
reproduce the free quark rate many more states must be included. 

Notice that the endpoint of the spectrum for $B\to X_u e\nu$ extends
beyond that of $B\to X_c e \nu$. This was the basis for early
determinations of $|V_{ub}|$, as mentioned above. The point is that
$|V_{ub}|\ll |V_{cb}|$ so the $b\to u e \nu$ transition hides under
$b\to c e \nu$ for most electron energies.  But the theoretical
determination of the spectrum constrained to the narrow region close
to the end of the spectrum is not accurate. Modern determinations of
$|V_{ub}|$ rely on summing over precise measurements of exclusive non-charm decay
exclusive  modes over the whole spectrum and using kinematic variables
other than $E_e$. 

Remarkably, quark-hadron duality for semileptonic heavy quark decays
can be established from first principles using HQS~\cite{Chay:1990da}.
Moreover, finite mass corrections  can be systematically
incorporated~\cite{Manohar:1993qn,Bigi:1992su}. Theory gives solid
predictions for moments of the spectrum in terns of few unknown
non-perturbative parameters that can be accurately fit to
experiment~\cite{Bauer:2004ve}, resulting in a determination at about
1\% precision. 

The green ring in Fig.~\ref{fig:CKMfitterTRIANGLES} shows the region of the $\bar\rho$-$\bar\eta$ plane allowed by the determination of $|V_{ub}|$. More precisely, note that  $\sqrt{\rho^2+\eta^2}=|V_{ub}/V_{us}V_{cb}|$ so that the ring requires the determination of the three KM elements. It is labeled ``$|V_{ub}|$'' because this is the least accurately determined of the three KM elements required. 

\subsubsection{Collecting results}
While we have not presented a full account of the measurements and
theory that are used in the determination of the KM magnitudes,  by now you
should have an idea of the variety of methods employed. 

The PDG gives for the full fit  of the magnitudes of the KM matrix elements
\[
|V|=\begin{pmatrix}
0.97427\pm0.00014&0.22536\pm0.00061&0.00355\pm0.00015\\
0.22522\pm0.00061&0.97343\pm0.00015&0.0414\pm0.0012\\
0.00886^{+0.00033}_{-0.00032}&0.0405^{+0.0011}_{-0.0012}&0.99914\pm0.00005
\end{pmatrix},
\]
or, in terms of the Wolfenstein parameters,
\begin{align*}
\lambda&=0.22537\pm0.00061,& A&=0.814^{+0.023}_{-0.024}\,,\\
\widebar{\rho}&=0.117\pm0.021,& \widebar{\eta}&=0.353\pm0.013\,.
\end{align*}
It also gives, for  the Jarlskog determinant,
$J=(3.06^{+0.21}_{-0.20})\times 10^{-5}$.

\subsection{Angles}
The angles of the unitarity triangle are associated with CP violation.
Next chapter is devoted to this. Here is a brief summary to  two routes to their determination:
\begin{enumerate}[(i)]
\item Neutral Meson Mixing.  It
  gives, for example, $V_{tb}^{\phantom{*}} V_{td}^*$ in the case of
  $B_d$ mixing and $V_{tb}^{\phantom{*}} V_{ts}^*$ for $B_s$
  mixing. The case of $K^0$ mixing is, as we will see,
  more complex. The yellow (``$\Delta m_d$'') and
  orange (``$\Delta m_d$ \& $\Delta m_s$'')  circular rings
  centered at $(1,0)$ in Fig.~\ref{fig:CKMfitterTRIANGLES} are
  determined by the rate of $B_d$ mixing and by the ratio of rates of
  $B_d$ and $B_s$ mixing, respectively. The ratio is used because in
  it some uncertainties cancel, hence yielding a thiner ring. The bright green region labeled
  $\varepsilon_K$ is determined by CP violation in
  $K^0$-$\widebar{K}^0$ mixing.

\item CP asymmetries. Decay asymmetries, measuring the difference in
  rates of a process and the CP conjugate process, directly probe
  relative phases of KM elements, and in particular the unitarity
  triangle angles $\alpha$, $\beta$ and $\gamma$. We will also study
  these, with particular attention to the  poster boy, the
  determination of $\sin(2\beta)$ from $B_d\to \psi K_S$, which is largely free from hadronic uncertainties. In Fig.~\ref{fig:CKMfitterTRIANGLES}  the blue and brown wedges labeled $\sin2\beta$ and $\gamma$, respectively, and the peculiarly shaped light blue region labeled $\alpha$ are all obtained from various CP asymmetries in decays of $B_d$ mesons.   
\end{enumerate}

\section{FCNC}
\label{sec:fcnc}
This stands for {\bf F}lavor {\bf C}hanging {\bf N}eutral {\bf
  C}urrents, but it is used more generally to mean Flavor Changing
Neutral transitions, not necessarily ``currents.''  By this we mean an
interaction that changes flavor but does not change electric
charge. For example, a transition from a $b$-quark to an $s$- or
$d$-quarks would be flavor changing neutral, but not  so a transition from a $b$-quark to a $c$- or $u$-quark.  Let's review flavor changing transitions in the SM:
\begin{enumerate}
\item Tree level. Only interactions with the charged vector bosons
  $W^\pm$ change flavor; {\it cf.} \eqref{eq:chargedCurr}. The photon and $Z$ coupe diagonally in flavor space, so these ``neutral currents'' are flavor conserving.   \\
\begin{tikzpicture} 
\coordinate[label=left:$d$] (d);
\coordinate[label=right:$u$,right=3 cm of d] (u);

\coordinate (v1) at ($0.5*(u)+0.5*(d)+(0,-0.5)$);
\coordinate (v2) at ($(v1)+ (2,-0.5)$);
\coordinate[label=right:$\widebar \nu$] (barnu) at ($(v2)+(1.5,0.5)$);
\coordinate[label=right:$e$] (e) at  ($(v2)+(1.5,-0.5)$);

\draw[particle] (d) -- (v1);
\draw[particle] (v1) -- (u);
\draw[particle] (barnu) -- (v2);
\draw[particle] (v2) -- (e);

\draw[photon] (v2) -- node[below=0.1]{$W^-$} (v1);

\node at ($(v1)+(-5.5,0)$) {For example, $n\to p e\widebar\nu$ is };

\end{tikzpicture}
 \item 1-loop. Can we have FCNCs at 1-loop? Say, $b\to s\gamma$? Answer: YES. Here is\break \raisebox{1.5cm}{a diagram:}\hspace{2cm}
\begin{tikzpicture} 
\coordinate[label=left:$b$] (b) at (-3,0);
\coordinate[label=right:$s$] (s) at (3,0);

\coordinate (v1) at (-1,0);
\coordinate (v2) at (1,0);
\coordinate (v3) at (0,-1);

\coordinate[label=right:$\gamma$] (g) at ($(v3)+(2,-0.3)$);

\draw[particle] (b) -- (v1);
\draw[particle]  (v1) arc (180:270:1) node[left=0.7cm]{$u,c,t$};
\draw[particle] (v3) arc (270:360:1);
\draw[particle] (v2) -- (s);

\draw[photon] (g) --  (v3);
\draw[photon] (v1) --  node[above]{$W$} (v2);

\end{tikzpicture}\\
Hence, FCNC are suppressed in the SM by a 1-loop factor of $\displaystyle \sim \frac{g_2^2}{16\pi^2}\sim\frac{\alpha}{4\pi c^2_W}$ relative to the flavor changing charged currents. 
\end{enumerate}

\newpage

\begin{exercises}
\begin{exercise}
Just in case you have never computed the $\mu$-lifetime, verify that 
\[\tau^{-1}_\mu\approx\Gamma(\mu\to e\nu_\mu\widebar\nu_e) = \frac{G_F^2m_\mu^5}{192\pi^3}\]
neglecting $m_e$, at lowest order in perturbation theory.
\end{exercise}
\begin{exercise}
Compute the amplitude for $Z\to b\widebar s$ in the SM to lowest
order in perturbation theory (in the strong and electroweak
couplings).  Don't bother to compute integrals explicitly, just make
sure they are finite (so you could evaluate them numerically if need
be). Of course, if you can express the result in closed analytic form,
you should. See~Ref.~\cite{Clements:1982mk}.
\end{exercise}
\end{exercises}

\section{GIM-mechanism: more suppression of FCNC}
\subsection{Old GIM}
\label{ssec:oldgim}
Let' s imagine a world with a light top and a hierarchy  $m_u< m_c <
m_t \ll M_W$. Just in case you forgot, the real world is not like
this, but rather it has $m_u\ll m_c \ll M_W \approx \tfrac12 m_t$. We can make a lot of progress towards the computation of the Feynman graph for $b\to s\gamma$ discussed previously without computing any integrals explicitly:\\[0.5cm]
\parbox[c]{6cm}{\begin{tikzpicture}[scale=0.7]
\coordinate[label=left:$b$] (b) at (-3,0);
\coordinate[label=right:$s$] (s) at (3,0);
\coordinate (v1) at (-1,0);
\coordinate (v2) at (1,0);
\coordinate (v3) at (0,-1);
\coordinate (g) at ($(v3)+(2,-0.3)$);
\draw[particle] (b) -- (v1);
\draw[particle]  (v1) arc (180:270:1) node[left=0.7cm]{$u,c,t$};
\draw[particle] (v3) arc (270:360:1);
\draw[particle] (v2) -- (s);
\draw[photon] (g) node[right] {$\gamma(q,\epsilon)$} --(v3);
\draw[photon] (v1) --  node[above]{$W$} (v2);
\end{tikzpicture}
}
$\displaystyle = e q_\mu\epsilon_\nu\widebar u(p_s)\sigma^{\mu\nu}{\textstyle \left(\frac{1+\gamma_5}{2}\right)}u(p_b)\frac{m_b}{M_W^2}\,\frac{g_2^2}{16\pi^2}\cdot I$\\
where
\[I=\sum_{i=u,c,t}V_{ib}^{\phantom{*}}V_{is}^*F({\textstyle\frac{m_i^2}{M_W^2}})
 \]
and $F(x)$ is some function that results form doing the integral
explicitly, and we expect it to be of order 1. The coefficient of this
unknown integral can be easily understood. First, it has the obvious
loop factor ($g_2^2/16\pi^2$), photon coupling constant ($e$) and KM
factors $V_{ib}^{\phantom{*}}V_{is}^*$ from the charged curent
interactions. 
Next, in order to produce a real (on-shell) photon the interaction has
to be
of the transition magnetic-moment form, $F_{\mu\nu}\widebar
s\sigma^{\mu\nu} b$, which translates into the Dirac spinors $u(p)$
for the quarks combining with the photon's momentum $q$ and
polarization vector ($\epsilon$) through $ q_\mu\epsilon_\nu\widebar
u(p_s)\sigma^{\mu\nu}u(p_b)$.\footnote{The other possibility, that the
  photon field $A_\mu$ couples to a flavor changing current, $A_\mu
  \widebar b \gamma^\mu s$, is forbidden by electromagnetic gauge
  invariance. Were you to expand the amplitude in powers of $q/M_Z$
  you could in principle obtain at  lowest order the contribution, $\epsilon^\mu
  \widebar u(p_s)\gamma^{\mu}u(p_b)$. But this should be invariant  (gauge invariance) under $\epsilon^\mu\to\epsilon^\mu+q^\mu$, where $q=p_b-p_s$.} Finally, notice that the external quarks interact with
the rest of the diagram through a weak interaction, which involves
only left-handed fields. This would suggest getting an amplitude
proportional to $\widebar
u(p_s)\left(\frac{1+\gamma_5}{2}\right)\sigma^{\mu\nu}{\textstyle
  \left(\frac{1-\gamma_5}{2}\right)}u(p_b)$ which, of course,
vanishes. So we need one or the other of the external quarks to flip
its  chirality,  and only then interact. A chirality flip produces a factor of the mass of the quark and we have chosen to flip the chirality of the $b$ quark because $m_b\gg m_s$. This explains both the factor of $m_b$ and the projector $\frac{1+\gamma_5}{2}$ acting on the spinor for the $b$-quark. The correct units (dimensional analysis) are made up by the factor of $1/M_W^2$. 

Now, since we are pretending $m_u< m_c < m_t \ll M_W$, let's expand in a Taylor series, $F(x)=F(0)+xF'(0)+\cdots$ 
\[
I=\left(\sum_{i=u,c,t} V_{ib}^{\phantom{*}}V_{is}^*\right) F(0)+\left(\sum_{i=u,c,t} V_{ib}^{\phantom{*}}V_{is}^*\frac{m_i^2}{M_W^2}\right) F'(0)+\cdots
\]
Unitarity of the KM matrix gives $\sum_{i=u,c,t} V_{ib}^{\phantom{*}}V_{is}^*=0$ so the first term vanishes. Moreover, we can rewrite the unitarity relation as giving one term as a combination of the other two, for example, 
\[ V_{tb}^{\phantom{*}}V_{ts}^* =- \sum_{i=u,c} V_{ib}^{\phantom{*}}V_{is}^*
\]
giving us
\[
I\approx  -F'(0) \sum_{i=u,c} V_{ib}^{\phantom{*}}V_{is}^*\frac{m_t^2-m_i^2}{M_W^2}
\]
We have uncovered additional FCNC suppression factors. Roughly, 
\[
I\sim V_{ub}^{\phantom{*}}V_{us}^*\frac{m_t^2-m_u^2}{M_W^2} + V_{cb}^{\phantom{*}}V_{cs}^*\frac{m_t^2-m_c^2}{M_W^2}\sim \epsilon^4\frac{m_t^2}{M_W^2} +\epsilon^2\frac{m_t^2}{M_W^2} .
\]
So in addition the 1-loop suppression, there is a mass suppression ($m_t^2/M_W^2$) and a mixing angle suppression ($\epsilon^2$).  This combination of suppression factors was uncovered by Glashow, Iliopoulos and Maiani (hence ``GIM'') \cite{Glashow:1970gm} back in the days when we only knew about the existence of three flavors, $u$, $d$ and $s$. 
They studied neutral kaon mixing, which involves a FCNC for $s$ to $d$ transitions and realized that theory would grossly over-estimate the mixing rate unless a fourth quark existed (the charm quark, $c$) that would produce the above type of cancellation (in the 2-generation case). Not only did they explain kaon mixing and predicted the existence of charm, they even gave a rough upper bound for the mass of the charm quark, which they could do since the contribution to the FCNC grows rapidly with the mass, as shown above. We will study kaon mixing in some detail later, and we will see that the top quark contribution to mixing is roughly as large as that of the charm quark: Glashow, Iliopoulos and Maiani were a bit lucky, the parameters of the SM-CKM could have easily favored top quark mediated dominance in kaon mixing and their bound could have been violated. As it turns out, the charm was discovered shortly after their work, and the mass turned out to be close to their upper bound.

\subsection{Modern GIM}
We have to revisit the above story, since $m_t\ll M_W$ is not a good approximation. Consider our example above, $b\to s\gamma$. The function $F(x)$ can not be safely Taylor expanded when the argument is the top quark mass. However, $I$ is invariant under $F(x)\to F(x)+\text{constant}$, so we may choose without loss of generality $F(0)=0$. Then  
\begin{align*}
I&= -V_{cb}^{\phantom{*}}V_{cs}^*\left(F({\textstyle\frac{m_t^2}{M_W^2}})-F'(0)\frac{m_c^2}{M_W^2}\right)
-V_{ub}^{\phantom{*}}V_{us}^*\left(F({\textstyle\frac{m_t^2}{M_W^2}})-F'(0)\frac{m_u^2}{M_W^2}\right)+\cdots\\
&=F({\textstyle\frac{m_t^2}{M_W^2}})V_{tb}^{\phantom{*}}V_{ts}^*
+F'(0)\sum_{i=u,c} V_{ib}^{\phantom{*}}V_{is}^*\frac{m_i^2}{M_W^2}+\cdots\\
&\sim \epsilon^2 F({\textstyle\frac{m_t^2}{M_W^2}})
\end{align*}
We expect $F(x)$ to be order 1. This is indeed the case, $F(x)$ is a slowly increasing function of $x$ that is of order $1$ at the top quark mass. The contributions from $u$ and $c$ quarks to $I$ are completely negligible, and virtual top-quark exchange dominates this amplitude. 

\begin{exercises}
\begin{exercise}
Consider $s\to d\gamma$. Show that the above type of analysis suggests that virtual top quark exchange no longer dominates, but that in fact the charm and top contributions are roughly equally important. {\it Note: For this you need to know the mass of charm relative to $M_W$. If you don't, look it up!}
\end{exercise}
\begin{solution}
For  $s\to d\gamma$ we now have\\[0.5cm]
\parbox[c]{6cm}{\begin{tikzpicture}[scale=0.7]
\coordinate[label=left:$s$] (b) at (-3,0);
\coordinate[label=right:$d$] (s) at (3,0);
\coordinate (v1) at (-1,0);
\coordinate (v2) at (1,0);
\coordinate (v3) at (0,-1);
\coordinate (g) at ($(v3)+(2,-0.3)$);
\draw[particle] (b) -- (v1);
\draw[particle]  (v1) arc (180:270:1) node[left=0.7cm]{$u,c,t$};
\draw[particle] (v3) arc (270:360:1);
\draw[particle] (v2) -- (s);
\draw[photon] (v3) --  (g) node[right] {$\gamma(q,\epsilon)$};
\draw[photon] (v1) --  node[above]{$W$} (v2);
\end{tikzpicture}
}
$\displaystyle = e q_\mu\epsilon_\nu\widebar u(p_d)\sigma^{\mu\nu}{\textstyle \left(\frac{1+\gamma_5}{2}\right)}u(p_d)\frac{m_s}{M_W^2}\,\frac{g_2^2}{16\pi^2}\cdot I$\\
where
\[I=\sum_{i=u,c,t}V_{is}^{\phantom{*}}V_{id}^*F({\textstyle\frac{m_i^2}{M_W^2}})
 \]
We still have 
\[
I=F({\textstyle\frac{m_t^2}{M_W^2}})V_{ts}^{\phantom{*}}V_{td}^*
+F'(0)\sum_{i=u,c} V_{is}^{\phantom{*}}V_{id}^*\frac{m_i^2}{M_W^2}+\cdots
\]
But now the counting of powers of $\epsilon$ is a bit different: $|V_{ts}^{\phantom{*}}V_{td}^*|\sim \epsilon^5$ while $|V_{is}^{\phantom{*}}V_{id}^*|\sim\epsilon$ for either $i=c$ or $i=u$. Since $m_u\ll m_c$ we neglect the $u$-quark contribution. Using $F\sim1$ at the top, then the ratio of top to charm contributions is $\sim \epsilon^5/(\epsilon m_c^2/M_W^2)=(\epsilon^2 M_W/m_c)^2$. Using $M_W/m_c\approx 80/1.5$ and $\epsilon\approx 0.1$ the ratio os 0.3, and we have every right to expect the two contributions are comparable in magnitude. 
\end{solution}
\end{exercises}

\section{Bounds on New Physics}

Now let's bring together all we have learned. Let's stick to the
process $b\to s \gamma$, which in fact places some of the most
stringent constraints on models of new physics (NP). Let's model the
contribution of NP by adding a dimension 6 operator to the
Lagrangian,\footnote{The field strength should be the one for weak hypercharge, and the coupling constant should be $g_1$. This is just a distraction and does not affect the result; in the interest of pedagogy I have been intentionally sloppy.} 
\[
\Delta \mathcal{L}=\frac{C}{\Lambda^2}e F_{\mu\nu} H \widebar q_L \sigma^{\mu\nu} b_R =
\frac{evC}{\sqrt{2}\Lambda^2} F_{\mu\nu} \widebar s_L \sigma^{\mu\nu} b_R+\cdots
\]
I have assumed the left handed doublet belongs in the second
generation. The coefficient of the
operator is $C/\Lambda^2$: $C$ is dimensionless and we assume it is of
order 1, while $\Lambda$ has dimensions of mass and indicates the
energy scale of the NP.  It is easy to compute this term's
contribution to the amplitude. It is even easier to roughly compare it to that of the SM, 
\[
\frac{\mathcal{A}_{\text{NP}}}{\mathcal{A}_{\text{SM}}}\sim \frac{\frac{vC}{\sqrt{2}\Lambda^2}}{|V_{tb}^{\phantom{*}}V_{ts}^*|\frac{\alpha}{4\pi s^2_W}\frac {m_b}{M_W^2}}
\]
Require this ratio be less than, say, 10\%, since the SM prediction agrees at that level with the measurement. This gives,
\[
C^{-1}\Lambda^2 \gtrsim \frac{vM_W^2s_W^2}{\sqrt{2}m_b|V_{tb}^{\phantom{*}}V_{ts}^*|\frac{\alpha}{4\pi}}\cdot\frac1{0.1}
\quad\Rightarrow\quad \Lambda\gtrsim 70~\text{TeV}.
\]
This bound is extraordinarily strong. The energy scale of 70~TeV is  much higher than that of any existing or planned particle physics accelerator facility. 

In the numerical bound above we have taken $C\sim1$, but clearly a small coefficient would help bring the scale of NP closer to experimental reach. The question is what would make the coefficient smaller. One possibility is that the NP is weakly coupled and the process occurs also at 1-loop but with NP mediators in the loop. Then we can expect $C\sim \alpha /4\pi s_W^2$, which brings the bound on the scale of new physics down to about 4~TeV.

\begin{figure}
\begin{center}
\includegraphics[width=0.7\textwidth]{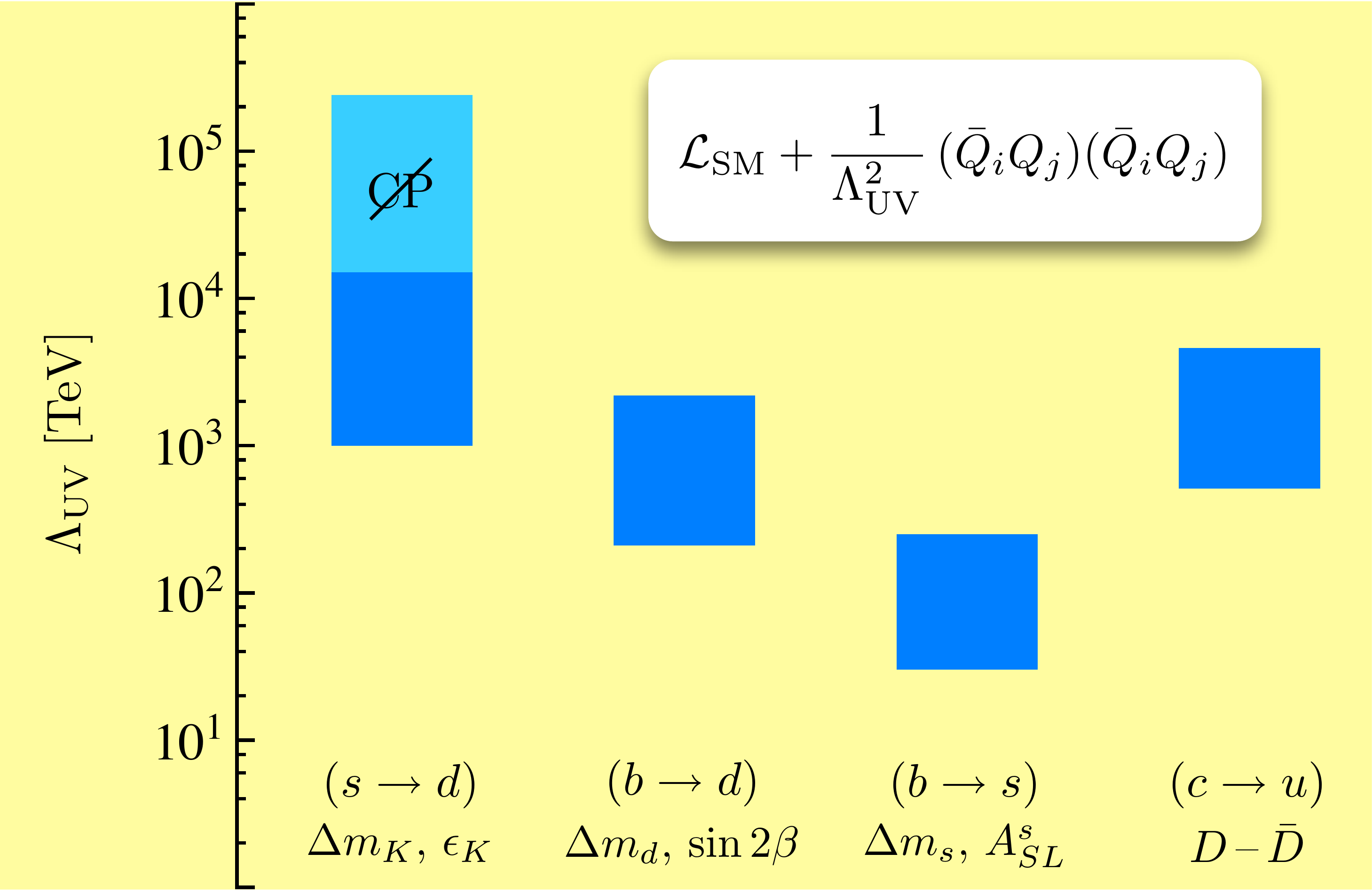}
\end{center}
\caption{\label{fig:neubertEPS2011}Bounds on the  NP scale from various processes. The NP is modeled as  dimension 6  operators. No accidental suppression of the coefficient (as in MFV) is included. The $b\to s$ case is consistent with the explicit $b\to s\gamma$ example worked out in these notes. The figure is taken from M.~Neubert's talk at EPS 2011. }
\end{figure}

Figure \ref{fig:neubertEPS2011} shows bounds on the scale of NP  from various processes. The NP is modeled as dimension 6  operators, just as in our discussion above. The coefficients of the operators $C/\Lambda^2$  are assumed to have $C\approx 1$.  The $b\to s$ case is consistent with our discussion above. 

\subsection{Minimal Flavor Violation}
Suppose we extend the SM by adding terms (local,\footnote{By ``local''
  we mean a product of fields all evaluated at the same spacetime
  point.} Lorentz invariant and gauge invariant) to the Lagrangian. Since the SM
already includes all possible monomials (``operators'')  of dimension
4 or smaller, we consider adding operators of dim~$\ge5$. We are going
to impose an additional constraint, and we will investigate its
consequence. We will require that these operators be invariant under
the flavor transformations, comprising the group $G_F$. We will include the Yukawa matrices as spurions:
\begin{equation}
\label{eq:mvf-trans}
q_L\to U_q\; q_L\;, \quad u_R\to U_u\; u_R \;, \quad d_R\to U_d\; d_R~,\quad \lambda_U\to U_q^{\phantom{\dagger}}\lambda_U U_u^\dagger, \quad\lambda_D\to U_q^{\phantom{\dagger}}\lambda_D U_d^\dagger.
\end{equation}
We add some terms to the Lagrangian
\[
\mathcal{L}\to\mathcal{L}+\Delta\mathcal{L}, \qquad \Delta\mathcal{L}=\sum_i c_iO_i
\]
with $O_i$ operators of dim~$\ge5$ invariant under \eqref{eq:mvf-trans}.  For example,
\begin{align*}
O_1&= G^a_{\mu\nu} H \widebar u_R T^a\sigma^{\mu\nu}\lambda_U q_L\,,\\
O_2 &= \widebar q_L \gamma^\mu  \lambda_U^\dagger\lambda_U^{\phantom{\dagger}}q_L\, \widebar d_R \gamma_\mu
\lambda_D^{\phantom{\dagger}}\lambda_D^\dagger d_R\,,
\end{align*}
where $ G^a_{\mu\nu}$ is the  field strength for the $SU(3)_c$ gauge field (which is quite irrelevant for our discussion, so don't be distracted).  Consider these operators when we rotate to the basis in which the mass matrices are diagonal. Start with the first:
\begin{align*}
O_1 &\to G^a_{\mu\nu} H \widebar u_R T^a\sigma^{\mu\nu}  V_{u_R}^\dagger\lambda_U\begin{pmatrix}V_{u_L}u_L\\V_{d_L}d_L\end{pmatrix}\\
&= G^a_{\mu\nu} H \widebar u_R T^a\sigma^{\mu\nu}  (V_{u_R}^\dagger\lambda_UV_{u_L}^{\phantom{\dagger}})\begin{pmatrix}u_L\\V_{u_L\phantom{d_L\!\!\!\!\!\!\!\!\!}}^\dagger V_{d_L}^{\phantom{\dagger}}d_L\end{pmatrix}\\
&= G^a_{\mu\nu} H \widebar u_R T^a\sigma^{\mu\nu}  \lambda_U^\prime \begin{pmatrix}u_L\\ V d_L\end{pmatrix}
\end{align*}
We see that the only flavor-changing interaction is governed by the off-diagonal components of $\lambda_U^\prime V$. Similarly
\[
O_2\to 
\widebar q'_L \gamma^\mu  (\lambda_U^\prime)^2 q'_L\, \widebar d_R \gamma_\mu
(\lambda_D^{\prime})^2 d_R, \quad\text{where}\quad q'_L=\begin{pmatrix}u_L\\ V d_L\end{pmatrix}.
\]

This construction, restricting the higher dimension operators by the flavor symmetry with the Yukawa couplings treated  as spurions, goes by the name of the {\it principle of Minimal Flavor Violation} (MFV). Extensions of the SM in which the  only breaking of $G_F$ is by $\lambda_U$ and $\lambda_D$ automatically satisfy  MFV. As we will see they are much less constrained by flavor changing and CP-violating observables than models with generic breaking of $G_F$. 

\begin{exercises}
\begin{exercise}
Had we considered an operator like $O_1$ but with $\Htilde \widebar d_R$ instead of $ H \widebar u_R$ the flavor off-diagonal terms would have been governed by $\lambda'_D V^\dagger$.   
Show this is generally true, that is, that flavor change in any operator is governed by $V$ and powers of $\lambda'$. 
\end{exercise}
\begin{solution}
In any operator use the inverse of \eqref{eq:massDtransf} to write $\lambda_{U,D}$ in terms of $\lambda'_{U,D}$ and the matrices $V_{u_{L,R}}$ and $V_{d_{L,R}}$. Now rotate quarks to go to the mass-diagonal basis. This would be a flavor symmetry transformation if $V_{u_{L}}=V_{d_{L}}=U_q$, so it fails to be a symmetry only because $V=V_{u_{L}}^\dagger V_{d_{L}}^{\phantom{\dagger}}\ne1$, which may appear in these operators. This is the only parameter that is off-diagonal in flavor space. 
\end{solution}
\begin{exercise}
Exhibit examples of operators of dimension 6 that produce flavor
change without involving $\lambda_{U,D}$. Can these be such that only
quarks of charge $+2/3$ are involved? (These would correspond to Flavor
Changing Neutral Currents; see Sec.~\ref{sec:fcnc} below).
\end{exercise}
\begin{solution}
The question is phrased loosely: the answer depends on whether we impose the flavor symmetry \eqref{eq:mvf-trans}. If we don't, then we can simply take an operator like $O_1$ but without the spurion $\lambda_U$ sandwiched between quarks. So, for example, the operator $G^a_{\mu\nu} H \widebar u_R T^a\sigma^{\mu\nu}\kappa q_L$, where $\kappa$ is some arbitrary matrix, when expressed in the mass eigenstate basis gives
\[
 G^a_{\mu\nu} H \widebar u_R T^a\sigma^{\mu\nu}  V_{u_R}^\dagger\kappa\begin{pmatrix}V_{u_L}u_L\\V_{d_L}d_L\end{pmatrix}
\]
Consider, instead, the case when we insist on the symmetry \eqref{eq:mvf-trans}. Now quark bilinears can only be of one SM-representation with itself, $\widebar q_L\gamma^\mu q_L$,  $\widebar q_L\tau^j \gamma^\mu q_L$,  $\widebar u_R\gamma^\mu u_R$  and $\widebar d_R\gamma^\mu d_R$. Of these, only  $\widebar q_L\tau^j \gamma^\mu q_L$ fails to be invariant under the transformation that takes the quarks to the mass eigenstate basis, and then only the terms involving $\tau^\pm$.  So, in the absence of factors of $\lambda_{U,D}$ we can only get charge changing flavor changing interactions. A simple example is the four quark operator $\widebar q_L\tau^j \gamma^\mu q_L\,\widebar q_L\tau^j \gamma_\mu q_L$.
\end{solution}
\end{exercises}

Now let's consider the effect of the principle of MFV on the process $b\to s \gamma$. Our first attempt is
\[
\Delta \mathcal{L}=\frac{C}{\Lambda^2}e F_{\mu\nu} H \widebar q_L \lambda_D \sigma^{\mu\nu} d_R\,.
\]
This gives no flavor changing
interaction when we go to the field basis that diagonalizes the mass
matrices (which can be seen from the analysis above, or simply by noting that this term has the same form, as far as flavor is concerned, as the mass
term in the Lagrangian).  To get around this we need to construct an
operator which either contains more fields, which will give a loop
suppression in the amplitude plus an additional suppression by powers of
$\Lambda$, or additional factors of spurions. We try the latter. Consider, then 
\[
\Delta \mathcal{L}=\frac{C}{\Lambda^2}e F_{\mu\nu} H \widebar q_L \lambda_U^{\phantom{\dagger}}\lambda_U^\dagger \lambda_D^{\phantom{\dagger}} \sigma^{\mu\nu} d_R.
\]
When you rotate the fields to diagonalize the mass matrix you get, for the charge neutral quark bi-linear, 
\begin{equation}
\label{eq:lambdacube}
\lambda_U^{\phantom{\dagger}}\lambda_U^\dagger \lambda_D^{\phantom{\dagger}} \to V_{d_L}^\dagger \lambda_U^{\phantom{\dagger}}\lambda_U^\dagger \lambda_D^{\phantom{\dagger}} V_{d_R}^{\phantom{\dagger}} =
V_{d_L}^\dagger V_{u_L}^{\phantom{\dagger}}  (\lambda'_U)^2V_{u_L}^\dagger V_{d_L}^{\phantom{\dagger}}   \lambda'_D=V^\dagger (\lambda'_U)^2 V  \lambda'_D,
\end{equation}
our estimate of the NP amplitude is suppressed much like in the SM, by
the mixing angles and the square of the ``small'' quark masses. Our  bound now reads
\[
C^{-1}\Lambda^2 \gtrsim \frac{M_W^2s_W^2}{\sqrt{2}\frac{\alpha}{4\pi}}\cdot\frac1{0.1}
\quad\Rightarrow\quad C^{-1/2}\Lambda\gtrsim 4~\text{TeV}
\]
This is within the reach of the LHC (barely), even if $C\sim1$ which
should correspond to a strongly coupled NP sector. If for a weakly
coupled sector $C$ is one loop suppressed, $\Lambda$ could be
interpreted as a mass $M_{\text{NP}}$ of the NP particles in the loop,
and the analysis gives $M_{\text{NP}}\gtrsim 200$~GeV. The moral is
that if you want to build a NP model to explain putative new phenomena
at the Tevatron or the LHC you can get around constraints from flavor
physics if your model incorporates the principle of MFV (or some other
mechanism that suppresses FCNC).

\begin{exercises}
\begin{exercise}
  Determine how much each of the bounds in
  Fig.~\ref{fig:neubertEPS2011} is weakened if you assume MFV. You may
  not be able to complete this problem if you do not have some idea of
  what the symbols $\Delta M_K$, $\epsilon_K$, etc, mean or what type
  of operators contribute to each process; in that case you should
  postpone this exercise until that material has been covered later in these
  lectures.
\end{exercise}
\end{exercises}

\subsection{Examples}
\label{Sec:FlavorOps}
This section may be safely skipped: it is not used elsewhere in these notes. The examples presented here require some background knowlede. Skip the first one if you have not studied supersymmetry yet. 

\begin{enumerate}
\item {\it The supersymmetrized SM.} I am not calling this the MSSM,
  because the discussion applies as well to the zoo of models in which the BEH sector has been extended, {\it e.g.}, the NMSSM. In the absence of SUSY breaking this model satisfies the principle of MFV. The Lagrangian is 
\[
\mathcal{L}
=\int \!\! d^4\theta\;\left[\widebar Q e^VQ+\widebar Ue^V U +\widebar D e^V D\right]+\text{gauge \& $H$ kinetic terms}+\int \!\! d^2\theta\,W +\text{h.c.}
\]
with superpotential 
\[
W=H_1 U y_UQ +H_2 D y_D Q+\text{non-quark-terms}
\]
Here $V$ stands for the vector superfields\footnote{Since I will not make explicit use of vector superfields, there should be no confusion with the corresponding symbol for the the KM matrix, which is used ubiquitously in these lectures.} and $Q$, $D$, $U$, $H_1$ and $H_2$ are  chiral superfields with the following quantum numbers:
\begin{equation*}
\begin{aligned}
Q & \sim (3,2)_{1/6}\\
U & \sim (\widebar 3,1)_{-2/3}\\
D & \sim (\widebar 3,1)_{1/3}\\
\end{aligned}\qquad
\begin{aligned}
H_1 & \sim (1,2)_{1/2}\\
H_2 & \sim (1,2)_{-1/2}\\
\end{aligned}
\end{equation*}
The fields on the left column come in three copies, the three generations we call flavor. We are again suppressing that index (as well as the gauge and Lorentz indices). Unlike the SM case, this Lagrangian is not the most general one for these fields  once renormalizability, Lorentz and gauge invariance are imposed. In addition one needs to impose, of course, supersymmetry. But even that is not enough. One has to impose an $R$-symmetry to forbid dangerous baryon number violating renormalizable interactions. 

When the Yukawa couplings are neglected, $y_U=y_D=0$, this theory has a $SU(3)^3$ flavor symmetry. The symmetry is broken only by the couplings and we can keep track of this again by treating the couplings as spurions. Specifically, under $SU(3)^3$, 
\[
Q\to U_q Q,\quad U \to S_UU , \quad D\to S_D D,\quad y_U\to S_U^*y_U U_q^\dagger, \quad y_D\to S_D^*y_D U_q^\dagger
\]
Note that this has both quarks and squarks transforming together. 
The transformations on quarks may look a little different than the transformation in the SM, Eq.~\eqref{eq:mvf-trans}. But they are the same, really. The superficial difference is that here the quark fields are all written as left-handed fields, which are obtained by charge-conjugation from the right handed ones in the standard representation of the SM. So in fact, the couplings are related by 
$y_U=\lambda_U^\dagger$ and $y_D=\lambda_D^\dagger$, and the transformations on the right handed fields by $S_U=U_u^*$ and $S_D=U_d^*$. While the relations are easily established, it is worth emphasizing that we could have carried out the analysis in the new basis without need to connect to the SM basis. All that matters is the way in which symmetry considerations restrict  certain interactions. 

Now let's add soft SUSY breaking terms. By ``soft'' we mean operators of dimension less than 4. Since we are focusing on flavor, we only keep terms that include fields that carry flavor:
\begin{multline}
\label{eq:SUSYbkg}
\Delta\mathcal{L}_{\text{SUSY-bkg}}=\phi_q^*\mathcal{M}^2_q\phi_q+\phi_u^*\mathcal{M}^2_u\phi_u+\phi_d^*\mathcal{M}^2_d\phi_d\\
+(\phi_{h_1}\phi_u g_U \phi_q +\phi_{h_2}\phi_d g_D \phi_q +\text{h.c.})
\end{multline}
Here $\phi_X$ is the scalar SUSY-partner of the quark $X$. 
This breaks the flavor symmetry unless
$\mathcal{M}^2_{q,u,d}\propto\mathbf{1}$ and $g_{U,D}\propto y_{U,D}$
(see, however, Exercise~\ref{ex:susy-bkg}). And unless these conditions are satisfied new flavor changing interactions are generically present and large. The qualifier ``generically'' is because the effects can be made small by lucky coincidences (fine tunings) or if the masses of scalars are large. 

This is the motivation  for gauge mediated SUSY-breaking~\cite{Dine:1993yw}:

\bigskip

\begin{centering}
\begin{tikzpicture}[scale=1]
\coordinate (A) at (-4,0);
\coordinate (B) at  (4,0); 

\path node at (A) [shape=ellipse,draw] {\parbox[c]{3cm}{\begin{centering}SUSY\\ breaking sector\\\end{centering}}}
node at (B)[shape=ellipse,draw] {SUSY SM} ;

\draw ($(A)+(2.5,0)$) -- node{\parbox[c]{3cm}{\begin{centering}gauge \\ interaction\\\end{centering}}} ($(B)+(-1.5,0)$);

\end{tikzpicture}
\end{centering}

\bigskip

The gauge interactions, {\it e.g.}, $\widebar Q e^V Q$, are diagonal
in flavor space. In theories of supergravity mediated supersymmetry
breaking the flavor problem is severe. To repeat, this is why gauge mediation and its variants were invented. 

\item {\it  MFV Fields.} Recently CDF and D0 reported a larger than
  expected forward-backward asymmetry in $t\widebar t$ pairs
  produced in $p\widebar p$ collisions~\cite{Aaltonen:2011kc}.  Roughly speaking, define
  the forward direction as the direction in which the protons move,
  and classify the outgoing particles of a collision according to
  whether they move in the forward or backward direction. You can be
  more careful and define this relative to the CM of the colliding
  partons, or better yet in terms of rapidity, which is invariant
  under boosts along the beam direction. But we need not worry about
  such subtleties: for our purposes we want to understand how flavor
  physics plays a role in this process that one would have guessed is
  dominated  by  SM interactions~\cite{Kuhn:2011ri}. Now, we take this as an educational example, but I should warn you that by the time you read this the reported effect may have evaporated. In fact, since the lectures were given D0 has revised its result and the deviation from the SM expected asymmetry is now much smaller~\cite{Leone:2014gwa}. 

There are two types of BSM models that explain this asymmetry, classified according to the the type of new particle exchange that produces the asymmetry:

\begin{enumerate}[(i)]
\item $s$-channel. For example an ``axi-gluon,'' much like a gluon but massive and coupling to axial currents of  quarks. 
The interference between vector and axial currents, 
\begin{tikzpicture}[scale=0.7]
\coordinate[label=left:$u$] (u);
\coordinate[label=right:$t$,right=3 cm of u] (t);
\coordinate[label=left:$\widebar u$,below=1.25cm of u] (baru);
\coordinate[label=right:$\widebar t$,below=1.25cm of t] (bart);
\coordinate (v1) at ($0.5*(u)+0.5*(baru)+(0.5,0)$);
\coordinate (v2) at ($0.5*(t)+0.5*(bart)+(-0.5,0)$);

\draw[particle] (u) -- (v1);
\draw[particle] (v1) -- (baru);
\draw[particle] (bart) -- (v2);
\draw[particle] (v2) -- (t);

\draw[gluon] (v1) -- node[above=2pt]{$g$} (v2);

\node at ($(v2)+(1.4,0)$) {$+$};

\begin{scope}[xshift=6cm]
\coordinate[label=left:$u$] (u);
\coordinate[label=right:$t$,right=3 cm of u] (t);
\coordinate[label=left:$\widebar u$,below=1.25cm of u] (baru);
\coordinate[label=right:$\widebar t$,below=1.25cm of t] (bart);
\coordinate (v1) at ($0.5*(u)+0.5*(baru)+(0.5,0)$);
\coordinate (v2) at ($0.5*(t)+0.5*(bart)+(-0.5,0)$);

\draw[particle] (u) -- (v1);
\draw[particle] (v1) -- (baru);
\draw[particle] (bart) -- (v2);
\draw[particle] (v2) -- (t);

\draw[photon] (v1) -- node[above=2pt]{$a$} (v2);
\end{scope}
\end{tikzpicture} 
produces a FB-asymmetry. It turns out that it is best to have the sign of the axigluon coupling to $t$-quarks be opposite that of the coupling to $u$ quarks,  in order to get the correct sign of the FB-asymmetry without violting  constraints from direct detection at the LHC. But different couplings to $u$ and $t$ means flavor symmetry violation and by now you should suspect that any complete model will be subjected to severe constraints from flavor physics. 
\item $t$-channel: for example, one may exchange a scalar, and the amplitude now looks like this:
\begin{tikzpicture}[scale=0.7]
\coordinate[label=left:$u$] (u);
\coordinate[label=right:$t$,right=3 cm of u] (t);
\coordinate[label=left:$\widebar u$,below=1.25cm of u] (baru);
\coordinate[label=right:$\widebar t$,below=1.25cm of t] (bart);
\coordinate (v1) at ($0.5*(u)+0.5*(baru)+(0.5,0)$);
\coordinate (v2) at ($0.5*(t)+0.5*(bart)+(-0.5,0)$);

\draw[particle] (u) -- (v1);
\draw[particle] (v1) -- (baru);
\draw[particle] (bart) -- (v2);
\draw[particle] (v2) -- (t);

\draw[gluon] (v1) -- node[above=2pt]{$g$} (v2);

\node at ($(v2)+(1.4,0)$) {$+$};

\begin{scope}[xshift=6cm]
\coordinate[label=left:$u$] (u);
\coordinate[label=right:$t$,right=3 cm of u] (t);
\coordinate[label=left:$\widebar u$,below=1.25cm of u] (baru);
\coordinate[label=right:$\widebar t$,below=1.25cm of t] (bart);
\coordinate (v1) at ($0.5*(u)+0.5*(t)$);
\coordinate (v2) at ($0.5*(baru)+0.5*(bart)$);

\draw[particle] (u) -- (v1);
\draw[particle] (v1) -- (t);
\draw[particle] (bart) -- (v2);
\draw[particle] (v2) -- (baru);

\draw[dashed] (v1) -- node[right]{$\phi$} (v2);
\end{scope}
\end{tikzpicture} 

This model has introduced a scalar $\phi$ with a coupling $\phi \widebar t u$ (plus its hermitian conjugate). 
This clearly violates flavor symmetry. Not only we expect that the
effects of  this flavor violating coupling would be directly
observable but, since the coupling is introduced in the mass
eigenbasis, we suspect there are also other couplings involving the
charge-$+2/3$  quarks, as in $\phi \widebar c u$ and $\phi
\widebar t u$  and flavor diagonal ones. This is because even if we
started with only one coupling  in some generic basis of fields, when we rotate the fields to go the mass eigenstate basis we will generate all the other couplings.  Of course this does not have to happen, but it will, generically, unless there is some underlying reason, like a symmetry. Moreover, since couplings to a scalar involve both right and left handed quarks, and the left handed quarks are in doublets of the electroweak group, we may also have flavor changing interactions involving the charge-$(-1/3)$ quarks in these models. 
\end{enumerate}

One way around these difficulties is to build the model so that it satisfies the principle of MFV, by design. Instead of having only a single scalar field, as above, one may include a multiplet of scalars transforming in some representation of $G_F$. So, for example,  
one can have a charged scalar multiplet $\phi$ transforming in the
$(\mathbf{3},\mathbf{\widebar{3}}, 1)$ representation of  $SU(3)_q\times
SU(3)_u\times SU(3)_d$, with gauge quantum numbers $(1,2)_{-1/2}$ and with interaction term 
\[
\lambda \widebar q_L\phi u_R\qquad\text{with}\quad \phi\to U_{q_L} \phi\, U^\dagger_{u_R}\,.
\]
Note that the coupling $\lambda$ is a single number (if we want invariance under flavor). This actually works! See \cite{Grinstein:2011yv}.
\begin{exercises}
\begin{exercise}
\protect\label{ex:susy-bkg}Below Eq.~\protect\eqref{eq:SUSYbkg} we said, ``This breaks the flavor
symmetry unless $\mathcal{M}^2_{q,u,d}\propto\mathbf{1}$ and
$g_{U,D}\propto y_{U,D}$.'' This is not strictly correct (or, more
bluntly, it is a lie). While not correct it is the simplest
choice. Why? Exhibit alternatives, that is, other forms for $\mathcal{M}^2_{q,u,d}$ and
$g_{U,D}$ that respect the symmetry. {\it Hint: See \protect\eqref{eq:lambdacube}.}
\end{exercise}
\begin{solution}
Flavor symmetry requires that $\mathcal{M}^2_q\to U_q\mathcal{M}^2_q U_q^\dagger$, $\mathcal{M}^2_u\to S_U\mathcal{M}^2_u S_U^\dagger$,  $\mathcal{M}^2_d\to S_D\mathcal{M}^2_d S_D^\dagger$, $g_U\to S_U^*g_U U_q^\dagger$ and $y_D\to S_D^*y_D U_q^\dagger$. 
\end{solution}
\begin{exercise}
Classify all possible dim-4 interactions of Yukawa form in the SM. To
this end list all possible Lorentz scalar combinations you can form
out of pairs of SM quark fields. Then give explicitly the
transformation properties of the scalar field, under the gauge and
flavor symmetry groups, required to make the Yukawa interaction
invariant. Do this first without including the SM Yukawa couplings as
spurions  and then including also one power of the SM Yukawa couplings.
\end{exercise}
\end{exercises}
\end{enumerate}

\chapter{Neutral Meson Mixing and CP Asymmetries}
\section{Why Study This?}
Yeah, why? In particular why bother with an old subject like neutral-$K$ meson mixing? I offer you an incomplete list of perfectly good reasons:
\begin{enumerate}[(i)]
\item CP violation  was discovered in neutral-$K$ meson mixing. 
\item Best constraints on NP from flavor physics are from meson mixing. Look at Fig.~\ref{fig:neubertEPS2011}, where the best constraint is from CP violation in neutral-$K$ mixing. In fact, other than $A^s_{SL}$, all of the other observables in the figure involve mixing. 
\item It's a really neat phenomenon (and that should be sufficient reason for wanting to learn about it, I hope you will agree).
\item It's an active field of research both in theory and in experiment. I may be just stating the obvious, but the LHCb collaboration has been very active and extremely successful, and even CMS and ATLAS have performed flavor physics analysis. And, of course, there are also several non-LHC experiments ongoing or planned; see, {\it e.g.}, \cite{Venditti:2008zz}.
\end{enumerate}

But there is another reason you should pay attention to this, and more generally to the ``phenomenology'' (as opposed to ``theory'' or ``model building'') part of these lectures.  Instead of playing with Lagrangians and symmetries we will use these to try to understand  dynamics, that is, the actual physical phenomena the Lagrangian and symmetries describe. As an experimentalist, or even as a model builder, you can get by without an understanding of this. 
Sort of. There are enough resources today where you can plug in the data from your model and obtain a prediction that can be tested against experiment. Some of the time. And all of the time without understanding what you are doing. You may get it wrong, you may miss effects. As a rule of thumb, if you are doing something good and interesting, it is novel enough that you may not want to rely on calculations you don't understand and therefore don't know if applicable. Besides, the more you know the better equipped you are to produce interesting physics. 

\begin{figure}
\begin{center}
\includegraphics{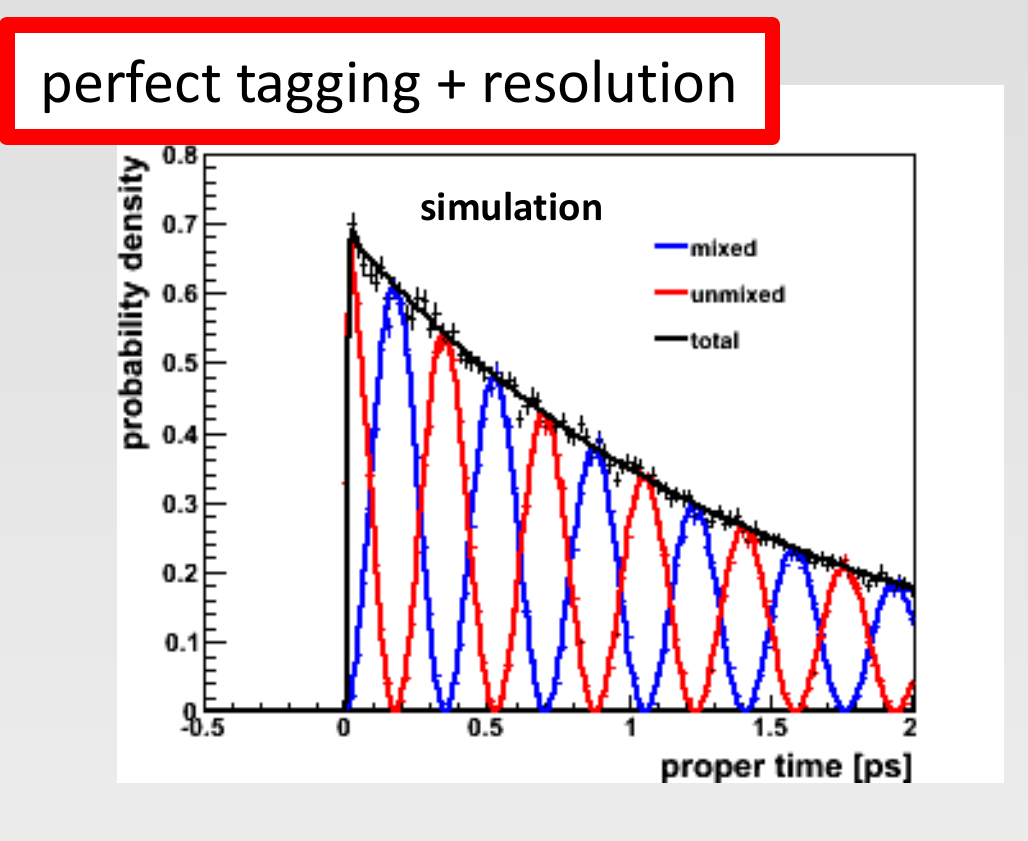}
\end{center}
\caption{\label{fig:mix1}Decay probability of a $\bar B_s$ meson as a function of
  proper time in a perfect world (perfect tagging and resolution) from
  Ref.~\cite{Wandernoth:2013dda}. The
  red and blue lines correspond to $D_s^+\pi^-$ and $D_s^-\pi^+$ final
  states, respectively, and the black is the sum. ``Unmixed'' refers
  to the  fact that the tagging determined that initially the state is
  $\bar B_s$.}
\end{figure}

\section{What is mixing?}
Suppose you have a $\widebar{B}_s$ meson with flavor quantum numbers
$\widebar sb$. If $b\to c\widebar ud$, so that $\widebar sb\to
\widebar s[c\widebar ud]=(\widebar sc)(\widebar ud)$ you can have a
decay $\widebar{B}_s\to D^+_s\pi^-$. Now, the decay is not immediate:
the $\widebar{B}_s$ meson has a non-zero lifetime. So if you somehow determined
that you produced a $\widebar B_s$ at $t=0$ and measure the
probability of decaying into $D^+_s\pi^-$ as a function of time you
get the oscillating function with an exponential envelope depicted by
the red line in Fig.~\ref{fig:mix1}. Moreover, if you measure its
decay probability into $D^-_s\pi^+$ you obtain the blue line in that
same figure. The sum of the two curves is the exponentially decaying
black curve.  The final state $D^-_s\pi^+$ is what you expect from a
decay of a $B_s$ meson, rather than a $\widebar B_s$. 

We guess that as $\widebar B_s$ evolves we have transmutations of
flavor, $\widebar B_s\to B_s \to \widebar B_s\to B_s \to \cdots$. We
can model this by assuming the  time evolution of the state is
\[
| \bar B_s(t)\rangle = e^{-\frac12\Gamma t}\left[\cos(\omega t)|\bar
  B_s\rangle +\sin(\omega t)|B_s\rangle\right]
\]
where the $\widebar B_s$ and $B_s$ states of the right hand side are defined as having the
quantum numbers $\widebar sb$ and $\widebar bs$, respectively. How can
a $\widebar B_s$ turn into a $B_s$? Weak interactions can do that:
Feynman graphs producing the transition are shown here:\\
\begin{center}\includegraphics[width=2in]{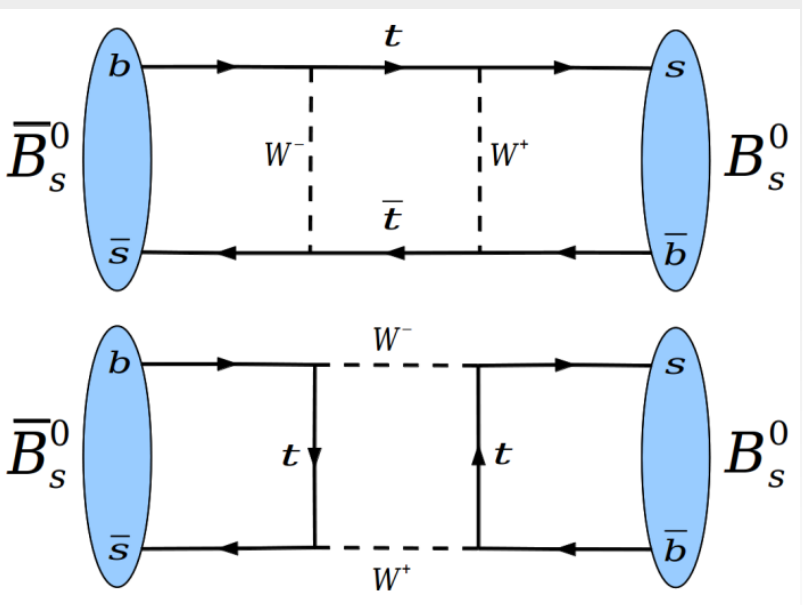}\end{center}
This must be a very small effect. It is a weak interaction. And it is
further suppressed by being a 1-loop effect and by CKM mixing angles
(modern GIM). 

Let's ignore the fact that there is a finite life-time for the moment
and concentrate on the mixing aspect of these states. In quantum
mechanics the state of a free $\widebar B_s$  at rest evolves according to
Schr\"odinger's equation, 
\[
i\frac{d}{dt} \widebar B_s(t)=M\widebar B_s(t)
\]
where I have used the mass, $M$, of the state as its energy at rest,
and similarly for the $B_s$ state which, incidentally,  has the same mass.
The small perturbation introduced by the Feynman diagrams above
couples the evolution of the two states. We can model this by coupling
the two Schr\"odinger equations as follows:
\[
i \frac{d}{dt}\begin{pmatrix}\bar B_s(t)\\B_s(t)\end{pmatrix} =M
\begin{pmatrix}1&\epsilon\\\epsilon & 1\end{pmatrix}\begin{pmatrix}\bar B_s(t)\\B_s(t)\end{pmatrix}
\]
The matrix $\begin{pmatrix}1&\epsilon\\ \epsilon & 1\end{pmatrix}$ has
eigenvalues $1\pm\epsilon$, but no matter how small $\epsilon$ is the
eigenvectors $\begin{pmatrix}1\\ \pm 1\end{pmatrix}$
are maximally mixed! The solution to the differential equation is
straightforward, 
\[
\bar B_s(t)=e^{-iMt}\left[\cos(\epsilon M t) \bar
  B_s(0)-i\sin(\epsilon Mt)B_s(0)\right].
\]
This is the magic of meson-mixing: a very small perturbation gives a
large effect (full mixing). The smallness of $\epsilon$ shows up in
the frequency of oscillation, but the oscillation turns the initial
$\widebar B_s$ into 100\%   $B_s$ in half a period of oscillation. 

\begin{figure}
\begin{center}
\includegraphics[width=0.7\textwidth,clip=true,trim=0 3.2in 0 3.2in]{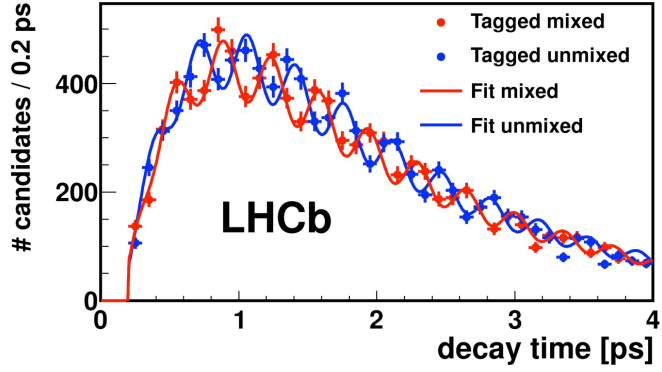}
\end{center}
\caption{\label{fig:mix2}As in Fig.~\ref{fig:mix1} but with
  finite resolution and imperfect tagging~\cite{Wandernoth:2013dda}.
  This time, however, the figure shows data measured at LHCb rather
  than a computer simulation.}
\end{figure}

\begin{figure}
\begin{center}
\includegraphics[width=0.7\textwidth,clip=true,trim=0 .1cm 0 .1cm]{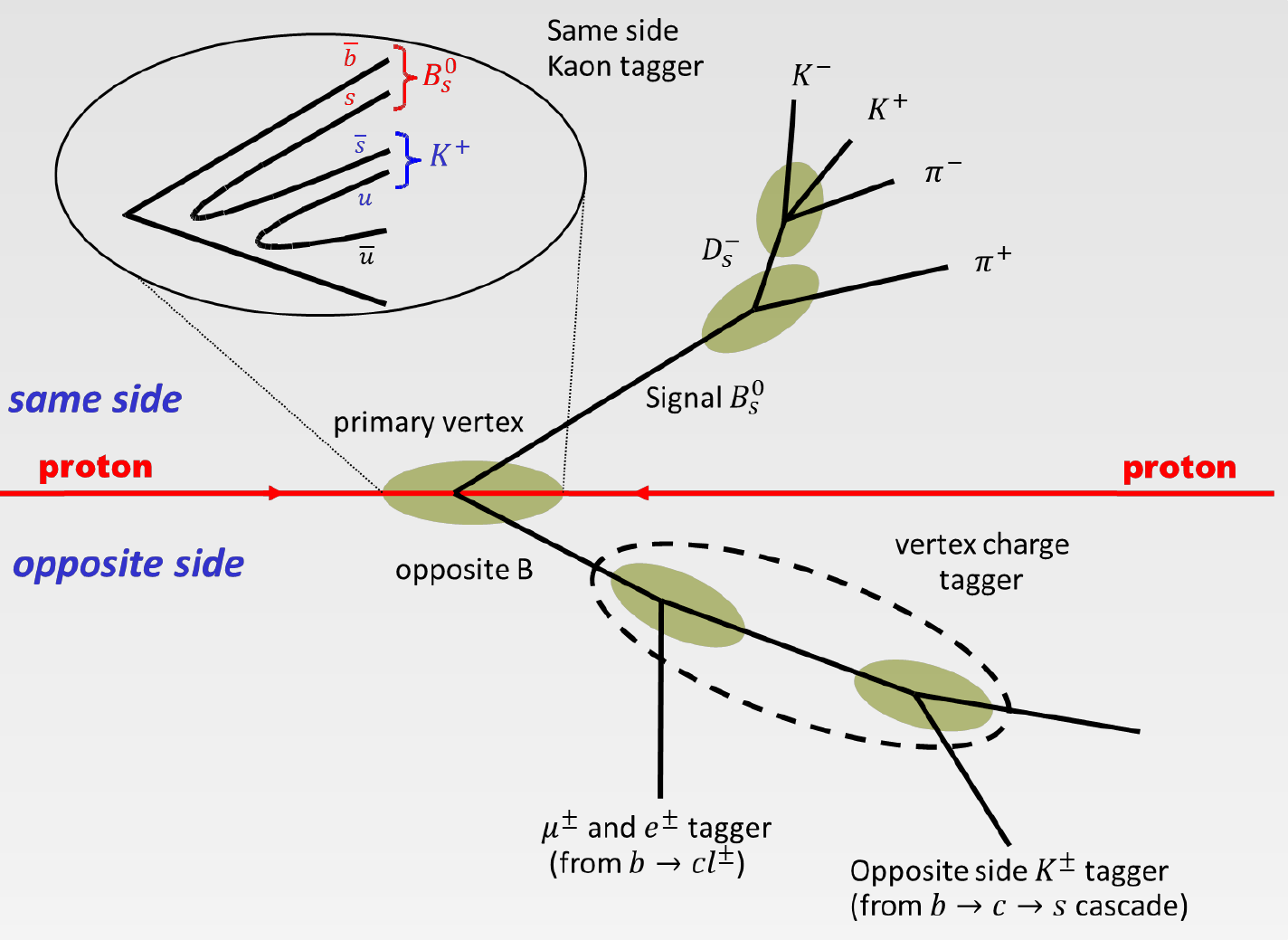}
\end{center}
\caption{\label{fig:tagging} Tagging from lepton charge or opposite  side $K$ charge for $ B_s\to D^-_s\pi^+$ decays. Figure  from Ref.~\cite{Wandernoth:2013dda}. }
\end{figure}

Before we go on to a more complete treatment of  this phenomenon
let's take a look at real data and understand how one can determine that the initial state is in
fact a $\widebar B_s$, as opposed to a $B_s$. Fig.~\ref{fig:mix2}
shows LHCb data that corresponds to the ideal case of
Fig.~\ref{fig:mix1}. The difference between the two figures is
well understood as arising from imperfect resolution and tagging.
Tagging is the method by which the experiment determines the initial
state is in fact a $\widebar B_s$. Figure~\ref{fig:tagging} is a
diagrammatic representation of  a
$B_s$ meson (with a $\widebar b$-quark) produced on the ``same side.''
At the primary vertex one may observe a $K^+$ signaling the presence
of the $\widebar s$ quark and hence a tag that the $B$-meson produced
contains an $s$-quark. The opposite side must
contain a state with a $b$ quark. If it decays semileptonically, $b\to
c\ell^-\nu$ it
will produce a negatively charged lepton; $e^-$ or $\mu^-$ also tag
the $B_s$. When the opposite side $b$
quark decays it is highly likely that it will produce a $c$-quark, and
this one, in turn, an $s$ quark, so a $K^-$ signales the presence of a
$b$ quark on the opposite side, giving a third tag. 
 
\section{Mixing: Formailsm}
We present the Weisskopf-Wigner mixing formalism for a generic neutral
meson-antimeson system, denoted by $X^0-\widebar X^0$. We can apply
this to the cases $X^0=K^0, D^0, B^0$ and $B_s$. Under
charge conjugation ($C$) and spatial inversions (or parity, $P$)
states with a single pseudoscalar meson at rest transform as
\begin{align*}
P|X^0\rangle&=-|X^0\rangle& P|\widebar X^0\rangle&=-|\widebar X^0\rangle\\
C| X^0\rangle&=|\widebar X^0\rangle& C|\widebar X^0\rangle&=| X^0\rangle
\end{align*}
Of course, there is an implicit tranformation of the momentum of the state under $P$. We will be interested in CP-violation. The combination of the above
transformations gives
\[
CP|\bar X^0\rangle=-| X^0\rangle\qquad\mbox{and}\qquad CP| \bar X^0\rangle = -|  X^0\rangle\,.
\]
As in our guess in the previous section we study this system allowing
for mixing between the two states in their rest frame. But now we want
to incorporate finite life-time effects. So for the time evolution we
need a Hamiltonian that contains a term that corresponds to the width.
In other words, since these one particle states may evolve into states that are not accounted for in the two state Hamiltonian, the evolution will not be unitary and the Hamiltonian will not be Hermitian. Keeping this in mind we write, for this effective Hamiltonian
\begin{equation}
\label{eq:CPK0}
\mathbf{H}=\mathbf{M}-\frac{i}{2}\mathbf{\Gamma}=\begin{pmatrix}
M-\frac{i}2\Gamma& M_{12}-\frac{i}2\Gamma_{12}\\
M_{12}^*-\frac{i}2\Gamma_{12}^*& M-\frac{i}2\Gamma\end{pmatrix}
\end{equation}
where $\mathbf{M}^\dagger = \mathbf{M}$ and   $
\mathbf{\Gamma}^\dagger =  \mathbf{\Gamma}$. Also we have taken
$|1\rangle=|X^0\rangle$ and $
|2\rangle=|\widebar X^0\rangle$. We have insisted on CPT:
$(CPT)^{-1}\,\mathbf{H}\, (CPT)=\mathbf{H}^\dagger~\Rightarrow
H_{11}=H_{22}$. Studies of CPT invariance relax this assumption; see Ref.~\cite{Huet:1994kr}.

\begin{exercises}
\begin{exercise}
Show that CPT implies $H_{11}=H_{22}$. 
\end{exercise}
\begin{solution}
Let $\Omega=CPT$. We have to be a bit careful in that this is an
anti-unitary operator. The bra-ket notation is somewhat confusing for
anti-linear operators, so we use old fashioned inner product notation
$(\psi,\eta)$ for $\langle\psi|\eta\rangle$. Anti-unitarity means
$(\Omega\psi,\Omega\eta)=(\eta,\omega)$, and anti-linearity means
$\Omega(a\psi+b\eta)= a^*\psi+b^*\eta$, where $a,b$ are constants and
$\psi,\eta$ wave-functions.  Now, the CPT theorem gives $\Omega H
\Omega^{-1}=H^\dagger$. So 
\begin{align*}
(\psi,H\eta)&=(\psi,H\Omega^{-1}\Omega\eta) && \\
&=(\Omega H\Omega^{-1}\Omega\eta, \Omega\psi) && \text{by
  anti-unitarity of $\Omega$}\\
&=( H^\dagger\Omega\eta, \Omega\psi) && \text{by CPT theorem}\\
&=( \Omega\eta, H \Omega\psi) && \text{by definition of adjoint of operator}
\end{align*}
The action of $\Omega$ on the one particle states at rest is just like
that of CP, $\Omega | X^0\rangle = -| \widebar X^0\rangle$ and 
$\Omega| \widebar X^0\rangle = -| X^0\rangle $. So taking $\psi$ and
$\eta$ above to be $| \widebar X^0\rangle $, we have
$H_{22}=(\psi,H\eta)=( \Omega\eta, H \Omega\psi)=H_{11}$.
Note that for $\psi=| X^0\rangle $ and $\eta= | \widebar X^0\rangle$
the same relation gives $H_{12}=H_{12}$. 
\end{solution}
\end{exercises}

CP invariance requires $M_{12}^*=M_{12}$ and $
\Gamma_{12}^*=\Gamma_{12}$. Therefore {\it either}
$\text{Im}M_{12}\ne0$ {\it or} $\text{Im} \Gamma_{12}\ne0$, or both,  signal that CP
is violated. Now, to study the time evolution of the system we solve
Schr\"odinger's equation. To this end we first solve the eigensystem for
the effective Hamiltonian. The physical eigenstates are labeled conventionally as Heavy and Light
\begin{equation}
\label{eq:XHLofX0}
 |X_H\rangle = p|X^0\rangle + q|\widebar{X}^0\rangle, \qquad  |X_L\rangle = p|X^0\rangle - q|\widebar{X}^0\rangle 
\end{equation}
and  the corresponding eigenvalues are defined as
\[
M_{X\!_{H\atop L}}-\tfrac{i}2\Gamma_{X\!_{H\atop L}}=M-\tfrac{i}2\Gamma\pm\tfrac12(\Delta M-\tfrac{i}2\Delta\Gamma).
\] 
Note that for $q=p$ these are $CP$-eigenstates: $CP|X\!_{H\atop
  L}\rangle =\mp|X\!_{H\atop L}\rangle $. 

We still have to give the eigenvalues and coefficients $p,q$ in terms
of the entries in the Hamiltonian. From the eigenstate equation we read off,
\[
\frac{p}{q}=2\frac{M_{12}-\frac{i}2\Gamma_{12}}{\Delta M-\frac{i}2\Delta\Gamma}=
\frac12\frac{\Delta M-\frac{i}2\Delta\Gamma}{M^*_{12}-\frac{i}2\Gamma^*_{12}}
\]
From this we can write simple non-linear equations giving $\Delta M$
and $\Delta\Gamma$:
\begin{equation}
\label{eq:deltaMandG}
\begin{aligned}
(\Delta M)^2-\frac14(\Delta\Gamma)^2&=
4|M_{12}|^2-|\Gamma_{12}|^2\\
\Delta M\Delta\Gamma&=
4\text{Re}(M_{12}\Gamma^*_{12})
\end{aligned}
\end{equation}

For Kaons it is standard practice to label the states differently, with Long
and Short instead of Heavy and Light: the
eigenvalues of the $2\times 2 $ Hamiltonian are 
\[
M_{K\!_{L\atop S}}-\tfrac{i}2\Gamma_{K\!_{L\atop S}}=M-\tfrac{i}2\Gamma\pm\tfrac12(\Delta M-\tfrac{i}2\Delta\Gamma)
\]
 and the corresponding eigenvectors are 
\begin{equation}\label{KLSdefd}
|K\!_{L\atop S}\rangle =\frac1{\sqrt{2(1+|\epsilon|^2)}}\left[(1+\epsilon)| K^0\rangle\pm (1-\epsilon)| \widebar K^0\rangle\right]
\end{equation}
If $\epsilon=0$  these are  $CP$-eigenstates: $CP |K_{L}\rangle=-|K_L\rangle$ and $CP |K_{S}\rangle=|K_S\rangle$. Since $CP |\pi\pi\rangle_{\ell=0}=|\pi\pi\rangle_{\ell=0}$ and  $CP |\pi\pi\pi\rangle_{\ell=0}=-|\pi\pi\pi\rangle_{\ell=0}$ we see that if CP were a good symmetry the decays $K_L\to\pi\pi\pi$ and $K_S\to\pi\pi$ are allowed, but not so the decays $K_L\to\pi\pi$ and $K_S\to\pi\pi\pi$. Barring CP violation in the decay amplitude, observation of $K_L\to\pi\pi$ or $K_S\to\pi\pi\pi$
indicates $\epsilon\ne0$, that is, CP-violation in mixing. 

This is very close to what is observed:
\begin{align}
\label{eq:KpipiBrs}
\text{Br}(K_S\to\pi\pi)&=100.00\pm0.24\%\nonumber\\
\text{Br}(K_L\to\pi\pi)&=0.297\pm0.023\%\\
\text{Br}(K_L\to\pi\pi\pi)&=33.9\pm1.2\%\nonumber
\end{align}
Hence, we conclude {\it (i)} $\epsilon$ is small, and {\it (ii)} CP is
not a symmetry. The longer life-time of $K_L$ is accidental. To
understand this notice that $3m_\pi\sim 3(140)~\text{MeV}=420~\text{MeV}$ while $m_K\sim490~\text{MeV}$, leaving little phase space for the decays $K\to\pi\pi\pi$. This explains why $K_L$ is much longer lived than $K_S$; the labels ``$L$'' and ``$S$'' stand for ``long'' and ``short,'' respectively:
\begin{align*}
\tau_{K_S} &= 0.59\times10^{-10}~\text{s}\\
\tau_{K_L} &= 5.18\times10^{-8}~\text{s}
\end{align*}
This is no longer the case for heavy mesons for which there is
a multitude of possible decay modes and only a few multi-particle decay
modes are phase-space suppressed. 

Eventually we will want to connect this effective $2\times2$  Hamiltonian to the underlying fundamental physics we are studying. This can be done using perturbation theory (in the weak interactions) and is an elementary exercise in Quantum Mechanics (see, {\it e.g.}, Messiah's textbook, p.994 -- 1001 \cite{Messiah}). With $| X^0\rangle =| 1\rangle $ and $| \widebar{X}^0\rangle =| 2\rangle $ one has
\begin{align}
\label{eq:MfromTheory}
M_{ij}&=M\delta_{ij}+\langle i|H|j\rangle+{\sum_n}'\text{PP}\frac{\langle i|H|n\rangle\langle n|H|j\rangle}{M-E_n}
+\cdots \\
\label{eq:GammafromTheory}
\Gamma_{ij}&=2\pi{\sum_n}'\delta(M-E_n)  \langle i|{H}|n\rangle\langle n|H|j\rangle 
+\cdots
\end{align}
Here the prime in the summation sign means that the states $ |
1\rangle $ and $| 2\rangle $  are excluded and PP stands for
``principal part.''  Beware the states are assume discrete and normalized
to unity. Also, $H$ is a Hamiltonian, not a
Hamiltonian density $\mathcal{H} $; $H=\int d^3x\,\mathcal{H} $. It is
the part of the SM Hamiltonian that can produce flavor changes. In the
absence of $H$ the states $| X^0\rangle =| 1\rangle $ and $|
\widebar{X}^0\rangle =| 2\rangle $ would be stable eigenstates of the
Hamiltonian and their time evolution would be by a trivial phase. It
is assumed that this flavor-changing interaction is weak, while there
may be other much stronger interactions (like the strong one that
binds the quarks together). The perturbative expansion is in powers of
the weak interaction while the matrix elements are computed
non-perturbatively with  respect to the remaining (strong) interactions. Of
course the weak flavor changing interaction is, well, the Weak
interaction of the electroweak model, and below we denote the Hamiltonian by
$H_w$.

\begin{figure}\begin{center}
\includegraphics[width=0.3\textwidth]{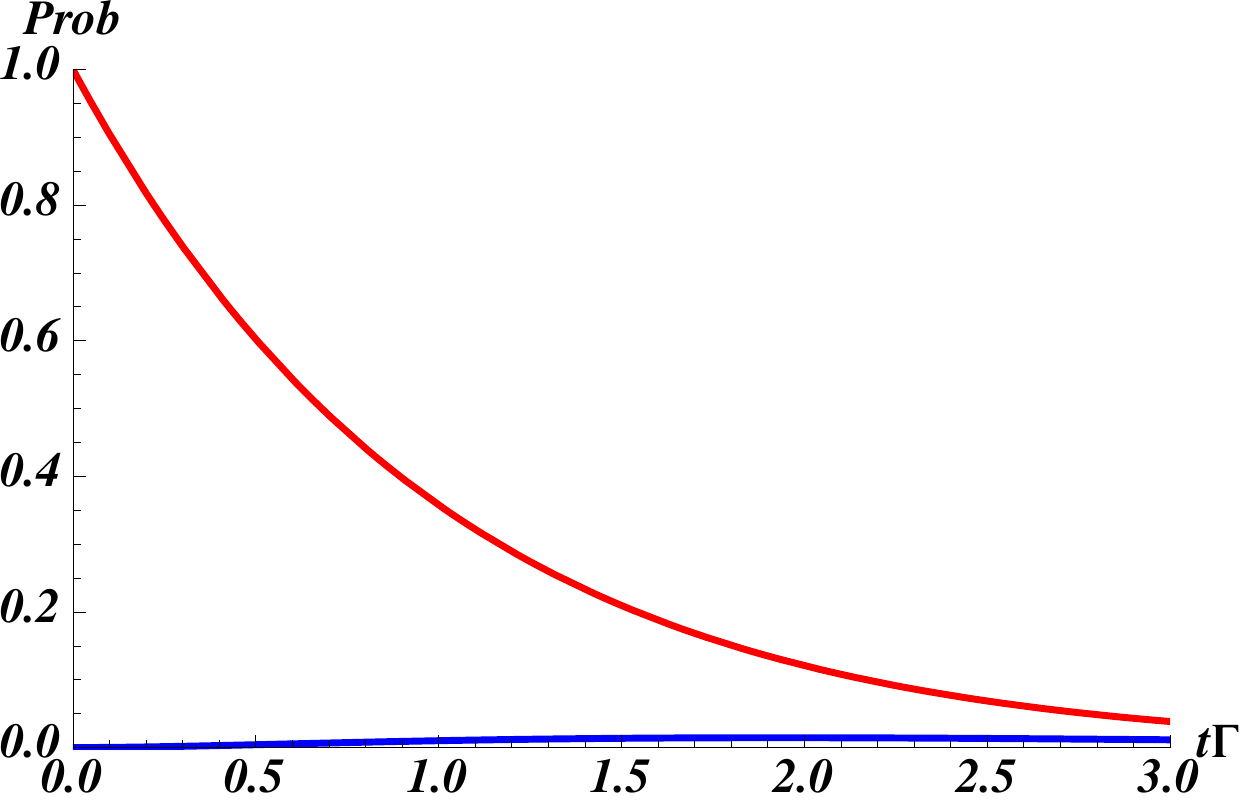}\hfill
\includegraphics[width=0.3\textwidth]{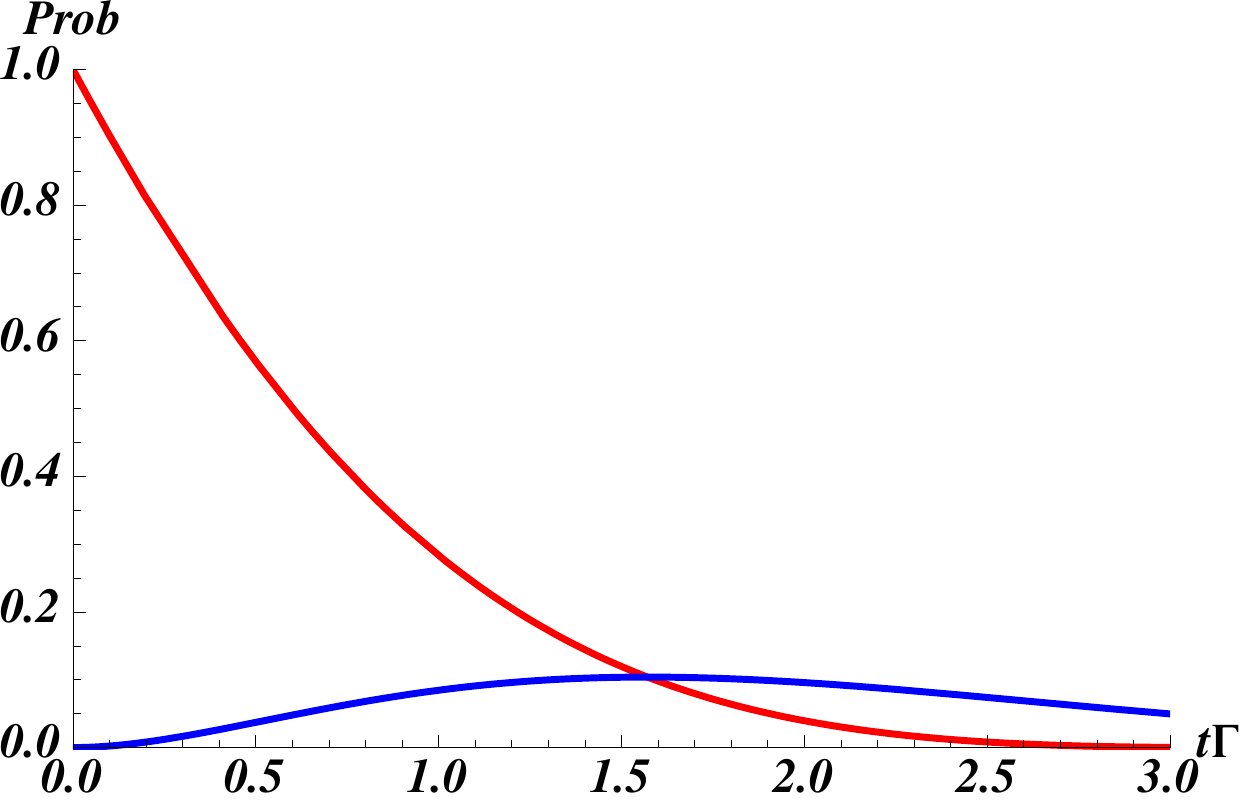}\hfill
\includegraphics[width=0.3\textwidth]{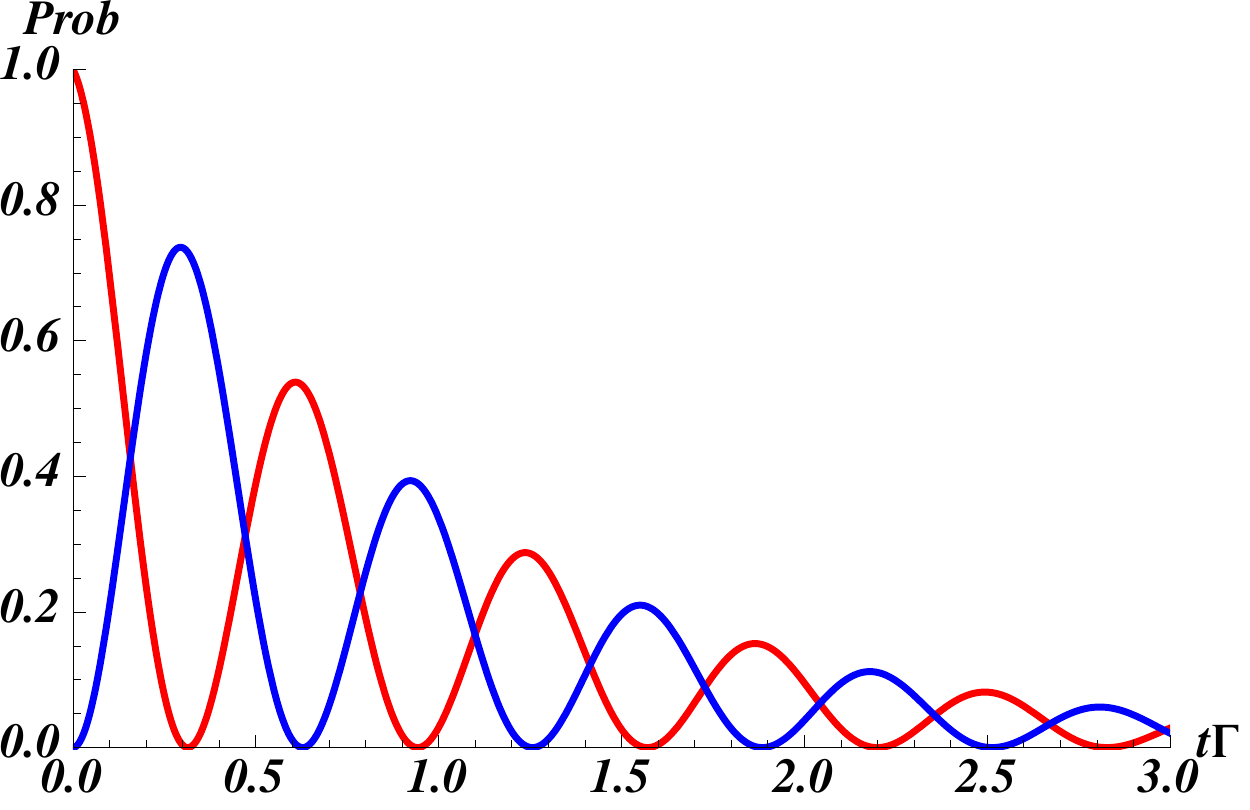}
\end{center}
\caption{\label{fig:slow-fast}Mixing probability in $X^0 - \protect\widebar
  X^0$ mixing as a function of
  $\Gamma t$ for $\Delta M/\Gamma= 1/3,1$ and 3  in left, center and right
  panels, respectively, assuming $\Delta \Gamma=0$ and $|p/q|=1$. In red is the
  probability for the unmixed state  and in blue for the mixed state.} 
\end{figure}

\section{Time Evolution in $X^0$-$\xoverline{X}^0$ mixing.}
We have looked at processes involving the `physical' states $K_L$ and $K_S$. As these are eigenvectors of $H$ their time evolution is quite simple 
\[
i\frac{d}{dt}|X_{\!{H,L}}\rangle =
(M_{\!{H,L}}-\tfrac{i}2\Gamma_{\!{H,L}})
|X_{\!{H,L}}\rangle\qquad
\Rightarrow\qquad 
|X_{\!{H,L}}(t)\rangle =
e^{-iM_{\!{H,L}}t}e^{-\tfrac{1}2\Gamma_{\!{H,L}}t}
|X_{\!{H,L}}(0)\rangle 
\]
Since $|X_{H,L}\rangle $ are eigenvectors of $\mathbf{H}$, they do not mix as they evolve. But often one creates $X^0$ or $\widebar{X}^0$ in the lab. These, of course, mix with each other since they are linear combinations of $X_H$ and $X_L$.

The time evolution of $X_{{H,L }}$ is trivially given by
\[
|X_{{H,L }}(t)\rangle=e^{-iM_{{H,L }}t}e^{-\frac12\Gamma_{{H,L}}t}|X_{{H,L }}(0)\rangle  . 
\]
Now we can invert, 
\begin{equation}
\label{eq:X0from XHL}
\begin{aligned}
|X^0\rangle&=\tfrac1{2p}\left(|X_H\rangle+|X_L\rangle\right),\\
|\widebar{X}^0\rangle&=\tfrac1{2q}\left(|X_H\rangle-|X_L\rangle\right).
\end{aligned}
\end{equation}
Hence, 
\[
|X^0(t)\rangle=\frac1{2p}\left[ e^{-iM_Ht}e^{-\frac12\Gamma_Ht}|X_H (0)\rangle  +
e^{-iM_Lt}e^{-\frac12\Gamma_Lt}|X_L(0)\rangle \right]
\]
and using \eqref{eq:XHLofX0} for the states at $t=0$ we obtain
\begin{equation}
|X^0(t)\rangle=f_+(t)|X^0\rangle+\tfrac{q}{p}f_-(t)|\widebar{X}^0\rangle
\end{equation}
where
\begin{equation}
\begin{aligned}
f_\pm(t)&=\tfrac12\left[e^{-iM_Ht}e^{-\frac12\Gamma_Ht}\pm e^{-iM_Lt}e^{-\frac12\Gamma_Lt}\right]\\
&=\tfrac12 e^{-iM_Ht}e^{-\frac12\Gamma_Ht}\left[1\pm e^{i\Delta Mt}e^{\frac12\Delta\Gamma t}\right]\\
&=\tfrac12 e^{-iM_Lt}e^{-\frac12\Gamma_Lt}\left[ e^{-i\Delta Mt}e^{-\frac12\Delta\Gamma t}\pm1\right]
\end{aligned}
\end{equation}
Similarly,
\begin{equation}
|\widebar{X}^0(t)\rangle=\tfrac{p}{q}f_-(t)|X^0\rangle+f_+(t)|\widebar{X}^0\rangle.
\end{equation}

\begin{figure}
\begin{center}
\includegraphics[width=0.6\textwidth,clip=true,trim=0 1.8in 0 1.6in]{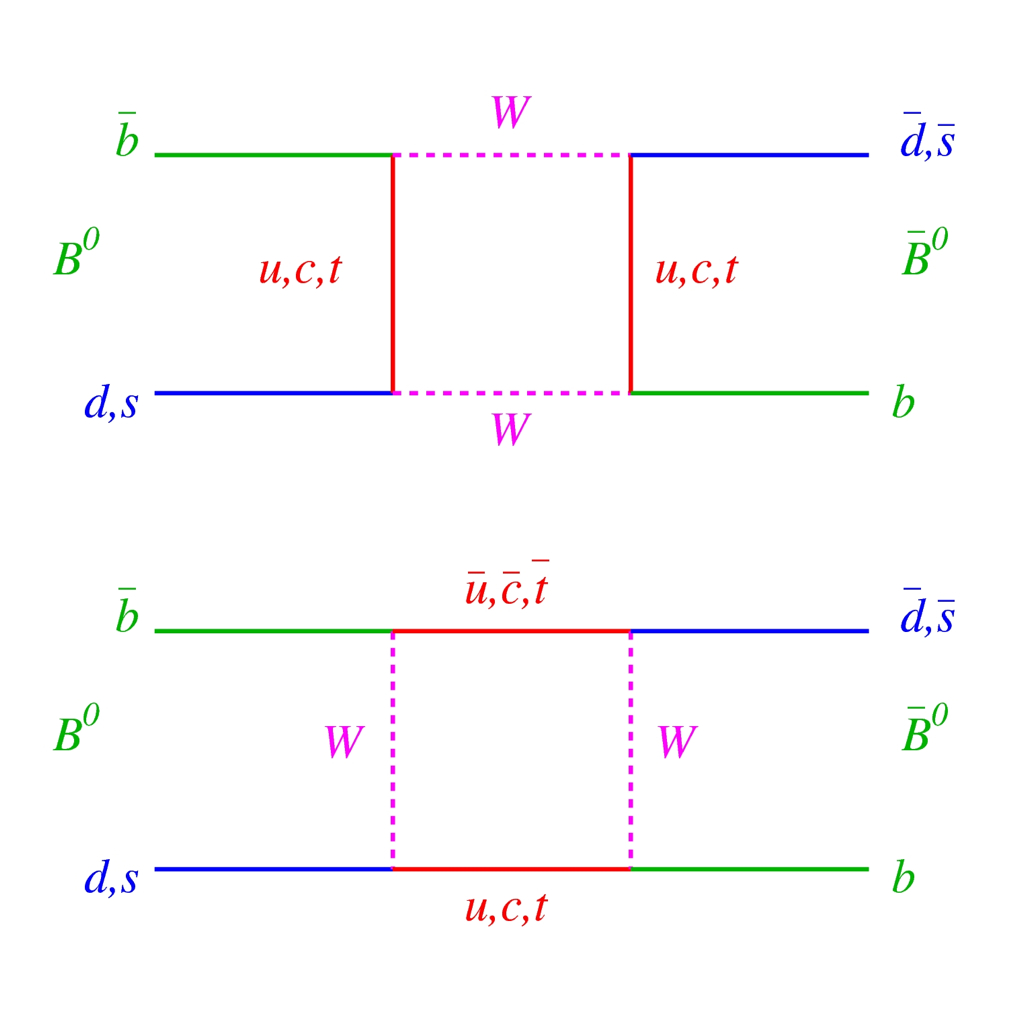}
\end{center}
\caption{\label{fig:boxdiag}Box diagrams contributing to
  $B_{d,s}$-mixing.}
\end{figure}

\subsection{Mixing: Slow {\it vs} Fast}
Fig.~\ref{fig:slow-fast} shows in red the probability of finding an
$X^0$ as a function of time (in units of lifetime, $1/\Gamma$) if the
starting state  is $X^0$. In blue is the probability of starting with
$X^0$ and finding $\widebar X^0$ at time $t$.  In all three panels
$\Delta \Gamma=0$ and $|p/q|=1$ is assumed. In the left panel $\Delta
M=\frac13\Gamma$ so the oscillation is slow, while in the right panel $\Delta
M=3\Gamma$,
the oscillation is fast. The middle panel is in-between, $\Delta
M=\Gamma$. The three panels  qualitatively show what is seen for
$D^0$, $B^0$ and $B_s$ as we go from left to right.

To understand how the SM accounts for the slow {\it versus} fast oscillation
behavior of the different neutral meson systems we need to look at the
underlying process. Consider the box diagrams in
Fig.~\ref{fig:boxdiag}. First note that each of the two fermion lines
in each diagram will produce a modern GIM: the diagrams come with a
factor of $(V_{qb}V_{qd,s}^*)^2$ with $q=u,c,t$, times $m_q^2$ dependent
functions. 

Next, let's recall the connection between the parameters of the
$2\times 2$ Hamiltonian and  fundamental theory,
Eqs.~\eqref{eq:MfromTheory} and \eqref{eq:GammafromTheory}. In
particular the presence of the delta function in Eq.~\eqref{eq:GammafromTheory} indicates that $\Gamma_{12}$ originates in
graphs where the intermediate states are on-shell. 
In the top box graph the intermediate states are $W^+W^-$ which are
much heavier than $B_{d,s}$ and therefore never on-shell. The upper panel box
cannot contribute to $\Gamma_{12}$. Then modern GIM dictates  the
graph is dominated by the top quark exchange.  The bottom panel box graph is a
little different. It does not contribute to $\Gamma_{12}$ when the
intermediate state is $t\bar t$, but it does for $c\bar c$ and $u\bar
u$. However, these contributions are much smaller than the ones with
$t\bar t$ or the ones in the upper panel graph. So we conclude that
$\Gamma_{12}$ is negligible (compared to $M_{12}$)  for $B^0=B_d$ and
$B_s$. From \eqref{eq:deltaMandG} we see that 
\[
\Gamma_{12}=0\quad\Rightarrow\quad \Delta M= 2 |M_{12}|
\quad\Rightarrow\quad \frac{p}{q}=\frac{M_{12}}{|M_{12}|}
\]
That is $p/q$ is a pure phase, $|p/q|=1$. Moreover, the phase
originates in the KM factors in the Feynman graph, because there is no
imaginary part produced by the loop integration  since intermediate
states cannot go on-shell (the very same reason $\Gamma_{12}=0$). So
we can read off the phase immediately:
\[
\left(\frac{p}{q}\right)_{\!\!B^0}=\frac{(V^{\phantom{*}}_{tb}V^*_{td})^2}{|V^{\phantom{*}}_{tb}V^*_{td}|^2},\qquad 
\left(\frac{p}{q}\right)_{\!\!B_s}=\frac{(V^{\phantom{*}}_{tb}V^*_{ts})^2}{|V^{\phantom{*}}_{tb}V^*_{ts}|^2}\,.
\]
Of course, we cannot compute $\Delta M$ fully, but we can compare this
quantity for $B^0$ and $B_s$. In particular, in the flavor-$SU(3)$
symmetry limit the strong interactions treat the $B^0$ and $B_s$
identically, so the only difference in the evaluation of $M_{12}$ stems
form the KM factors. So to the accuracy that SU(3) may hold
(typically 20\%), we have 
\[
\frac{(\Delta M)_{B_s}}{(\Delta
  M)_{B^0}}=\left|\frac{V_{ts}}{V_{td}}\right|^2
\]

Let's look back at Fig.~\ref{fig:neubertEPS2011}. We can understand a
lot of it now. For example, the most stringent bound is from CP
violation in $K^0-\widebar K^0$ mixing. We have seen that this
requires $\text{Im}M_{12}\ne0$ or $\text{Im}\Gamma_{12}\ne0$. Now we
can write, roughly, that the imaginary part of the box diagram for
$K^0$ mixing gives
\[
\text{Im}M_{12}\approx\text{Im}\left(
\hbox to 5.2cm{\hskip-0.9cm\vbox to 1.3cm {\begin{tikzpicture} 
\coordinate (A) at (0,0);
\coordinate (x) at (1,0);
\coordinate (y) at (0,-1);

\coordinate (v1) at ($(A)+(x)$);
\coordinate (v2) at  ($(v1)+(x)$);
\coordinate (v3) at  ($(v2)+(x)$);
\coordinate (v4) at ($(A)+(y)$);
\coordinate (v5) at  ($(v4)+(x)$);
\coordinate (v6) at  ($(v5)+(x)$);
\coordinate (v7) at  ($(v6)+(x)$);

\draw[particle] (A) -- node[above=0.1cm]{$s$} (v1);
\draw[particle] (v1) -- node[above]{$u,c,t$} (v2);
\draw[particle] (v2) -- node[above=0.1cm]{$d$} (v3);
\draw[particle] (v7) -- node[below=0.1cm]{$s$} (v6);
\draw[particle] (v6) -- node[below]{$u,c,t$} (v5);
\draw[particle] (v5)  -- node[below=0.0cm]{$d$} (v4);

\draw[photon] (v1) -- node[left]{$W$} (v5);
\draw[photon] (v2) -- node[right]{$W$} (v6);

\draw[fill] ($(A)+0.5*(y)$) ellipse (0.1cm and 0.5cm) ;
\draw[fill] ($(A)+0.5*(y)+3*(x)$) ellipse (0.1cm and 0.5cm) ;

\draw[line width=3pt] ($(A)+0.5*(y)$) -- +($-0.5*(x)$)node[left]{$\widebar{K}^0$} ;
\draw[line width=3pt] ($(A)+0.5*(y)+3*(x)$) -- +($0.5*(x)$)node[right]{$K^0$} ;

\end{tikzpicture} 
}}
\right)\sim\hskip4cm
\]
\[
\text{Im}\left[\frac{G_F^2M_W^2}{4\pi^2}\sum_{q,q'=u,c,t}V^*_{qd}V^{\phantom{*}}_{qs}V^*_{q'd}V^{\phantom{*}}_{q's}\,f\!\left(m_q,m_{q'}\right)
\langle K^0|\widebar{d}_L\gamma^\mu s_L\,\widebar{d}_L\gamma_\mu s_L|\widebar{K}^0\rangle\right]
\]
Here $f$ is a dimensionless function that is computed from a Feynman integral of the
box diagram and depends on $M_W$ implicitly. Note that the diagram has a double GIM, one per quark
line. In the second line above, the non-zero imaginary part is from the phase in the
KM-matrix. In the standard parametrization $V_{ud}$ and $V_{us}$ are
real, so we need at least one heavy quark in the Feynman diagram to
get a non-zero imaginary part. One can show that  the diagram
with one $u$ quark and one heavy, $c$ or $t$, quark is suppressed. We
are left with $c$ and $t$ contributions only. Notice also that
KM-unitarity gives $\sum_q V^*_{qd}V^{\phantom{*}}_{qs}=0$, and since
$\text{Im}V^*_{ud}V^{\phantom{*}}_{us}=0$, we have a single common
coefficient,
$\text{Im}V^*_{cd}V^{\phantom{*}}_{cs}=-\text{Im}V^*_{td}V^{\phantom{*}}_{ts}=A^2\lambda^5\eta$
in terms of the Wolfenstein parametrization. Taking only the top
contribution we can compare with the contribution from new phsyics
which we parametrize as
\[
\frac1{\Lambda^2}\langle K^0|\bar d_L \gamma^\mu s_L\bar d_L
\gamma_\mu s_L|\bar K^0\rangle
\]
Comparing to the SM results and assuming the SM approximately accounts for the observed quantity,  this gives
\[
\Lambda^2\gtrsim
\frac{4\pi^2}{G_F^2M_W^2}\frac{1}{|V_{td}^*V^{\phantom{*}}_{ts}|^2}\approx\left[\frac{6}{(10^{-5})(
    10^2)}\frac{1}{(0.04)(0.004)}\text{GeV}\right]^2\approx [4\times
10^4\text{TeV}]^2
\]

\begin{exercises}
\begin{exercise}
Challenge: Can you check the other three mixing “bounds” in
Fig.~\ref{fig:neubertEPS2011} (assuming the SM gives about the right result).
\end{exercise}
\end{exercises}

\section{CPV}
We now turn our attention to CP violation, or CPV for short. There are
several ways of measuring CPV. Some of them are associated with
mixing, some with decay and some with both at once. We will take a
look at each of these.

\begin{figure}
\begin{center}
\includegraphics{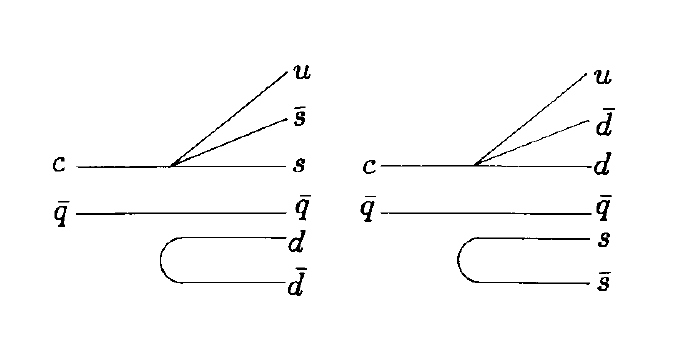}
\end{center}
\caption{\label{fig:Ddecay}Sample Feynman diagrams for some $D$-meson
  decay.}
\end{figure}

\subsection{CPV in Decay}
We begin by looking at CPV in decay. This has nothing to do with mixing
per-se. It is conceptually simple but the price we pay for this
simplicity is that they are hard to compute from first principles. We
will see later that in some cases CPV in interference between mixing
and decay can be accurately predicted. 

Very generally we define an asymmetry as
\[
\mathcal{A}=\frac{\Gamma-\widebar\Gamma}{\Gamma+\widebar\Gamma}
\]
where $\Gamma$ is some rate for some process and $\widebar\Gamma$ is the rate for the
process conjugated under something, like $C$, or $P$ or
$\theta\to\pi-\theta$ (Forward-backward asymmetry). For a CP decay
asymmetry in the decay $X\to f$ we have
\[
\mathcal{A}=\frac{|\langle f|X\rangle|^2-|\langle \bar f|\bar X\rangle|^2}{|\langle f|X\rangle|^2+|\langle \bar f|\bar X\rangle|^2}
\]
where the $\widebar X$ and $\widebar f$ are the CP conjugates of $X$
and $f$ respectively. 

Fig.~\ref{fig:Ddecay} shows diagrams for a $D$-meson decay. The two
diagrams produce the same final state, so they both contribute to the
decay amplitude. The $W$ exchange is shown as a 4-fermion point
vertex. The first diagram contains a KM factor of
$V^*_{cs}V^{\phantom{*}}_{us}$ while the second has a factor of
$V^*_{cd}V^{\phantom{*}}_{ud}$.  So in preparation for a computation
of the CPV decay asymmetry we write
\begin{align*}
\langle f|X\rangle &= aA+bB\\
\langle \bar f|\bar X\rangle &= a^*\bar A+b^*\bar B
\end{align*}
where $a=V^*_{cs}V^{\phantom{*}}_{us}$ and
$b=V^*_{cd}V^{\phantom{*}}_{ud}$ and the rest are matrix elements
computed in the presence of strong interactions
\begin{align*}
A&=\langle f|(\bar u_L\gamma^\mu s_L)(\bar s_L\gamma_\mu c_L)|D\rangle\\
B&=\langle f|(\bar u_L\gamma^\mu d_L)(\bar d_L\gamma_\mu c_L)|D\rangle\,.
\end{align*}
While we cannot compute these, we can say something useful about them.
Assuming the strong interactions are invariant under CP we have
$\widebar A=A$ and $\widebar B = B$. This is easy to show:
\begin{align*}
A&=\langle f|(\bar u_L\gamma^\mu s_L)(\bar s_L\gamma_\mu c_L)|D\rangle\\
&=\langle f|(CP)^{-1}(CP)(\bar u_L\gamma^\mu s_L)(\bar s_L\gamma_\mu c_L)(CP)^{-1}(CP)|D\rangle\\
&=\langle \bar f|(\bar s_L\gamma^\mu u_L)(\bar c_L\gamma_\mu s_L)|\bar D\rangle\\
&=\bar A
\end{align*}
Using this and plugging into the above definition of the asymmetry
$\mathcal{ A}$ we have 
\begin{equation}
\label{eq:genericAsym}
\mathcal{A}=\frac{2\text{Im}(a^*b)\text{Im}(A^*B)}{|aA|^2+|bB|^2+2\text{Re}(a^*b)\text{Re}(A^*B)}
\end{equation}
In order that CP be violated in the decay it is necessary that we have
a relative phase between $a$ and $b$ and also between $A$ and $B$. The
fist one is from the KM matrix, but the second requires computation of 
non-trivial strongly interaction matrix elements. Note that
\[
\text{Im}(a^*b)=\text{Im}((V_{cs}^*V_{us})^*
V_{cd}^*V_{ud})=\text{Im}(V_{cs} V_{cd}^*V_{ud}V_{us}^*)=J
\]
so, as promised, the Jarlskog determinant must be non-zero in order to
see CPV. 

There are numerous CPV decay asymmetries listed in the PDG. It is too
bad we cannot use them to extract the KM angles precisely, let alone
test for new physics (because of our inability to compute the strong
interaction matrix elements).

\subsection{CPV in Mixing}
We will look at the case of kaons first and come back to heavy mesons
later. This is partly because CPV  was discovered through CPV in
mixing in kaons. But also because it offers a special condition not
found in other neutral meson mixing: the vast difference in lifetimes
between eigenstates allows clean separation between them. 

This allows us to meaningfully define the $K_L$ semileptonic decay charge-asymmetry, which is  a measure of CP violation:
\[\delta =\frac{\Gamma(K_L\to\pi^-e^+\nu)-
\Gamma(K_L\to\pi^+e^-\widebar\nu)}{\Gamma(K_L\to\pi^-e^+\nu)+
\Gamma(K_L\to\pi^+e^-\widebar\nu)}
\]
In order to compute this we use the expansion of $K_L$ in terms of flavor eigenstates $K^0$ and $\widebar K^0$ of Eq.~\eqref{KLSdefd}, and note that the underlying process is $s\to u e^-\bar\nu$ (or $\bar s \to \bar u e^+\nu$) so that we assume
 $\langle
\pi^-e^+\nu|H_W|\widebar{K}^0(t)\rangle=0=\langle
\pi^+e^-\nu|H_W|{K}^0(t)\rangle$. Moreover, we assume  CPV is in the mixing only (through the parameter $\epsilon$) and therefore assume that CP is a good symmetry of the decay amplitude: $\langle\pi^-e^+\nu|H_W|K^0(t)\rangle=\langle
\pi^+e^-\nu|H_W|\widebar{K}^0(t)\rangle$. 
\begin{exercises}
\begin{exercise}
With these assumptions show
\[\delta =\frac{|1+\epsilon|^2-|1-\epsilon|^2}{|1+\epsilon|^2+|1-\epsilon|^2}\approx2\text{Re}\epsilon
\]
\end{exercise}
\end{exercises}
Experimental measurement gives $\delta_{\text{exp}}=0.330\pm0.012\%$, from which $\text{Re}\epsilon\simeq1.65\times10^{-3}$. 

\paragraph{Example: Time dependent asymmetry in semileptonic $K$ decay (``$K_{\ell3}$ decay'').}
This is the time dependent analogue of $\delta$ above. The experimental set-up is as follows:

\begin{tikzpicture}[scale=1.3] 
\coordinate (A) at (0,0);
\coordinate (x) at (1,0);
\coordinate (y) at (0,1);

\draw[->,>=stealth,thick] (A) --node[below]{$p$ beam} ++($3*(x)$);
\draw[fill] ($(A)+4*(x)$) ellipse (0.1cm and 0.6cm) node[below=0.7cm]{target};

\node at ($(A)+5.*(x)$) [rectangle, fill=black!50]{};
\node at ($(A)+5.*(x)$) [below=0.1cm]{``magic box''};

\draw[->,>=stealth,thick] ($(A)+6.5*(x)$) -- node[above]{monochromatic beam of $K^0$ and $\widebar{K}^0$} +($3*(x)$);

\draw[->,>=stealth,thick] ($(A)+7.2*(x)-0.1*(y)$) -- +($0.3*(x)-0.5*(y)$) node[below]{$e^-\pi^+\widebar\nu$};
\draw[->,>=stealth,thick] ($(A)+8.2*(x)-0.1*(y)$) -- +($0.3*(x)-0.5*(y)$) node[below]{$e^+\pi^-\nu$};

\node[rectangle,draw,fill=black!10] at  ($(A)+8*(x)-1.3*(y)$){detector array};
\end{tikzpicture} 

The proton beam hits a target, and the magic box produces a clean monochromatic beam of neutral $K$ mesons. These decay in flight and the semileptonic decays are registered in the detector array. We denote by $N_{K^0}$  the number of  $K^0$-mesons, and by $N_{\widebar{K}^0}$  that of $\widebar{K}^0$-mesons, from the beam. Measure
\[
\delta(t)=\frac{N^+-N^-}{N^++N^-}
\]
 as a function of distance from the beam (which can be translated into time from production at the magic box). Here $N^\pm$ refers to the total number of $K_{\ell3}$ events observed with charge $\pm$ lepton.  In reality ``$\pi^\pm$'' really stands for ``hadronic stuff'' since only the electrons are detected. We have then,
\begin{equation*}
\delta(t)=\frac{\begin{aligned}N_{K^0}\!\left[\Gamma(K^0(t)\to\pi^-e^+\nu)\right.&-\left.\Gamma(K^0(t)\to\pi^+e^-\widebar\nu)\right]\\
&+N_{\widebar{K}^0}\!\left[\Gamma(\widebar{K}^0(t)\to\pi^-e^+\nu)-\Gamma(\widebar{K}^0(t)\to\pi^+e^-\widebar\nu)\right]\end{aligned}}%
{\begin{aligned}N_{K^0}\!\left[\Gamma(K^0(t)\to\pi^-e^+\nu)\right.&+\left.\Gamma(K^0(t)\to\pi^+e^-\widebar\nu)\right]\\
&+N_{\widebar{K}^0}\!\left[\Gamma(\widebar{K}^0(t)\to\pi^-e^+\nu)+\Gamma(\widebar{K}^0(t)\to\pi^+e^-\widebar\nu)\right]\end{aligned}}
\end{equation*}
The calculation of $\delta(t)$ in terms of the mixing parameters $q$ and $p$ and the mass and width differences is much like the calculation of $\delta$ above so, again,  I leave it as an exercise:
 
\begin{exercises}
\begin{exercise}
Use $\Gamma(K^0(t)\to\pi^-e^+\nu)\propto|\langle
\pi^-e^+\nu|H_W|K^0(t)\rangle|^2$ and the assumptions that
\begin{enumerate}[(i)]
\item $\langle
\pi^-e^+\nu|H_W|\widebar{K}^0(t)\rangle=0=\langle
\pi^+e^-\nu|H_W|{K}^0(t)\rangle$
\item $\langle\pi^-e^+\nu|H_W|K^0(t)\rangle=\langle
\pi^+e^-\nu|H_W|\widebar{K}^0(t)\rangle$
\end{enumerate}
to show that
\[
\delta(t)=\frac{(N_{K^0}-N_{\widebar{K}^0})\left[|f_+(t)|^2-|f_-(t)|^2\tfrac12\left(\left|\frac{q}{p}\right|^2+
      \left|\frac{p}{q}\right|^2\right)\right] 
+\tfrac12 (N_{K^0}+N_{\widebar{K}^0})|f_-(t)|^2\left(\left|\frac{p}{q}\right|^2-
                         \left|\frac{q}{p}\right|^2\right)}%
{(N_{K^0}+N_{\widebar{K}^0})\left[|f_+(t)|^2+|f_-(t)|^2\tfrac12\left(\left|\frac{q}{p}\right|^2+
      \left|\frac{p}{q}\right|^2\right)\right] 
-\tfrac12 (N_{K^0}-N_{\widebar{K}^0})|f_-(t)|^2\left(\left|\frac{p}{q}\right|^2-
                         \left|\frac{q}{p}\right|^2\right)}
\]
Justify assumptions (i) and (ii).
\end{exercise}
\end{exercises}

\begin{figure}
\begin{center}
\includegraphics[width=4in]{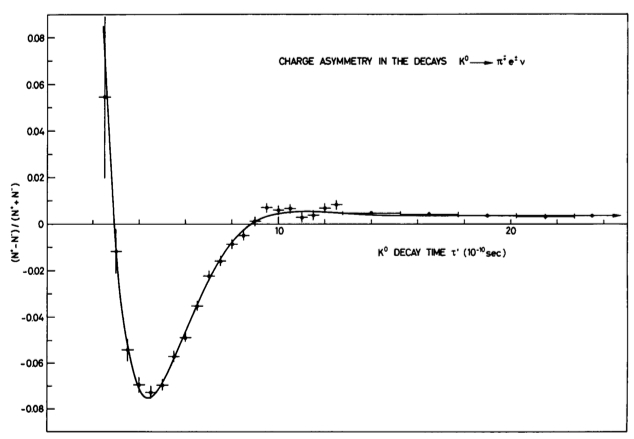}
\end{center}
\caption{\label{fig:deltaKl3} Charge asymmetry in semi-leptonic
  neutral kaon decays, from an experiment by Gjesdal {\it et al},~\cite{Gjesdal:1974yj}.
  The solid curve is a fit to the formula
  \eqref{eq:deltaKl3-timeDep} from which the parameters $\Gamma_S$,
  $\Delta M$, $a$ and $\text{Re}(\epsilon)$ are extracted. }
\end{figure}

The formula in the exercise is valid for any $X^0$-$\widebar{X}^0$
system. We can simplify further for kaons, using
$p/q=(1+\epsilon)/(1-\epsilon)$, $a\equiv
(N_{K^0}-N_{\widebar{K}^0})/(N_{K^0}+N_{\widebar{K}^0})$ and
$\Delta\Gamma\approx-\Gamma_S$. Then 
\begin{align}
\delta(t)&=
\frac{a\left[|f_+(t)|^2-|f_-(t)|^2\right]  +4\text{Re}(\epsilon) |f_-(t)|^2}%
{\left[|f_+(t)|^2+|f_-(t)|^2\right] -4a\text{Re}(\epsilon) |f_-(t)|^2}\nonumber\\
\label{eq:deltaKl3-timeDep}
&\approx\frac{2ae^{-\frac12\Gamma_St}\cos(\Delta  Mt)
+\big(1+e^{-\Gamma_St}-2e^{-\frac12\Gamma_St}\cos(\Delta Mt)\big)2\left(1+\tfrac{a}2\right)\text{Re}(\epsilon)}%
{1+e^{-\Gamma_St}}
\end{align}

Figure~\ref{fig:deltaKl3}  shows the experimental measurement of the
asymmetry~\cite{Gjesdal:1974yj}. The solid curve is a fit to the formula
  \eqref{eq:deltaKl3-timeDep} from which the parameters $\Gamma_S$,
  $\Delta M$, $a$ and $\text{Re}(\epsilon)$ are extracted. The fit to
  this figure gives $\Delta M_K=(0.5287\pm0.0040)\times10^{10}\,\text{s}^{-1}$. The current value, from the \href{http://pdg.lbl.gov}{PDG} is $\Delta M_K=(0.5293\pm 0.0009)\times10^{10}\,\text{s}^{-1}$.

\section{CP-Asymmetries: Interference of Mixing and Decay}
We have seen in \eqref{eq:genericAsym} that in order to generate a non-vanishing CP-asymmetry we need two amplitudes that can interfere. 
One way to get an interference is to have two ``paths'' from
$|\text{in}\rangle$  to $|\text{out}\rangle$. For example, consider an
asymmetry constructed from  $\Gamma=\Gamma(X^0\to
f)$ and $\widebar\Gamma=\Gamma(\widebar X^0\to\widebar f)$,
where $f$ stands for some final state and $\widebar f$ for its CP
conjugate. Then $\Gamma$ may get contributions either from a direct
decay $X^0\to f$ or it may first oscillate into $\widebar X^0$ and
then decay $\widebar X^0\to f$. Note that this requires that both
$X^0$ and its antiparticle, $\overline X^0$, decay to the same common
state. Similarly for $\widebar \Gamma$ we may get contributions from
both $\widebar X^0\to \widebar f$  and the oscillation of
$\widebar X^0$ into $X^0$ followed by a decay into $\widebar f$. In
pictures,

\begin{tikzpicture}[scale=1.] 

\node (center) {};
\node (A)  [above= of center] {$X^0$};
\node (B)  [below=of center] {$\widebar X^0$};
\node (C)  [left=of center]{$X^0$} ;
\node (D) [right=of center]{$f$} ;

\draw[->,>=stealth,thick] (C) -- (A);
\draw[->,>=stealth,thick] (C) -- (B);\draw[->,>=stealth,thick] (A) -- (D);\draw[->,>=stealth,thick] (B) -- (D);

\begin{scope}[xshift=5cm]
\node (center) {};
\node (A)  [above= of center] {$X^0$};
\node (B)  [below=of center] {$\widebar X^0$};
\node (C)  [left=of center]{$\widebar X^0$} ;
\node (D) [right=of center]{$\widebar f$} ;

\draw[->,>=stealth,thick] (C) -- (A);
\draw[->,>=stealth,thick] (C) -- (B);\draw[->,>=stealth,thick] (A) -- (D);\draw[->,>=stealth,thick] (B) -- (D);

\end{scope}
\end{tikzpicture}

Concretely,
\begin{align*}
\Gamma(X^0(t)\to f)&\propto |f_+(t)\langle f|H_w|X^0\rangle
+f_-(t)\tfrac{q}{p}\langle f|H_w|\widebar X^0\rangle|^2\\
&\equiv|f_+(t)A_f+f_-(t)\tfrac{q}{p}\widebar A_f|^2\\
\Gamma(\widebar X^0(t)\to \widebar f)&\propto
|f_-(t)\tfrac{p}{q}\langle \widebar f|H_w|X^0\rangle
+f_+(t)\langle \widebar f|H_w|\widebar X^0\rangle|^2\\
&\equiv|\tfrac{p}{q}f_-(t)A_{\overline f}+f_+(t)\widebar
A_{\overline f}|^2
\end{align*}
I hope the notation, which is pretty standard, is not just
self-explanatory, but fairly explicit. The bar over an amplitude $A$ refers to the
decaying state being $\widebar X^0$, while the decay product is
explicitly given by the subscript, {\it e.g.},  $\widebar
A_{\overline f} = \langle \widebar f|H_w|\widebar X^0\rangle$. 

\begin{exercises}
\begin{exercise}
\label{ex:CPTeigenstates}
If $f$ is an eigenstate of the strong interactions, show that CPT
implies $|A_f|^2=|\widebar A_{\overline f}|^2$ and $|A_{\overline f} |^2=|\widebar A_{ f}|^2$ 
\end{exercise}
\end{exercises}

The time dependent asymmetry is 
\[
\mathcal{A}(t)=\frac{\Gamma(\widebar X^0(t)\to \widebar f)-\Gamma(X^0(t)\to f)}{\Gamma(\widebar X^0(t)\to \widebar f)+\Gamma(X^0(t)\to f)}
\]
and the time integrated asymmetry is
\[
a=\frac{\Gamma(\widebar X^0\to \widebar
  f)-\Gamma(X^0\to f)}{\Gamma(\widebar X^0\to \widebar f)+\Gamma(X^0\to f)}
\]
where $\Gamma(X^0\to f)\equiv\int_0^\infty dt\,\Gamma(X^0(t)\to f)$,
and likewise for the CP conjugate. These are analogs of
the quantities we called $\delta(t)$ and $\delta$ we studied for
kaons. 

\subsection{Semileptonic}
We take $f=e^-+\mbox{any}$. Note that we are taking the wrong sign decay of $X^0$. That is, $\widebar b\to \widebar c e^+\nu$ implies $X^0\to e^++\mbox{any}$ so that $A_f=0$. Similarly, $ b\to c e^-\widebar \nu$ implies $\widebar X^0\to e^-+\mbox{any}$ so that $\widebar A_{\widebar f}=0$. Therefore we have $\Gamma(X^0(t)\to f) = | \tfrac{q}{p}f_-(t)\bar A_f|^2$ and $\Gamma(\bar X^0(t)\to \bar f) = |\tfrac{p}{q} f_-(t)A_{\bar f}|^2$. We obtain
\[
\mathcal{A}_{\text{SL}}(t)=\frac{\left|\frac{p}{q}\right|^2-\left|\frac{q}{p}\right|^2}{\left|\frac{p}{q}\right|^2+\left|\frac{q}{p}\right|^2}
\]
Comments:
\begin{enumerate}[(i)]
\item This is useful because it directly probes $|q/p|$ without contamination from other quantities, in particular from those that require knowledge of strong interactions. 
\item We started off with an {\it a priori}  time dependent quantity, but discovered it is time independent.
\item We already saw that in the SM this is expected to vanish to high accuracy for $B$ mesons, because $\Gamma_{12}$ is small. 
\item It is not expected to vanish identically because $\Gamma_{12}$ while small is non-vanishing. We can guesstimate,
\[
B^0:~~\mathcal{A}^d_{\rm SL}=\mathcal{O}\left[(m_c^2/m_t^2)\sin\beta\right]\lesssim 10^{-3},\qquad
B^s:~~\mathcal{A}^s_{\rm SL}=\mathcal{O}\left[(m_c^2/m_t^2)\sin\beta_s\right]\lesssim 10^{-4}. 
\]
\item Experiment:
\begin{align*}
\mathcal{A}^d_{\rm SL}&=(+0.7\pm 2.7)\times10^{-3}&\Rightarrow\qquad
|q/p|&=0.9997\pm0.0013\\
\mathcal{A}^s_{\rm SL}&=(-17.1\pm 5.5)\times10^{-3}&\Rightarrow\qquad
|q/p|&=1.0086\pm0.0028
\end{align*}
\end{enumerate}

For the rest of this section we will make the approximation that $|q/p|=1$. In addition, we will assume $\Delta\Gamma$ is negligible. We have seen why this is a good approximation. In fact, for the case of $B^0$,  $\Delta\Gamma/\Gamma\sim10^{-2}$, while for $B_s$ the ratio is about 10\%. This simplifies matters because in this approximation
\[
f_\pm(t)\approx e^{-iM t}e^{-\frac12\Gamma t}\begin{cases}\cos(\tfrac12\Delta M t)\\-i\sin(\tfrac12\Delta M t)\end{cases}
\]

\subsection{CPV in interference between a decay with mixing and a decay without mixing}
Assume $\widebar f=\pm f$. Such self-conjugate states are easy to come by. For example $D^+D^-$ or, to good approximation, $J/\psi K_S$. Now, in this case we have $A_{\widebar f}=\pm A_f$ and $\widebar  A_{\overline f}=\pm \widebar A_f$. Our formula for the asymmetry now takes the form
\[
\mathcal{A}_{f_{CP}}=\frac{|\tfrac{p}{q} f_-(t)A_{ f}+f_+(t)\bar A_{ f}|^2
-| f_+(t)A_f+\tfrac{q}{p}f_-(t)\bar A_f|^2}{
|\tfrac{p}{q} f_-(t)A_{ f}+f_+(t)\bar A_{ f}|^2
+| f_+(t)A_f+\tfrac{q}{p}f_-(t)\bar A_f|^2}
\]
Now, dividing by $A_f|^2$ and defining
\[
\lambda_{ f}=\frac{q}{p}\frac{\bar A_f}{A_f}
\]
we have
\begin{align*}
\mathcal{A}_{f_{\rm CP}} &=\frac{| f_-(t)+f_+(t)\lambda_{ f}|^2
-| f_+(t)+f_-(t)\lambda_f|^2}{| f_-(t)+f_+(t)\lambda_{ f}|^2
-| f_+(t)+f_-(t)\lambda_f|^2}\\
&=-\frac{1-|\lambda_f|^2}{1+|\lambda_f|^2}{\cos(\Delta Mt)}+\frac{2\text{Im}\lambda_f}{1+|\lambda_f|^2}{\sin(\Delta Mt)}\\
&\equiv-C_f\,{\cos(\Delta Mt)}+S_f\,{\sin(\Delta Mt)}
\end{align*}

Here is what is amazing about this formula, for which Bigi and Sanda~\cite{Bigi:1981qs} were awarded the Sakurai Prize for Theoretical Particle Physics: the coefficients $C_f$ and $S_f$ can be computed  in terms of KM elements only. They are independent of non-computable, non-perturbative matrix elements. The point is that what most often frustrates us in extracting fundamental parameters from experiment is our inability to calculate in terms of  the parameters to be measured and, at most, other known parameters. I now explain the claim that $C_f$ and $S_f$ are calculable and its range of validity.

The leading contributions to the processes $B^0\to f$ and $\widebar{B}^0\to f$ in the case $f=D^+D^-$ are shown in the following figures:
\begin{center}
\begin{tikzpicture}[scale=1.3] 
\coordinate (A) at (0,0);
\coordinate (x) at (1,0);
\coordinate (y) at (0,1);

\coordinate (v1) at ($(A)+0.5*(y)$); 
\coordinate (v2) at  ($(v1)+1.3*(x)$); 
\coordinate (v3) at  ($(v2)+(x)+0.25*(y)$); 
\coordinate (v4) at ($(v3)+(x)+0.5*(y)$); 
\coordinate (v5) at  ($(v3)+(x)-0.5*(y)$); 
\coordinate (v6) at  ($(A)-0.5*(y)$);
\coordinate (v7) at  ($(v6)+2.3*(x)-0.5*(y)$);
\coordinate (v8) at  ($(v7)+(y)$); 

\draw[particle] (v2) -- node[above]{$b$} (v1);
\draw[particle] (v3) -- node[above=0.1cm]{$c$} (v4);
\draw[particle] (v5) -- node[below]{$d$} (v3);
\draw[particle] (v8) -- node[above right]{$c$}  (v2);
\draw[particle] (v6) .. controls ($(v6)+(x)$) and ($(v7)-(x)+0.2*(y)$)  .. node[below]{$d$} (v7);

\draw[photon] (v2) -- node[above]{$W$} (v3);

\draw[fill] ($(A)$) let \p1=($(v1)-(A)$) in ellipse ({0.2*veclen(\x1,\y1)} and {veclen(\x1,\y1)}) node[left=0.1cm]{${B}^0$} ;
\draw[fill] ($0.5*(v5)+0.5*(v4)$) let \p1= ($0.5*(v5)-0.5*(v4)$) in ellipse ({0.2*veclen(\x1,\y1)} and {veclen(\x1,\y1)}) node[right=0.1cm]{$D^+$} ;
\draw[fill] ($0.5*(v8)+0.5*(v7)$) let \p1= ($0.5*(v8)-0.5*(v7)$) in ellipse ({0.2*veclen(\x1,\y1)} and {veclen(\x1,\y1)})  node[right=0.1cm]{$D^-$};
\begin{scope}[yshift=-1.5cm,xshift=1cm]
\node at (0,0) {$A_{D^+D^-}\propto V_{cb}^*V^{\phantom{*}}_{cd}$};
\end{scope}

\end{tikzpicture} 
\begin{tikzpicture}[scale=1.3] 
\coordinate (A) at (0,0);
\coordinate (x) at (1,0);
\coordinate (y) at (0,1);

\coordinate (v1) at ($(A)+0.5*(y)$); 
\coordinate (v2) at  ($(v1)+1.3*(x)$); 
\coordinate (v3) at  ($(v2)+(x)+0.25*(y)$); 
\coordinate (v4) at ($(v3)+(x)+0.5*(y)$); 
\coordinate (v5) at  ($(v3)+(x)-0.5*(y)$); 
\coordinate (v6) at  ($(A)-0.5*(y)$);
\coordinate (v7) at  ($(v6)+2.3*(x)-0.5*(y)$);
\coordinate (v8) at  ($(v7)+(y)$); 

\draw[particle] (v1) -- node[above]{$b$} (v2);
\draw[particle] (v4) -- node[above=0.1cm]{$c$} (v3);
\draw[particle] (v3) -- node[below]{$d$} (v5);
\draw[particle] (v2) -- node[above right]{$c$}  (v8);
\draw[particle] (v7) .. controls ($(v7)-(x)+0.2*(y)$)  and  ($(v6)+(x)$)   .. node[below]{$d$} (v6);

\draw[photon] (v2) -- node[above]{$W$} (v3);

\draw[fill] ($(A)$) let \p1=($(v1)-(A)$) in ellipse ({0.2*veclen(\x1,\y1)} and {veclen(\x1,\y1)}) node[left=0.1cm]{$\widebar{B}^0$} ;
\draw[fill] ($0.5*(v5)+0.5*(v4)$) let \p1= ($0.5*(v5)-0.5*(v4)$) in ellipse ({0.2*veclen(\x1,\y1)} and {veclen(\x1,\y1)}) node[right=0.1cm]{$D^-$} ;
\draw[fill] ($0.5*(v8)+0.5*(v7)$) let \p1= ($0.5*(v8)-0.5*(v7)$) in ellipse ({0.2*veclen(\x1,\y1)} and {veclen(\x1,\y1)})  node[right=0.1cm]{$D^+$};
\begin{scope}[yshift=-1.5cm,xshift=1cm]
\node at (0,0) {$\widebar{A}_{D^+D^-}\propto V^{\phantom{*}}_{cb}V_{cd}^*$};
\end{scope}

\end{tikzpicture} 
\end{center}
\noindent Either using CP symmetry of the strong interactions or noting that as far as the strong
interactions are concerned the two diagrams are identical, we have 
\[
\frac{\widebar A_{D^+D^-}}{A_{D^+D^-}}=\frac{V_{cb}^{\phantom{*}}V_{cd}^*}{V_{cb}^*V_{cd}^{\phantom{*}}}.
\]
Since $|\widebar A_{D^+D^-}/A_{D^+D^-}|=1$, this is a pure phase, and we see that the phase is given purely in terms of KM elements. 

To complete the argument we need $q/p$. But we have already seen that $\Gamma_{12}$ is negligible. Hence
\[
\frac{p}{q}=\frac{2M_{12}}{\Delta M}=\frac{\Delta
  M}{2M_{12}^*}=\frac{M_{12}}{|M_{12}|}
=\frac{V_{tb}^*V_{td}^{\phantom{*}}}{V_{tb}^{\phantom{*}}V_{td}^*}.
\]
Collecting results
\[
\text{Im}\left(\lambda_{D^+D^-}\right) =
\text{Im}\left(\frac{V_{cb}^{\phantom{*}}V_{cd}^*}{V_{cb}^*V_{cd}^{\phantom{*}}}
\frac{V_{tb}^*V_{td}^{\phantom{*}}}{V_{tb}^{\phantom{*}}V_{td}^*}\right)
=\text{Im}(e^{2i\beta})=\sin(2\beta)
\]
and the asymmetry parameters are $C_{D^+D^-}=0$ and $S_{D^+D^-}=\sin(2\beta)$. Measurements of the asymmetry  gives (twice the sine of) one of the angles of the unitarity triangle
without hadronic uncertainties!

\begin{figure}
\begin{center}
\includegraphics[width=0.4\textwidth]{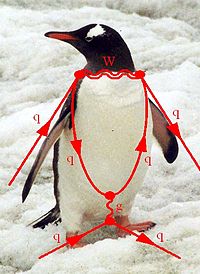}
\end{center}
\caption{\label{fig:penguin}Penguin Feynman diagram. }
\end{figure}

More generally, precisely as in the case of direct CPV we can have several terms contributing to $A_f$, each with different combinations of KM elements:
\begin{align*}
A_f&=aT+bP,\\
\widebar A_f&=a^*T+b^*P,
\end{align*}
where $a$ and $b$ are KM elements and $T$ and $P$ are matrix elements. A word about notation. $T$ stads for ``tree'' because we have in mind a contribution that at the quark level and before dressing up with gluons is a Feynman diagram at tree level. $P$ stands for ``penguin'' and represents a contribution that at the quark level starts at 1-loop. {\it Digression:}
 I do not know why this is called so. I have heard many stories. It was certainly first introduced in the context we are studying. Fig.~\ref{fig:penguin} shows a penguin-like depiction of the diagram. {\it End digression}.
The trick is to find processes where the penguin contribution is expected to be suppressed. Suppose $|P|$=0. Then 
\[
\lambda_f=\frac{q}{p}\frac{a^*}{a}.
\]
This is the same result as above, only emphasizing the hidden assumption.

The most celebrated case is $B\to J/\psi K_S$. Here are the leading diagrams:
\vskip0.2cm
\begin{center}
\begin{tikzpicture}[scale=1.3] 
\coordinate (A) at (0,0);
\coordinate (x) at (1,0);
\coordinate (y) at (0,1);

\coordinate (v1) at ($(A)+0.5*(y)$); 
\coordinate (v2) at  ($(v1)+(x)$);
\coordinate (v3) at  ($(v2)+(x)+0.25*(y)$);
\coordinate (v4) at ($(v3)-0.5*(y)$);
\coordinate (v5) at  ($(A)+(x)$);
\coordinate (v6) at  ($(v4)-0.5*(y)$);
\coordinate (v7) at  ($(v6)-0.5*(y)$);
 
\coordinate (v8) at  ($(A)-0.5*(y)$);

\draw[particle] (v1) -- node[above=1mm]{$b$} (v2);
\draw[particle] (v2) -- node[above=1mm]{$c$} (v3);
\draw[particle] (v4) --node[above right]{$c$}  (v5);
\draw[particle] (v5) --node[below=0.5mm]{$s$}  (v6);
\draw[particle] (v7) .. controls ($(v7)-0.8*(x)+0.2*(y)$) and ($(v8)+(x)$)  .. node[below]{$d$} (v8);

\draw[photon] (v2) -- node[left]{$W$} (v5);

\draw[fill] ($(A)$) let \p1=($(v1)-(A)$) in ellipse ({0.2*veclen(\x1,\y1)} and {veclen(\x1,\y1)}) node[left=0.1cm]{$\widebar{B}^0$} ;
\draw[fill] ($0.5*(v3)+0.5*(v4)$) let \p1= ($0.5*(v3)-0.5*(v4)$) in ellipse ({0.2*veclen(\x1,\y1)} and {veclen(\x1,\y1)}) node[right]{$J/\psi$} ;
\draw[fill] ($0.5*(v6)+0.5*(v7)$) let \p1= ($0.5*(v6)-0.5*(v7)$) in ellipse ({0.2*veclen(\x1,\y1)} and {veclen(\x1,\y1)})  node[right]{$\widebar K^0(K_S)$};
\begin{scope}[xshift=4.5cm]
\coordinate (A) at (0,0);
\coordinate (x) at (1,0);
\coordinate (y) at (0,1);

\coordinate (v1) at ($(A)+0.5*(y)$); 
\coordinate (v2) at  ($(v1)+(x)$);
\coordinate (v3) at  ($(v2)+(x)+0.25*(y)$);
\coordinate (v4) at ($(v3)-0.5*(y)$);
\coordinate (v5) at  ($(A)+(x)$);
\coordinate (v6) at  ($(v4)-0.5*(y)$);
\coordinate (v7) at  ($(v6)-0.5*(y)$);
 
\coordinate (v8) at  ($(A)-0.5*(y)$);

\draw[particle] (v2) -- node[above=1mm]{$b$} (v1);
\draw[particle] (v3) -- node[above=1mm]{$c$} (v2);
\draw[particle] (v5) -- node[above right]{$c$} (v4);
\draw[particle] (v6) -- node[below=0.5mm]{$s$}  (v5);
\draw[particle] (v8) .. controls ($(v8)+(x)$) and ($(v7)-0.8*(x)+0.2*(y)$)  .. node[below=0.5mm]{$d$} (v7);

\draw[photon] (v2) -- node[left]{$W$} (v5);

\draw[fill] ($(A)$) let \p1=($(v1)-(A)$) in ellipse ({0.2*veclen(\x1,\y1)} and {veclen(\x1,\y1)}) node[left=0.1cm]{${B}^0$} ;
\draw[fill] ($0.5*(v3)+0.5*(v4)$) let \p1= ($0.5*(v3)-0.5*(v4)$) in ellipse ({0.2*veclen(\x1,\y1)} and {veclen(\x1,\y1)}) node[right]{$J/\psi$} ;
\draw[fill] ($0.5*(v6)+0.5*(v7)$) let \p1= ($0.5*(v6)-0.5*(v7)$) in ellipse ({0.2*veclen(\x1,\y1)} and {veclen(\x1,\y1)})  node[right]{$ K^0(K_S)$};
\end{scope}

\end{tikzpicture} 
\end{center}
Generally we should write
\[
\frac{\widebar{A}_{\psi K_S}}{A_{\psi K_S}}
= -\frac{(V_{cb}^{\phantom{*}}V_{cs\vphantom{d}}^*)T+(V_{ub}^{\phantom{*}}V_{us\vphantom{d}}^*)P}{(V_{cb}^*V_{cs\vphantom{d}}^{\phantom{*}})T+(V_{ub}^*V_{us\vphantom{d}}^{\phantom{*}})P}\times\frac{V_{cd}^*V_{cs\vphantom{d}}^{\phantom{*}}}{V_{cd}^{\phantom{*}}V^*_{cs\vphantom{d}}}
\]
The novelty here is the last factor which arises from  projecting the $K^0$ and $\widebar{K}^0$ states onto $K_S$. Using Using \eqref{eq:X0from XHL} with $L$ and $S$ for $H$ and $L$, respectively, this is just $-q/p=-V_{cd}^*V_{cs\vphantom{d}}^{\phantom{*}}/V_{cd}^{\phantom{*}}V^*_{cs\vphantom{d}}$. Now in this case the penguin contribution is suppressed by a 1-loop factor relative to the tree level contribution and in addition the KM factor of the penguin contribution is very suppressed relative to that in the tree contribution: counting powers of Wolfenstein's $\lambda$ parameter   $|V_{ub}^{\phantom{*}}V_{us\vphantom{d}}^*|/|V_{cb}^{\phantom{*}}V_{cs\vphantom{d}}^*|
\sim \lambda^2$. Safely neglecting  $P$ we have
\[
\lambda_{\psi K_S}=-e^{-2i\beta}\qquad S_{\psi K_S}=\sin(2\beta),~~C_{\psi K_S}=0
\]
The PDG values are
\[
S_{\psi K_S}=+0.682\pm0.019, \qquad C_{\psi K_S}=(0.5\pm2.0)\times10^{-2}.
\]
The vanishing of $C_{\psi K_S}$ is reassuring, we must know what we are doing!

How about other angles? We can get $\sin(2\alpha)$ from $B\to\pi\pi$ if the penguin can be neglected in 
\[
\frac{\widebar{A}_{\pi\pi}}{A_{\pi\pi}}
= \frac{(V_{ub}^{\phantom{*}}V_{ud}^*)T+(V_{tb}^{\phantom{*}}V_{td}^*)P}{(V_{ub}^*V_{ud}^{\phantom{*}})T+(V_{tb}^*V_{td}^{\phantom{*}})P}
\]
It was realized well before the experiment was performed that the penguin here cannot be expected to be negligible\cite{Gronau:1989ia}. The PDG gives the measured value $C_{\pi^+\pi^-}=-0.31\pm0.05$ confirming this expectation. This can be fixed by determining $P/T$ from an isospin analysis and measurement of several rates and asymmetries \cite{Gronau:1990ka}. But the analysis is difficult and compromises the precision in the determination of $\alpha$. The moral is that you must have a good reason to neglect $P$ before you can claim a clean determination of the angles of the unitarity triangle. 

\newpage

\begin{exercises}
\begin{exercise}
The following table is reproduced from the PDG. \\
\includegraphics[width=\textwidth]{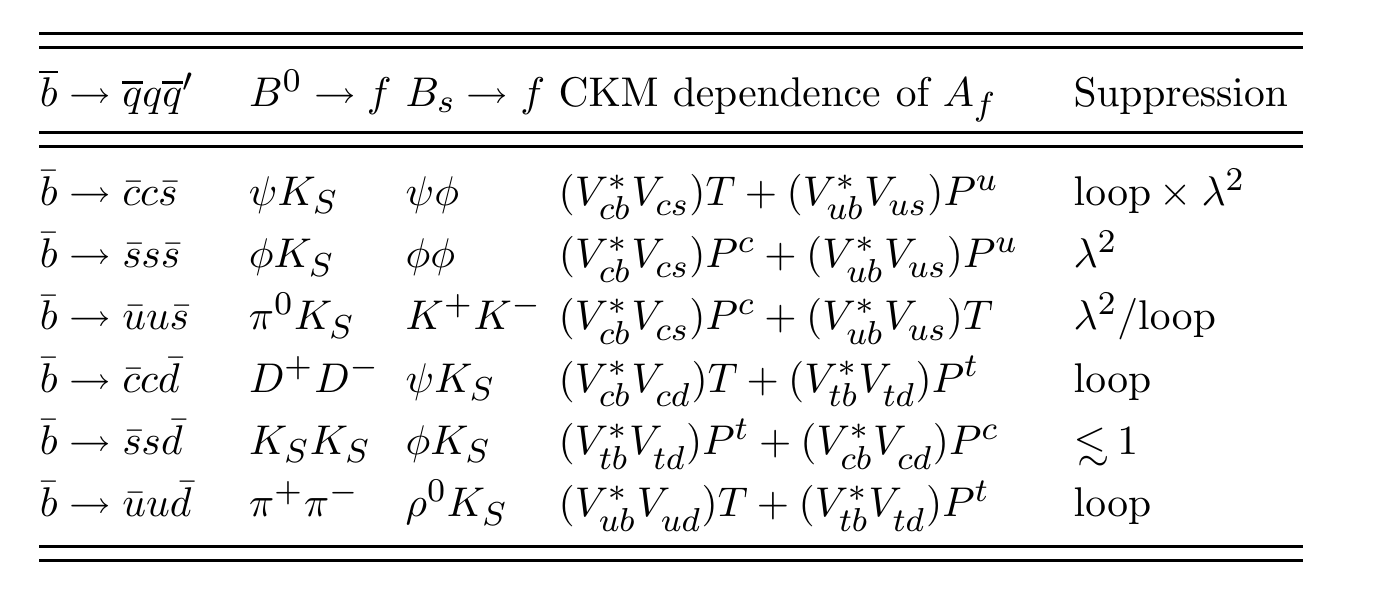}

The columns from left to right give the underlying quark process, the final state in $B^0$ decay, the final state in $B_s$ decay, an expression for the amplitude including KM factors and $T$ or $P$ for whether the  underlying process is tree level or penguin, and lastly, suppression factor of the sub-leading contribution to the amplitude relative to the leading one. Note that in some cases both contributions to the amplitude are from 1-loop diagrams, so they are both labeled $P$. Reproduce the last column (we have done the first line already). Find $S_f$ in each case, assuming you can neglect the suppressed amplitude.     
\end{exercise}
\end{exercises}

\addcontentsline{toc}{chapter}{Bibliography}

\clearpage
\hypertarget{chap:sols}{{\huge Solutions to selected Problems}}
\addcontentsline{toc}{chapter}{Solutions to selected Problems}
\section*{Flavor Theory}
\addcontentsline{toc}{section}{Chapter 1: Flavor Theory}
\par---\newline\textbf{Exercise{} 1.3-1}
Show that this can always be done. That is, that an arbitrary matrix $M$ can be transformed  into a real, positive diagonal matrix $M'=P^\dagger MQ$ by a pair of unitary matrices, $P$ and $Q$.
~\newline
\textbf{Solution{}}
\newline
I'll give you a physicist's proof. If you want to be a mathematician, and use Jordan Normal forms, be my guest. Consider the matrices $M^\dagger M $ and $ M M^\dagger$. They are both hermitian so they can each be diagonalized by a unitary transformation. Moreover, they both obviously have real non-negative  eigenvalues. And they  have the same eigenvalues: using the properties of the determinant you can see that the characteristic polynomial is the same, $\det(M^\dagger M -x)=\det(MM^\dagger-x)$. So we have matrices $P$ and $Q$ such that $P^\dagger ( MM^\dagger) P=Q^\dagger ( M^\dagger M) Q=D=$ real, non-negative, diagonal. We can rewrite $D= P^\dagger ( MM^\dagger) P= ( P^\dagger  M Q)( P^\dagger  M Q)^\dagger= XX^\dagger$, where $X=  P^\dagger  M Q$. Similarly, we also have $D=X^\dagger X$, and multiplying this by X on the left we combine the two results into $XD=DX$. Let's assume all the entries in $D$ are all different and non-vanishing  (I will leave out the special cases, you can feel in the details). Then $DX-XD=0$ means, in components $(D_{ii}-D_{jj})X_{ij}=0$ which means that $X_{ij}=0$ for $j\ne i$. So $X$ is diagonal, with $|X_{ii}| = \sqrt{D_{ii}}$. We can always take $ P^\dagger  M Q=\sqrt{D}=M'$, by further transformation by a diagonal unitary matrix on the  left or right.
\par---\newline\textbf{Exercise{} 1.3-2}
In QED, charge conjugation is $\widebar e\gamma^\mu e\to -\widebar e \gamma^\mu e$ and $A^\mu \to - A^\mu$.  So $\widebar e \slashed {A} e$ is invariant under $C$.\\
So what about QCD? Under charge conjugation  $\widebar q T^a \gamma^\mu q \to \widebar q (-T^a)^T \gamma^\mu q$, but $ (-T^a)^T=(-T^a)^*$ does not equal $-T^a$ (nor $T^a$). So what does charge conjugation mean in QCD? How does the gluon field, $A^a_\mu$, transform?
~\newline
\textbf{Solution{}}
\newline
I have never seen this discussed in a textbook, or elsewhere. Maybe one of the readers will write a nice article for AJP (don't forget to include me!). If you think of the ``transformation arrow'' more properly as the action by a unitary operator on the Hilbert space, $C$,
so that $\widebar e\gamma^\mu e\to -\widebar e \gamma^\mu e$ really means $C(\widebar e\gamma^\mu e)C^{-1}= -\widebar e \gamma^\mu e$, then it is clear that $T^a$ is not changed since it is a $c$-number that commutes with $C$. What we need is
$A_\mu^a T^a\to A_\mu^a (-T^a)^T$. This is accomplished by a transformation $ A_\mu^a\to R^{ab}A_\mu^b$ with a real matrix $R$ that must take $T^a$ into minus its transpose: $R^{ba} T^b = -T^{aT}$. Since the matrices $T^a$ are in the fundamental representation of $SU(3)$ we have $R^{ca}=2\text{Tr}[T^c(R^{ba} T^b)] =-2\text{Tr}( T^c T^{aT})$. $R$ is indeed real:
$(R^{ca})^* =-2\text{Tr}( T^{c*} T^{a\dagger})$, then using  $T^{a\dagger}=T^a$, $\text{Tr}(A^T)=\text{Tr}(A)$ and cyclicity of trace, it follows that $R$ is a real symmetric matrix. Notice that $R^2=1$ for consistency (the negative transpose of the negative transpose is the identity). You can check this using the identity $2T^a_{ij}T^a_{mn}=\delta_{in}\delta_{mj}-\frac13\delta_{ij}\delta_{mn}$.

In physical terms this means that under charge conjugation the, say, blue-antigreen gluon is transformed into minus the green-antiblue gluon, and so on.

I have seen in places  an explanation for charge conjugation in QCD along these lines: first take the quark field $q$ and rewrite in terms of a left- and a right-handed fields, $q_L$ and $q_R$. Then replace $q_R$ by its charge-conjugate, which is also a left-handed field, $q^c_L$. Now $q_L$ is a triplet under color while $q_L^c$ is an antitriplet under color. So charge conjugation is simply $q_L\leftrightarrow q^c_L$. This is incomplete (and therefore wrong). If you were to ignore the transformation of the gluon field the resulting Lagrangian would not be gauge invariant since now the covariant  derivative acting on $q_L$ has a generator for an anti-triplet, $-T^{aT}$, while the covariant derivative acting on $q_L^c$ has generator $T^a$ appropriate for a triplet. It is only after you transform the gluon field that everything works as it should!
\par---\newline\textbf{Exercise{} 1.3-3}
If two entries  in $m_U$ (or in $m_D$) are equal show that $V$ can be brought into a real matrix and hence is an  orthogonal transformation (an element of $O(3)$).
~\newline
\textbf{Solution{}}
\newline
Without loss of generality we may assume the first two entries in
$m_U$ are equal. This means that the remnant freedom to redefine
quark fields without changing neither the kinetic nor the mass terms
is not just by individual phases on all flavors but also by a
$2\times2$ unitary matrix acting on the degenerate quarks. Let
$u_{L,R}\to Uu_{L,R} $, then $U$ is of the form
\[\left(\begin{array}{c|c}A & 0\\ \hline \\[-2.5ex] 0 & e^{i\alpha_3}\end{array}
\right)\]
where $A$ is a $2\times2$ unitary matrix and ``0'' stands for a
2-component zero vector.
Let also $W$ be the diagonal matrix with entries $e^{i\beta_i}$,
$i=1,2,3$, and redefine $d_{L,R}\to W d_{L,R}$. This has the effect of
redefining $V\to U^\dagger VW$. To see what is going on let's write
$V$ in terms of a $2\times2$ submatrix, $X$, two 2-component column
vectors, $\psi$ and $\eta$, and a complex number, $z$:
\[\left(\begin{array}{c|c}X & \psi\\ \hline \\[-2.5ex]\eta^T & z\end{array}.
\right)\]
Then $V$ is transformed into
\begin{equation}
\label{eqsol:vtrans}
V=\left(\begin{array}{c|c}A^\dagger X\begin{pmatrix}e^{i\beta_1}&0\\0& e^{i\beta_2}\end{pmatrix}  & e^{i\beta_3}A^\dagger\psi\\[1.8ex]
    \hline \\[-1.9ex] e^{-i\alpha_3}\eta^T\begin{pmatrix}e^{i\beta_1}&0\\0& e^{i\beta_2}\end{pmatrix} & e^{i(\beta_3-\alpha_3)}z\end{array}
\right).\end{equation}
Now we can choose $A^\dagger$ so that $e^{i\beta_3}A^\dagger\psi$ has
vanishing lower component and real upper component. This still leaves
freedom in $A$ to make a rotation by a phase of the (vanishing) lower
component. So we may take
\[
\psi= \begin{pmatrix}|\psi|\\0\end{pmatrix}\qquad\text{and}\qquad
e^{i\beta_3}A^\dagger=
\begin{pmatrix}1 &0\\0&e^{i(\gamma+\beta_3)}\end{pmatrix}.
\]
At this point it is worth making the trivial observation that for
fixed $\beta_3$ one can make the third row of the new $V$ matrix in
\eqref{eqsol:vtrans} real by choosing $\beta_1,\beta_2$
and~$\alpha_3$. We are left with the $2\times2$ block,
\[
A^\dagger X\begin{pmatrix}e^{i\beta_1}&0\\0& e^{i\beta_2}\end{pmatrix} =\begin{pmatrix}e^{-i\beta_3} &0\\0&e^{i\gamma}\end{pmatrix}X\begin{pmatrix}e^{i\beta_1}&0\\0& e^{i\beta_2}\end{pmatrix}
\]
Now choose $\beta_3$ and $\gamma$ to make real the first column. This
means that the only entries of $V$ with a phase are the top two
entries of the second column, $V_{21}$ and $V_{22}$. But unitarity of
$V$ requires $V_{2i}V_{3i}^*=0$. Since $V_{32}=0$ this can only be
satisfied if $V_{21}$ is real. Then $V_{2i}V_{1i}^*=0$ can only be
satisfied if $V_{22}$ is real. Hence all elements in $V$ are real.
\par---\newline\textbf{Exercise{} 1.3-4}
\label{eqsol:alphabetagamma}
Show that
\begin{enumerate}[(i)]
\item $\displaystyle \beta= \text{arg}\left(- \fracup{V_{cd}^{\phantom{*}}V_{cb}^*}{V_{td}^{\phantom{*}}V_{tb}^*}\right)$,
 $\displaystyle \alpha= \text{arg}\left(- \fracup{V_{td}^{\phantom{*}}V_{tb}^*}{V_{ud}^{\phantom{*}}V_{ub}^*}\right)$ and
 $\displaystyle \gamma= \text{arg}\left(- \fracup{V_{ud}^{\phantom{*}}V_{ub}^*}{V_{cd}^{\phantom{*}}V_{cb}^*}\right)$.
\item These are invariant under phase redefinitions of quark fields (that is, under the remaining arbitrariness). Hence these are candidates for observable quantities.
\item The area of the triangle is $-\frac12\,\text{Im}\frac{\raisebox{0.45ex}{$\scriptstyle V_{ud}^{\phantom{*}}V_{ub}^*$}}{V_{cd}^{\phantom{*}}V_{cb}^*}
=-\frac12\,\frac1{|V_{cd}^{\phantom{*}}V_{cb}^*|^2}\text{Im}\left(V_{ud}^{\phantom{*}}V_{cd}^*V_{cb}^{\phantom{*}}V_{ub}^*\right)$.
\item The product $J=
\text{Im}\left(V_{ud}^{\phantom{*}}V_{cd}^*V_{cb}^{\phantom{*}}V_{ub}^*\right)$ (a ``Jarlskog invariant'') is also invariant under  phase redefinitions of quark fields.
\end{enumerate}
Note that
$\text{Im}\left(V_{ij}^{\phantom{*}}V_{kl}^{\phantom{*}}V_{il}^*V_{kj}^*\right)=J(\delta_{ij}\delta_{kl}-\delta_{il}\delta_{kj})$
is the common area of all the un-normalized triangles. The area of a
normalized triangle is $J$ divided by the square of the magnitude of the side that is normalized to unity.
~\newline
\textbf{Solution{}}
\newline
\begin{enumerate}[(i)]
\item Take the equation that defines the triangle
\[
\frac{V_{ud}^{\phantom{*}}V_{ub}^*}{V_{cd}^{\phantom{*}}V_{cb}^*}+1+\frac{V_{td}^{\phantom{*}}V_{tb}^*}{V_{cd}^{\phantom{*}}V_{cb}^*}=0
\]
and depict it as a triangle in the complex plane:
\begin{center}
\begin{tikzpicture}[scale=1.7,>=triangle 90]
\coordinate (A) at (0,0);
\coordinate (B) at  (4,0);
\coordinate (C) at  (40:3cm) ;

\draw[->,>=stealth,thick] ($(A)+(-0.7,0)$) -- ($(B)+(1,0)$) node[below right]{$\widebar\rho$};
\draw[->,>=stealth,thick] ($(A)+(0,-0.7)$) -- +(0,3.3) node[left]{$\widebar\eta$};

\draw[->,very thick]   (A) -- (B);
\draw[->,very thick] (B) -- node[above right]{$\displaystyle\frac{V_{td}^{\phantom{*}}V_{tb}^*}{V_{cd}^{\phantom{*}}V_{cb}^*}$} (C);
\draw[->,very thick] (C)  node[below=0.9cm]{ $\alpha$} -- node[above left]{$\displaystyle\frac{V_{ud}^{\phantom{*}}V_{ub}^*}{V_{cd}^{\phantom{*}}V_{cb}^*}$}(0,0);

\draw[black]  ($(B)+(155:0.5cm)$)  node[left]{$\beta$};
\draw[->] ($(A)+(0.5,0)$) arc (0:40:0.5cm)  ;
\draw  ($(A)+(20:0.5cm)$)  node[right]{$\gamma$};
\draw[->] ($(C)+(220:0.5)$) arc (220:310:0.5);
\draw[->] ($(B)+(130:0.5)$) arc (130:180:0.5);
\end{tikzpicture}
\end{center}
Note that the vector from the origin to $(\bar\rho,\bar\eta)$ is the opposite of $\frac{V_{ud}^{\phantom{*}}V_{ub}^*}{V_{cd}^{\phantom{*}}V_{cb}^*}$, so the angle $\gamma$ is the argument of minus this, $ \gamma= \text{arg}\left(- \frac{V_{ud}^{\phantom{*}}V_{ub}^*}{V_{cd}^{\phantom{*}}V_{cb}^*}\right)$. Next, the angle that $\frac{V_{td}^{\phantom{*}}V_{tb}^*}{V_{cd}^{\phantom{*}}V_{cb}^*}$ makes with the $\bar\rho$ axis is $\pi-\beta= \text{arg}\left( \frac{V_{td}^{\phantom{*}}V_{tb}^*}{V_{cd}^{\phantom{*}}V_{cb}^*}\right)$, from which $\beta= \text{arg}\left(- \frac{V_{cd}^{\phantom{*}}V_{cb}^*}{V_{td}^{\phantom{*}}V_{tb}^*}\right)$ follows. $\alpha$ is most easily obtained from $\alpha+\beta+\gamma=\pi$ using the two previous results and the fact that $\text{arg}(z_1)+\text{arg}(z_2)=\text{arg}(z_1z_2)$.
\item In the numerator or denominator of these expressions, the re-phasing of the charge-$+\frac23$ quarks cancel; for example, $V_{td}^{\phantom{*}}V_{tb}^*\to
(e^{i\phi}V_{td}^{\phantom{*}})(e^{i\phi}V_{tb}^{\phantom{*}})^*=V_{td}^{\phantom{*}}V_{tb}^*$. The re-phasing of the charge-$-\frac13$ quarks cancel between numerator and denominator; for example, for the $d$ quark  $\frac{V_{ud}^{\phantom{*}}V_{ub}^*}{V_{cd}^{\phantom{*}}V_{cb}^*}\to \frac{e^{i\phi}V_{ud}^{\phantom{*}}V_{ub}^*}{e^{i\phi}V_{cd}^{\phantom{*}}V_{cb}^*}=\frac{V_{ud}^{\phantom{*}}V_{ub}^*}{V_{cd}^{\phantom{*}}V_{cb}^*}$.
\item From question (i) we see that $\bar\eta=-\text{Im}\frac{V_{ud}^{\phantom{*}}V_{ub}^*}{V_{cd}^{\phantom{*}}V_{cb}^*}$, and this is the height of the triangle of unit base. The area is $1/2$ base time height from which the first result follows. The second expression is obtained from the first by multiplying  by
$1=\frac{V_{cd}^*V_{cb}^{\phantom{*}}}{V_{cd}^*V_{cb}^{\phantom{*}}}$.
\item In $J=
\text{Im}\left(V_{ud}^{\phantom{*}}V_{cd}^*V_{cb}^{\phantom{*}}V_{ub}^*\right)$ each $V_{ix}$ appears with one, and only one, other factor of $V_{iy}^*$, and one, and only one, factor of $V_{jx}^*$.
\end{enumerate}
\par---\newline\textbf{Exercise{} 1.3-5}
\begin{enumerate}[(i)]
\item Show that  \[
\widebar\rho+i\widebar \eta = - \fracup{V_{ud}^{\phantom{*}}V_{ub}^*}{V_{cd}^{\phantom{*}}V_{cb}^*},
\]
hence $\widebar\rho$ and $\widebar \eta$ are indeed the coordinates of the apex of the unitarity triangle and are invariant under quark phase redefinitions.
\item Expand in $\lambda\ll1$ to show
\[
V=\begin{pmatrix} 1-\tfrac12\lambda^2 &\lambda& A\lambda^3(\rho-i\eta)\\ -\lambda&  1-\tfrac12\lambda^2 & A\lambda^2\\
A\lambda^3(1-\rho-i\eta)& - A\lambda^2 &1\end{pmatrix} +\mathcal{O}(\lambda^4)
\]

\end{enumerate}
~\newline
\textbf{Solution{}}
\newline
\begin{enumerate}[(i)]
\item  We are not looking for a graphical representation solution, as was done in  Exercise \ref{eqsol:alphabetagamma}. Instead, we want to show this form the definitions of the parameters $\lambda, A, \bar\rho$ and~$\eta$.  This is just plug in and go. First,
\[V=
\begin{pmatrix}
 c_{12} c_{13} & c_{13} s_{12} & s_{13}e^{-i\delta} \\
 -c_{12} s_{23} s_{13}e^{i\delta}-c_{23}
   s_{12} & c_{12} c_{23}-s_{12} s_{23}
   s_{13}e^{i\delta} & c_{13} s_{23} \\
 s_{12} s_{23}-c_{12} c_{23} s_{13}e^{i\delta}
   & -c_{12} s_{23}-c_{23} s_{12}
   s_{13}e^{i\delta} & c_{13} c_{23}
\end{pmatrix}.
\]
You can break the computation into
  smaller steps. For example,
  $V_{ud}=c_{12}c_{13}=\sqrt{1-\lambda^2}c_{13}$ and
  $V_{c,b}=c_{13}s_{23}=A\lambda^2 c_ {13}$ so that
\[ \frac{V_{ud}}{V^*_{cb}}=\frac{\sqrt{1-\lambda^2}}{A\lambda^2}.\]
Similarly,
\[ \frac{V^*_{ub}}{V_{cd}}=-\frac{A\lambda^2\sqrt{1-A^2\lambda^4}z}{\sqrt{(1-\lambda^2)(1-A^2\lambda^4)}},\]
where $z=\bar\rho+i\bar\eta$. The result follows.
\item Again plug in and go. But you can be clever about it. For
  example, since $s_{13}\sim\lambda^3$, we have
  $c_{13}=\sqrt{1-s_{13}^2}=1+\mathcal{O}(\lambda^6)$. Similarly
  $c_{23}=1+\mathcal{O}(\lambda^4)$ and
  $c_{12}=1-\frac12\lambda^2+\mathcal{O}(\lambda^4)$.  Plugging these,
  and \eqref{eq:Wpar} into the explicit form of $V$ above the result follows.
  and the
\end{enumerate}
\par---\newline\textbf{Exercise{} 1.4.1-1}
Show  that $q\cdot (V-A)\sim m_\ell$ for the leptonic charged
current. Be more
precise than ``$\sim$.''
~\newline
\par---\newline\textbf{Exercise{} 1.4.1-2}
For $B\to D\ell\nu$ write the form factors $f_\pm(q^2)$ in terms of the Isgur-Wise function. What does $\xi(1)=1$ imply for $f_\pm$? Eliminate the Isgur-Wise function to obtain a relation between $f_+$ and $f_-$.
~\newline
\textbf{Solution{}}
\newline
In
\[
\vev{\pvec{v}'|V^\mu|\pvec v}=\xi(v\cdot v')(v+v')^\mu
\]
we need to (i) write $v=p/m_b$ and $v'=p'/m_c$ and (ii) replace $|\pvec v\rangle\to (1/\sqrt{m_b})|\pvec p\rangle $ and $|\pvec{v}'\rangle\to (1/\sqrt{m_c})|\pvec{p}'\rangle $, thus:
\[
\frac1{\sqrt{m_bm_c}}\vev{\pvec{p}'|V^\mu|\pvec p}=\xi(v\cdot
v')\left(\frac{p^\mu}{m_b}+\frac{p^{\prime\mu}}{m_c}\right)
\]
Comparing with \eqref{eq:ffsdefd}, we read off
\[
f_\pm(q^2)=\frac12\sqrt{m_bm_c}\left(\frac1{m_b}\pm\frac1{m_c}\right)\xi((q^2-m_b^2-m_c^2)/2m_bm_c).
\]
The relation between form factors is $f_-/f_+=(m_c-m_b)/(m_c+m_b)$
and we note that this correctly gives $f_-=0$ when the two quarks are
identical. Finally, at $q^2=q^2_{\rm max}$ we have
\[
f_\pm(q_{\rm max}^2)=\frac12\sqrt{m_bm_c}\left(\frac1{m_b}\pm\frac1{m_c}\right).
\]
\par---\newline\textbf{Exercise{} 1.5-1}
Just in case you have never computed the $\mu$-lifetime, verify that
\[\tau^{-1}_\mu\approx\Gamma(\mu\to e\nu_\mu\widebar\nu_e) = \frac{G_F^2m_\mu^5}{192\pi^3}\]
neglecting $m_e$, at lowest order in perturbation theory.
~\newline
\par---\newline\textbf{Exercise{} 1.5-2}
Compute the amplitude for $Z\to b\widebar s$ in the SM to lowest
order in perturbation theory (in the strong and electroweak
couplings).  Don't bother to compute integrals explicitly, just make
sure they are finite (so you could evaluate them numerically if need
be). Of course, if you can express the result in closed analytic form,
you should. See~Ref.~\cite{Clements:1982mk}.
~\newline
\par---\newline\textbf{Exercise{} 1.6.2-1}
Consider $s\to d\gamma$. Show that the above type of analysis suggests that virtual top quark exchange no longer dominates, but that in fact the charm and top contributions are roughly equally important. {\it Note: For this you need to know the mass of charm relative to $M_W$. If you don't, look it up!}
~\newline
\textbf{Solution{}}
\newline
For  $s\to d\gamma$ we now have\\[0.5cm]
\parbox[c]{6cm}{\begin{tikzpicture}[scale=0.7]
\coordinate[label=left:$s$] (b) at (-3,0);
\coordinate[label=right:$d$] (s) at (3,0);
\coordinate (v1) at (-1,0);
\coordinate (v2) at (1,0);
\coordinate (v3) at (0,-1);
\coordinate (g) at ($(v3)+(2,-0.3)$);
\draw[particle] (b) -- (v1);
\draw[particle]  (v1) arc (180:270:1) node[left=0.7cm]{$u,c,t$};
\draw[particle] (v3) arc (270:360:1);
\draw[particle] (v2) -- (s);
\draw[photon] (v3) --  (g) node[right] {$\gamma(q,\epsilon)$};
\draw[photon] (v1) --  node[above]{$W$} (v2);
\end{tikzpicture}
}
$\displaystyle = e q_\mu\epsilon_\nu\widebar u(p_d)\sigma^{\mu\nu}{\textstyle \left(\frac{1+\gamma_5}{2}\right)}u(p_d)\frac{m_s}{M_W^2}\,\frac{g_2^2}{16\pi^2}\cdot I$\\
where
\[I=\sum_{i=u,c,t}V_{is}^{\phantom{*}}V_{id}^*F({\textstyle\frac{m_i^2}{M_W^2}})
 \]
We still have
\[
I=F({\textstyle\frac{m_t^2}{M_W^2}})V_{ts}^{\phantom{*}}V_{td}^*
+F'(0)\sum_{i=u,c} V_{is}^{\phantom{*}}V_{id}^*\frac{m_i^2}{M_W^2}+\cdots
\]
But now the counting of powers of $\epsilon$ is a bit different: $|V_{ts}^{\phantom{*}}V_{td}^*|\sim \epsilon^5$ while $|V_{is}^{\phantom{*}}V_{id}^*|\sim\epsilon$ for either $i=c$ or $i=u$. Since $m_u\ll m_c$ we neglect the $u$-quark contribution. Using $F\sim1$ at the top, then the ratio of top to charm contributions is $\sim \epsilon^5/(\epsilon m_c^2/M_W^2)=(\epsilon^2 M_W/m_c)^2$. Using $M_W/m_c\approx 80/1.5$ and $\epsilon\approx 0.1$ the ratio os 0.3, and we have every right to expect the two contributions are comparable in magnitude.
\par---\newline\textbf{Exercise{} 1.7.1-1}
Had we considered an operator like $O_1$ but with $\Htilde \widebar d_R$ instead of $ H \widebar u_R$ the flavor off-diagonal terms would have been governed by $\lambda'_D V^\dagger$.
Show this is generally true, that is, that flavor change in any operator is governed by $V$ and powers of $\lambda'$.
~\newline
\textbf{Solution{}}
\newline
In any operator use the inverse of \eqref{eq:massDtransf} to write $\lambda_{U,D}$ in terms of $\lambda'_{U,D}$ and the matrices $V_{u_{L,R}}$ and $V_{d_{L,R}}$. Now rotate quarks to go to the mass-diagonal basis. This would be a flavor symmetry transformation if $V_{u_{L}}=V_{d_{L}}=U_q$, so it fails to be a symmetry only because $V=V_{u_{L}}^\dagger V_{d_{L}}^{\phantom{\dagger}}\ne1$, which may appear in these operators. This is the only parameter that is off-diagonal in flavor space.
\par---\newline\textbf{Exercise{} 1.7.1-2}
Exhibit examples of operators of dimension 6 that produce flavor
change without involving $\lambda_{U,D}$. Can these be such that only
quarks of charge $+2/3$ are involved? (These would correspond to Flavor
Changing Neutral Currents; see Sec.~\ref{sec:fcnc} below).
~\newline
\textbf{Solution{}}
\newline
The question is phrased loosely: the answer depends on whether we impose the flavor symmetry \eqref{eq:mvf-trans}. If we don't, then we can simply take an operator like $O_1$ but without the spurion $\lambda_U$ sandwiched between quarks. So, for example, the operator $G^a_{\mu\nu} H \widebar u_R T^a\sigma^{\mu\nu}\kappa q_L$, where $\kappa$ is some arbitrary matrix, when expressed in the mass eigenstate basis gives
\[
 G^a_{\mu\nu} H \widebar u_R T^a\sigma^{\mu\nu}  V_{u_R}^\dagger\kappa\begin{pmatrix}V_{u_L}u_L\\V_{d_L}d_L\end{pmatrix}
\]
Consider, instead, the case when we insist on the symmetry \eqref{eq:mvf-trans}. Now quark bilinears can only be of one SM-representation with itself, $\widebar q_L\gamma^\mu q_L$,  $\widebar q_L\tau^j \gamma^\mu q_L$,  $\widebar u_R\gamma^\mu u_R$  and $\widebar d_R\gamma^\mu d_R$. Of these, only  $\widebar q_L\tau^j \gamma^\mu q_L$ fails to be invariant under the transformation that takes the quarks to the mass eigenstate basis, and then only the terms involving $\tau^\pm$.  So, in the absence of factors of $\lambda_{U,D}$ we can only get charge changing flavor changing interactions. A simple example is the four quark operator $\widebar q_L\tau^j \gamma^\mu q_L\,\widebar q_L\tau^j \gamma_\mu q_L$.
\par---\newline\textbf{Exercise{} 1.7.1-3}
  Determine how much each of the bounds in
  Fig.~\ref{fig:neubertEPS2011} is weakened if you assume MFV. You may
  not be able to complete this problem if you do not have some idea of
  what the symbols $\Delta M_K$, $\epsilon_K$, etc, mean or what type
  of operators contribute to each process; in that case you should
  postpone this exercise until that material has been covered later in these
  lectures.
~\newline
\par---\newline\textbf{Exercise{} 1.7.2-1}
\protect\label{ex:susy-bkg}Below Eq.~\protect\eqref{eq:SUSYbkg} we said, ``This breaks the flavor
symmetry unless $\mathcal{M}^2_{q,u,d}\propto\mathbf{1}$ and
$g_{U,D}\propto y_{U,D}$.'' This is not strictly correct (or, more
bluntly, it is a lie). While not correct it is the simplest
choice. Why? Exhibit alternatives, that is, other forms for $\mathcal{M}^2_{q,u,d}$ and
$g_{U,D}$ that respect the symmetry. {\it Hint: See \protect\eqref{eq:lambdacube}.}
~\newline
\textbf{Solution{}}
\newline
Flavor symmetry requires that $\mathcal{M}^2_q\to U_q\mathcal{M}^2_q U_q^\dagger$, $\mathcal{M}^2_u\to S_U\mathcal{M}^2_u S_U^\dagger$,  $\mathcal{M}^2_d\to S_D\mathcal{M}^2_d S_D^\dagger$, $g_U\to S_U^*g_U U_q^\dagger$ and $y_D\to S_D^*y_D U_q^\dagger$.
\par---\newline\textbf{Exercise{} 1.7.2-2}
Classify all possible dim-4 interactions of Yukawa form in the SM. To
this end list all possible Lorentz scalar combinations you can form
out of pairs of SM quark fields. Then give explicitly the
transformation properties of the scalar field, under the gauge and
flavor symmetry groups, required to make the Yukawa interaction
invariant. Do this first without including the SM Yukawa couplings as
spurions  and then including also one power of the SM Yukawa couplings.
~\newline
\section*{Neutral Meson Mixing and CP Asymmetries}
\addcontentsline{toc}{section}{Chapter 2: Neutral Meson Mixing and CP Asymmetries}
\par---\newline\textbf{Exercise{} 2.3-1}
Show that CPT implies $H_{11}=H_{22}$.
~\newline
\textbf{Solution{}}
\newline
Let $\Omega=CPT$. We have to be a bit careful in that this is an
anti-unitary operator. The bra-ket notation is somewhat confusing for
anti-linear operators, so we use old fashioned inner product notation
$(\psi,\eta)$ for $\langle\psi|\eta\rangle$. Anti-unitarity means
$(\Omega\psi,\Omega\eta)=(\eta,\omega)$, and anti-linearity means
$\Omega(a\psi+b\eta)= a^*\psi+b^*\eta$, where $a,b$ are constants and
$\psi,\eta$ wave-functions.  Now, the CPT theorem gives $\Omega H
\Omega^{-1}=H^\dagger$. So
\begin{align*}
(\psi,H\eta)&=(\psi,H\Omega^{-1}\Omega\eta) && \\
&=(\Omega H\Omega^{-1}\Omega\eta, \Omega\psi) && \text{by
  anti-unitarity of $\Omega$}\\
&=( H^\dagger\Omega\eta, \Omega\psi) && \text{by CPT theorem}\\
&=( \Omega\eta, H \Omega\psi) && \text{by definition of adjoint of operator}
\end{align*}
The action of $\Omega$ on the one particle states at rest is just like
that of CP, $\Omega | X^0\rangle = -| \widebar X^0\rangle$ and
$\Omega| \widebar X^0\rangle = -| X^0\rangle $. So taking $\psi$ and
$\eta$ above to be $| \widebar X^0\rangle $, we have
$H_{22}=(\psi,H\eta)=( \Omega\eta, H \Omega\psi)=H_{11}$.
Note that for $\psi=| X^0\rangle $ and $\eta= | \widebar X^0\rangle$
the same relation gives $H_{12}=H_{12}$.
\par---\newline\textbf{Exercise{} 2.4.1-1}
Challenge: Can you check the other three mixing “bounds” in
Fig.~\ref{fig:neubertEPS2011} (assuming the SM gives about the right result).
~\newline
\par---\newline\textbf{Exercise{} 2.5.2-1}
With these assumptions show
\[\delta =\frac{|1+\epsilon|^2-|1-\epsilon|^2}{|1+\epsilon|^2+|1-\epsilon|^2}\approx2\text{Re}\epsilon
\]
~\newline
\par---\newline\textbf{Exercise{} 2.5.2-2}
Use $\Gamma(K^0(t)\to\pi^-e^+\nu)\propto|\langle
\pi^-e^+\nu|H_W|K^0(t)\rangle|^2$ and the assumptions that
\begin{enumerate}[(i)]
\item $\langle
\pi^-e^+\nu|H_W|\widebar{K}^0(t)\rangle=0=\langle
\pi^+e^-\nu|H_W|{K}^0(t)\rangle$
\item $\langle\pi^-e^+\nu|H_W|K^0(t)\rangle=\langle
\pi^+e^-\nu|H_W|\widebar{K}^0(t)\rangle$
\end{enumerate}
to show that
\[
\delta(t)=\frac{(N_{K^0}-N_{\widebar{K}^0})\left[|f_+(t)|^2-|f_-(t)|^2\tfrac12\left(\left|\frac{q}{p}\right|^2+
      \left|\frac{p}{q}\right|^2\right)\right]
+\tfrac12 (N_{K^0}+N_{\widebar{K}^0})|f_-(t)|^2\left(\left|\frac{p}{q}\right|^2-
                         \left|\frac{q}{p}\right|^2\right)}%
{(N_{K^0}+N_{\widebar{K}^0})\left[|f_+(t)|^2+|f_-(t)|^2\tfrac12\left(\left|\frac{q}{p}\right|^2+
      \left|\frac{p}{q}\right|^2\right)\right]
-\tfrac12 (N_{K^0}-N_{\widebar{K}^0})|f_-(t)|^2\left(\left|\frac{p}{q}\right|^2-
                         \left|\frac{q}{p}\right|^2\right)}
\]
Justify assumptions (i) and (ii).
~\newline
\par---\newline\textbf{Exercise{} 2.6-1}
\label{ex:CPTeigenstates}
If $f$ is an eigenstate of the strong interactions, show that CPT
implies $|A_f|^2=|\widebar A_{\overline f}|^2$ and $|A_{\overline f} |^2=|\widebar A_{ f}|^2$
~\newline
\par---\newline\textbf{Exercise{} 2.6.2-1}
The following table is reproduced from the PDG. \\
\includegraphics[width=\textwidth]{pdg-table.pdf}

The columns from left to right give the underlying quark process, the final state in $B^0$ decay, the final state in $B_s$ decay, an expression for the amplitude including KM factors and $T$ or $P$ for whether the  underlying process is tree level or penguin, and lastly, suppression factor of the sub-leading contribution to the amplitude relative to the leading one. Note that in some cases both contributions to the amplitude are from 1-loop diagrams, so they are both labeled $P$. Reproduce the last column (we have done the first line already). Find $S_f$ in each case, assuming you can neglect the suppressed amplitude.
~\newline



\begin{thebibliography}{99}
\bibitem{ckmfitter}
CKMfitter Group (J. Charles et al.), Eur. Phys. J. C41, 1-131 (2005) [hep-ph/0406184], updated results and plots available at: \href{http://ckmfitter.in2p3.fr}{http://ckmfitter.in2p3.fr}

\bibitem{pdg}
K.A. Olive et al. (Particle Data Group), Chin. Phys. C, {\bf }, 090001 (2014). 

\bibitem{Grinstein:1992ss}
  B.~Grinstein,
  Ann.\ Rev.\ Nucl.\ Part.\ Sci.\  {\bf 42} (1992) 101.

\bibitem{Luke:1990eg} 
  M.~E.~Luke,
  Phys.\ Lett.\ B {\bf 252}, 447 (1990).


\bibitem{Isgur:1988gb} 
  N.~Isgur, D.~Scora, B.~Grinstein and M.~B.~Wise,
  Phys.\ Rev.\ D {\bf 39}, 799 (1989).

\bibitem{Chay:1990da} 
  J.~Chay, H.~Georgi and B.~Grinstein,
  Phys.\ Lett.\ B {\bf 247}, 399 (1990).

\bibitem{Manohar:1993qn} 
  A.~V.~Manohar and M.~B.~Wise,
  Phys.\ Rev.\ D {\bf 49}, 1310 (1994)
  [hep-ph/9308246].

\bibitem{Bigi:1992su} 
  I.~I.~Y.~Bigi, N.~G.~Uraltsev and A.~I.~Vainshtein,
  Phys.\ Lett.\ B {\bf 293}, 430 (1992)
  [Erratum-ibid.\ B {\bf 297}, 477 (1993)]
  [hep-ph/9207214].

\bibitem{Bauer:2004ve} 
  C.~W.~Bauer, Z.~Ligeti, M.~Luke, A.~V.~Manohar and M.~Trott,
  Phys.\ Rev.\ D {\bf 70}, 094017 (2004)
  [hep-ph/0408002].

\bibitem{Clements:1982mk} 
  M.~Clements, C.~Footman, A.~S.~Kronfeld, S.~Narasimhan and D.~Photiadis,
  Phys.\ Rev.\ D {\bf 27}, 570 (1983).

\bibitem{Glashow:1970gm} 
  S.~L.~Glashow, J.~Iliopoulos and L.~Maiani,
  Phys.\ Rev.\ D {\bf 2}, 1285 (1970).


\bibitem{Dine:1993yw}
  M.~Dine and A.~E.~Nelson,
  Phys.\ Rev.\ D {\bf 48} (1993) 1277
  [hep-ph/9303230].
  M.~Dine, A.~E.~Nelson and Y.~Shirman,
  Phys.\ Rev.\ D {\bf 51} (1995) 1362
  [hep-ph/9408384].
  M.~Dine, A.~E.~Nelson, Y.~Nir and Y.~Shirman,
  Phys.\ Rev.\ D {\bf 53} (1996) 2658
  [hep-ph/9507378].

\bibitem{Aaltonen:2011kc} 
  T.~Aaltonen {\it et al.}  [CDF Collaboration],
  Phys.\ Rev.\ D {\bf 83}, 112003 (2011)
  [arXiv:1101.0034 [hep-ex]].
  V.~M.~Abazov {\it et al.}  [D0 Collaboration],
  Phys.\ Rev.\ D {\bf 84}, 112005 (2011)
  [arXiv:1107.4995 [hep-ex]].

\bibitem{Kuhn:2011ri} 
  J.~H.~Kuhn and G.~Rodrigo,
  JHEP {\bf 1201}, 063 (2012)
  [arXiv:1109.6830 [hep-ph]].
  W.~Hollik and D.~Pagani,
  Phys.\ Rev.\ D {\bf 84}, 093003 (2011)
  [arXiv:1107.2606 [hep-ph]].

\bibitem{Leone:2014gwa} 
  S.~Leone [CDF and D0 Collaboration],
  Nuovo Cim.\ C {\bf 037}, no. 02, 40 (2014).

\bibitem{Grinstein:2011yv} 
  B.~Grinstein, A.~L.~Kagan, M.~Trott and J.~Zupan,
  Phys.\ Rev.\ Lett.\  {\bf 107}, 012002 (2011)
  [arXiv:1102.3374 [hep-ph]].
{\it idem},   JHEP {\bf 1110}, 072 (2011)
  [arXiv:1108.4027 [hep-ph]].

\bibitem{Venditti:2008zz} 
  S.~Venditti [NA62 Collaboration],
  Nuovo Cim.\ B {\bf 123}, 844 (2008).
  M.~Akashi-Ronquest [KTeV Collaboration],
  arXiv:1003.5574 [hep-ex].
  A.~V.~Artamonov {\it et al.}  [BNL-E949 Collaboration],
  Phys.\ Rev.\ D {\bf 79}, 092004 (2009)
  [arXiv:0903.0030 [hep-ex]].
 T.~Abe {\it et al.}  [Belle-II Collaboration],
  arXiv:1011.0352 [physics.ins-det].


\bibitem{Wandernoth:2013dda} 
  S.~Wandernoth,
  ``Precision measurement of the oscillation frequency in the $B_{s}^{0}-\bar{B}_{s}^{0}$  system,''
 Proceedings of the 48th Rencontres de Moriond on Electroweak Interactions and Unified Theories. La Thuile, Italy, March 2-9, 2013,






\bibitem{Huet:1994kr} 
  P.~Huet and M.~E.~Peskin,
  Nucl.\ Phys.\ B {\bf 434}, 3 (1995)
  [hep-ph/9403257].

\bibitem{Messiah}
A.~Messiah, \href{https://archive.org/details/QuantumMechanicsVolumeIi}{{\it Quantum Mechanics}, Volume II, North Holland Publishing Company, 1965.}

\bibitem{Gjesdal:1974yj} 
  S.~Gjesdal, G.~Presser, T.~Kamae, P.~Steffen, J.~Steinberger, F.~Vannucci, H.~Wahl and F.~Eisele {\it et al.},
  Phys.\ Lett.\ B {\bf 52}, 113 (1974).

\bibitem{Bigi:1981qs} 
  I.~I.~Y.~Bigi and A.~I.~Sanda,
  Nucl.\ Phys.\ B {\bf 193}, 85 (1981).

\bibitem{Gronau:1989ia} 
  M.~Gronau,
  Phys.\ Rev.\ Lett.\  {\bf 63}, 1451 (1989).
  D.~London and R.~D.~Peccei,
  Phys.\ Lett.\ B {\bf 223}, 257 (1989).
  B.~Grinstein,
  Phys.\ Lett.\ B {\bf 229}, 280 (1989).

\bibitem{Gronau:1990ka} 
  M.~Gronau and D.~London,
  Phys.\ Rev.\ Lett.\  {\bf 65}, 3381 (1990).







\end{thebibliography}
\end{document}